\documentclass[prb,twocolumn,amsmath,amssymb,floatfix,nofootinbib,bibnotes,longbibliography,]{revtex4-2}

\pdfoutput=1

\usepackage[colorlinks=true,citecolor=blue,linkcolor=magenta]{hyperref}
\usepackage{multirow}
\usepackage{amsmath}
\usepackage{bbm} 
\usepackage{graphicx}
\usepackage{dsfont}
\usepackage{changepage}
\usepackage{hyperref}
\usepackage{fancyhdr}
\usepackage{amsthm, amssymb}
 \usepackage{floatflt}
 \usepackage{float}
\usepackage{array}
\newcolumntype{P}[1]{>{\centering\arraybackslash}p{#1}}
\usepackage[usenames,dvipsnames]{color}
\usepackage[section]{placeins}
\usepackage[T1]{fontenc}
\usepackage[active]{srcltx}
\usepackage{color}
\usepackage{esint}
\usepackage[version=4]{mhchem}
\usepackage[normalem]{ulem}

\usepackage{mathtools}

\def\bm{\boldsymbol}

\newcommand {\apgt} {\ {\raise-.5ex\hbox{$\buildrel>\over\sim$}}\ }
\newcommand {\aplt} {\ {\raise-.5ex\hbox{$\buildrel<\over\sim$}}\ }

\def\titlename{Phase-Space Approach to Wannier Pairing and Bogoliubov Orbitals in Square-Octagon Lattices}
\def \authornames{Rajesh O. Sharma, and Tanmoy Das}
\def \affiliations{Department of Physics, Indian Institute of Science, Bangalore 560012, India}

\begin{document}

\title{\titlename}
\author{\authornames}
\affiliation{\affiliations}

\begin{abstract}
Low-energy lattice models are the cornerstone for studying many-body physics and interactions between the system and measurement fields. A key challenge is identifying appropriate quasiparticle states that canonically transform between momentum and real space while retaining the correlation, entanglement, and geometric properties — generally called the Wannier obstruction. Here, we introduce a phase-space approach to bypass these obstructions. Instead of treating the phase space as a manifold, we embed a real space through a Bloch vector space at each momentum. Orbital and spin states are introduced through product states with the Bloch vector, while quantum statistics, correlations, topology, and entanglements are inherited from the Hamiltonian. We apply this framework to explore the unconventional pairing symmetry and the Bogoliubov-de Gennes (BdG) equation in phase space. Our findings demonstrate that while superconductivity exhibits global coherence, the local Wannier orbital symmetry primarily determines the pairing symmetry. We analytically solve the spin-fluctuation-mediated pairing symmetry on the phase space by engineering a flat band with artificial gauge fields. We validate the model on the square-octagon superconductor Lu$_2$Fe$_3$Si$_5$ using density functional theory (DFT), revealing the coexistence of nodeless $s^{\pm}$ and nodal $s_{z^2}$ pairing symmetries. This phase-space framework provides a robust, obstruction-free lattice model for complex many-body systems and their exotic excitations. 
\end{abstract}

\keywords{Wannier Orbitals, Phase-space framework, Unconventional superconductivity, Square-octagon superconductors}

\maketitle

\section{Introduction}
Constructing low-energy lattice models for correlated quantum systems remains a major challenge. While tight-binding and Wannier orbital models, despite certain obstructions, work well for quasiparticle states\cite{Marzari2012, Marzari1997, Souza2001, Soluyanov2011, Wu2012, Winkler2016, Gresch2017, Cornean2017, Po2018}, they encounter difficulties for fractional or entangled states in intermediate to strongly correlated systems.\cite{Anisimov2005, Lechermann2006, Das2014, Freimuth2023, Vollhardt2020}  In such cases, long-wavelength field theories near critical points are often used, but they are prone to ultraviolet and infrared divergences. A lattice theory, which is free from these obstructions and divergences, would be invaluable for describing exotic excitations in many-body ground states $-$ such as magnetic order, spin liquids, fractional quantum Hall, or superconducting (SC) states $-$ and for evaluating the matrix-element effects of the system-measurement couplings.

In a lattice model of quasiparticle excitations, linearly independent Wannier states must exist for each unit cell, matching the number of Bloch states in momentum space and connected via Fourier transformation. However, global topology or degeneracies can obstruct this transformation.\cite{Soluyanov2011, Winkler2016, Gresch2017, Cornean2017, Po2018, Boyack2024, Read2017} For exotic excitations, additional challenges arise to obtain canonical transformation between position and momentum space for properties such as quantum statistics, fractional quantum numbers, correlations, and geometric properties.\cite{Wu2012,Schindler2020, Wahl2013, Qi2011, Nakagawa2020, Gupta2022, Li2024, Wang2024}. 

The conventional variational method works by constructing a many-body ground state to serve as the vacuum state for low-energy excitations defined in either real or momentum space.  Here, we extend this approach to the position-momentum \textit{phase space}. Rather than treating phase space as a manifold, we embed position space into a Bloch vector space at each momentum (or vice versa) and construct a variational wavefunction as a product of a Bloch vector and orbital-like excitations. Quantum statistics, correlations, and geometric properties are introduced via projection operators, while non-local unitary transformations generate entanglement between particle-hole pairs, spin-orbit/spin-momentum locking, and/or topology. Operators, Hamiltonians, order parameters, and mean-field theories are represented in phase space, while expectation values are transformed into momentum or real space as desired. 

We apply this phase-space framework to study mean-field superconductivity and spin-fluctuation-mediated unconventional pairing symmetries. Spin-fluctuation theory and pairing eigenvalues are derived in phase space, allowing us to solve the pairing problem analytically. Our results show that while superconductivity exhibits global coherence in momentum space, the local lattice symmetry constrained Wannier orbital dictates the pairing symmetry. By expressing the Bogoliubov-de Gennes (BdG) Hamiltonian in the irreducible representation (irrep) of the point group symmetry, we find that only the Wannier orbital sharing the same symmetry as the pairing order parameter undergoes fractionalization—splitting into two Bogoliubov orbitals per unit cell—while all other orbitals remain intact.  

As a case study, we examine a square-octagon (SO) lattice. We first study a flat-band scenario in a tight-binding model with artificial gauge fields. Wannier orbitals and pairing states are constructed in phase space, enabling analytical solutions for pairing symmetry. We recast the BdG equation in phase space by expanding both the SC gap and the dispersion terms in the same irrep of the Bloch vector space. This gives insights into how the Wannier orbitals and pairing symmetry conspire in real space to manifest global coherence in the SC ground state. The method also allows us to obtain local Bogoliubov excitations of the SC condensate.   

Next, we consider a recently discovered superconductor in the SO material Lu$_2$Fe$_3$Si$_5$ within a DFT calculation, followed by the Wannier orbital calculation. Lu$_2$Fe$_3$Si$_5$ exhibits characteristics of a two-gap superconductor with $T_c\sim $6.1 K. The specific heat measurements\cite{Nakajima2012, Machida2011, Tamegai2009, Nakajima2008, Vining1983} and penetration depth experiments \cite{Biswas2011, Gordon2008} indicate the presence of nodeless pairing symmetry, leaving open questions about whether the pairing mechanism is conventional (attractive) or unconventional (repulsive). On the other hand, the rapid suppression of $T_c$ caused by non-magnetic impurities \cite{Hidaka2010, Watanabe2009, Hidaka2009, Sasame2009}, and atomic disorder induced by neutron irradiation \cite{Karkin2011}, suggests that Lu$_2$Fe$_3$Si$_5$ is an unconventional superconductor exhibiting superconductivity \cite{Braun1980}. 

Using DFT-derived Wannier orbitals, we solve the spin-fluctuation-mediated pairing eigenvalue equation for Lu$_2$Fe$_3$Si$_5$. The material’s strong three-dimensionality stabilizes compact Wannier orbitals with 
$s_{z^2}$ and $d_{z^2}$ wave symmetries. Dominant spin fluctuations arise in both inter- and intraband channels, mediating a coexistence of an unconventional nodeless, isotropic $s^{\pm}$ -wave pairing and a nodal $s_{z^2}$ channel. This $s+s$ pairing symmetry is in agreement with the temperature dependence of the magnetic penetration depth measurements, which is well-described by an $s+s$ pairing channel\cite{Biswas2011}. 

The rest of the paper is arranged as follows. We devote Sec. \ref{sec-2} to developing the full formalism of phase-space orbital theory. We present the single particle state in Sec.~\ref{Sec:WannFields} many-body state in Sec.~\ref{Sec:2WannFields} and phase space operators Sec.~\ref{Sec:Operator}. In Sec.~\ref{Sec:RPA}, we discuss the random-phase approximation (RPA) to evaluate spin and charge fluctuation mediated pairing interaction. We follow it by the construction of the mean fields, the inclusion of quantum statistics, the orbital constructions of Bogoliubov orbitals, and the self-consistent pairing gap equation in the following four subsections. We present the result of a tight-binding model and flat band superconductivity in Sec.~\ref{Sec:Flatband}. We consider the pairing symmetry calculations starting from the DFT result for Lu$_2$Fe$_3$Si$_5$ in Sec.~\ref{sec-3.1}. Finally, we conclude in Sec.~\ref{sec-4}.

\section{The Phase Space Model}\label{sec-2}

\begin{figure}[!htbp]
\centering
\includegraphics[width=0.49\textwidth]{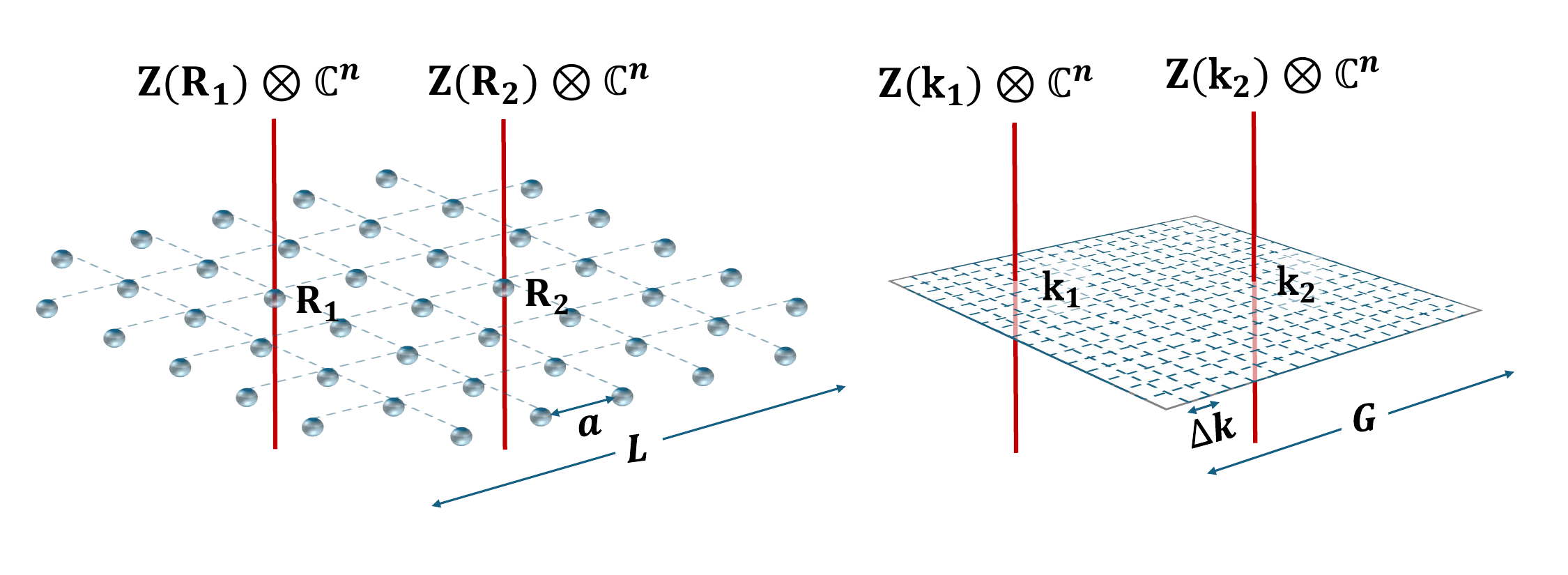}
\caption{\label{fig:blochbundle} The real-space lattice of length $L=Na$ is split into unit cells of length $a$. Similarly, the BZ is split with a grid size of $\Delta{\bf k}=2\pi/L$ and repeated by the reciprocal lattice vector ${\bf G}=2\pi/a$. At each lattice site ${\bf R}$, we introduce a product state of the Bloch phase vector ${\bf Z}({\bf R})$ and Wannier orbital spinor $\mathbb{C}^{\bm{n}}$, where $\bm{n}=n\times s$ for $n$ bands and $s$ spins. We introduce a product state  ${\bf Z}({\bf k})\otimes$ $\mathbb{C}^{\bm n}$ at each ${\bf k}$ point. Note that the Wannier orbitals $\mathbb{C}^{\bm n}$ are assumed to have a maximum uncertainty within a unit cell in real space maximum uncertainty of $\Delta {\bf k}$ within a BZ. We consider a trivial bundle where the same orbitals are included at all ${\bf k}$ and ${\bf R}$ points, where additional gauge obstruction, correlation, and entanglement arise from the Hamiltonian.}
\end{figure}

Let us start with ${\bm r}$ and ${\bm k}$ being continuous position and momentum variables spanning the entire system. We calibrate ${\bm r}$ with respect to the unit cell (UC) as ${\bm r} ={\bf r}+{\bf R}$, where ${\bf R}\in \mathbb{Z}_{L^d}$ are the positions of the UCs and  ${\bf r}\in$UC is the position within a UC. $L^d$ is the total number of UCs in a $d$-dimensional lattice (lattice constant $a=1$) with a periodic boundary condition: $L+1\sim 1$ (see Fig. \ref{fig:blochbundle}). Similarly, we split the momentum space as ${\bm k} ={\bf k}+{\bf G}$, where ${\bf k}\in \mathbb{T}^d$ lies in a $d$-dimensional torus, and ${\bf G} $ is the reciprocal lattice vector. The finite lattice of $L^d$ sites discretizes the Brillouin zone (BZ) $\mathbb{T}^d$ with a grid size of $\Delta k=2\pi/L$, such that ${\bf k}\in \mathbb{Z}_{L^d}$. This makes the ${\bf k}$ space a Pontryagin dual of the ${\bf R}$ space. The Fourier transformation between the two spaces is obtained by a polynomial of degree $L^d$ with the basis function of $z_{{\bf R}}({\bf k})=e^{i{\bf k}\cdot{\bf R}}$. In fact, if we associate Hilbert spaces of $\mathcal{L}^2(\mathbb{T}^d)$ and $\mathcal{L}^2(\mathbb{Z}_{L^d})$ in the corresponding momentum and position spaces, respectively, then $z_{{\bf R}}({\bf k})$ serves as the component of a unitary transformation between them. This gives us the idea to construct a linearly independent Bloch basis vector at each ${\bf k}$ as ${\bf Z}({\bf k}):=1/\sqrt{L^d}\big(z_{1}({\bf k})~...~z_{L^d}({\bf k})\big)^T$. Similarly, a same dimensional vector space ${\bf Z}({\bf R})$ can be defined at each ${\bf R}$. 

Therefore, ${\bf Z}({\bf k})\in \mathcal{L}^2(\mathbb{T}^d)$ can be considered as sections of a vector bundle $\mathbb{E}_{\bf k}\rightarrow\mathbb{T}^d$ (where $\mathbb{E}_{\bf k}$ is the total space). Similarly, ${\bf Z}({\bf R})\in \mathcal{L}^2(\mathbb{Z}_{L^d})$ is a section of the real space vector bundle $\mathbb{E}_{\bf R}\rightarrow \mathbb{Z}_{L^d}$. Note that we have not made any assumption about the orientability of the bundles, and a bundle-isomorphism (Fourier transformation) exists between them. Next, we introduce a Hilbert space $\mathbb{C}^{(n\times s)}$ for the internal degrees of freedom $n$, $s$ for orbital and spin indices at each base point ${\bf k}$ and ${\bf R}$ of their respective manifolds. Without loss generality, we start with a trivial vector bundle by considering a tensor product state as $\mathcal{H}_{{\bf k}/{\bf R}}:= {\bf Z}({{\bf k}/{\bf R}})\otimes \mathbb{C}^{(n\times s)}$. The existence of Wannier orbitals for the Bloch eigenstates is encoded in the existence of the bundle-isomorphism between them.\footnote{This trivialization of the vector bundle in the momentum (and real) space does not exclude the Wannierization of the topological insulators and certain flat band systems, as the topology or frat-band degeneracy will be included in the Hamiltonian. In fact, since the lattice symmetry and (non-/) compact nature of the orbital can be encoded in the Bloch basis, one can choose $w({\bf r})$ to be simple tight-binding or atomic orbitals.}

We now discuss how to express fields and operators in the above representations. Let $f({\bm r})$ be a local field in real space. It can be defined in the periodic lattice as $\int d{\bm r} f({\bm r})$ =$\sum_{{\bf R}}\int_{\rm UC} d{\bf r}f_{\bf R}({\bf r})$, where $f_{{\bf R}}({\bf r}):=$$f({\bf r}-{\bf R})$. We can define a vector field ${\bf f}({\bf r}) :=$$ (f_1({\bf r})~f_2({\bf r})~...)^T$, such that its projection on the periodic plane wave basis becomes
\begin{equation}
f({\bf k},{\bf r})={\bf Z}^{\dagger}({\bf k}){\bf f}({\bf r})=\sum_{{\bf R}}z^*_{{\bf R}}({\bf k})f_{\bf R}({\bf r}),
\label{eq:func}
\end{equation}
$\forall {\bf r}\in{\rm UC}, \forall {\bf k}\in{\rm BZ}$. We use this protocol to define the fields and operators in phase space. 

\subsection{Single particle Wannier fields}\label{Sec:WannFields}

Suppose we are interested in constructing Wannier states of the $2{N}$ ($2$ for spins) bands across the Fermi level. These states are obtained by the projector $\mathcal{P}({\bf k})=\sum_{{\bm n}} |{\bf k},{\bm n}\rangle\langle {\bf k},{\bm n}|$, where $|{\bf k},{\bm n}\rangle$ are eigenstates of the full Hamiltonian. ${\bm n}=(n,s)$ is assumed to be a composite index for band ($n=1,..,N$) and spin ($s=\uparrow,\downarrow$). According to the prescription outlined above, we start with a tensor product basis $|{\bf k},{\bm n}\rangle=|{\bf k}\rangle\otimes|n\rangle \otimes |s\rangle\in\mathcal{H}_{\bf k}$. 

Next, we project the single-particle states to real space $|{\bm r}_{n},{\bm n}\rangle$, where ${\bm r}_{n}$ is the position of the ${n}^{\rm th}$ Wannier orbital. We write ${\bm r}_{ n}={\bf r}_{n}-{\bf R}$, where ${\bf r}_{n}\in$ unit cell are the corresponding Wyckoff positions. We expand $|{\bf R},n,s_z\rangle\in \mathcal{H}_{\bf R}$ on the basis of $|{\bf k},n,s\rangle$, and define the expansion coefficients as $e^{i{\bf k}\cdot{\bf r}}\Psi_{ns}({\bf k},{\bf r})$, where $\Psi_{{\bf R}ns}({\bf k},{\bf r})$ is a product of three parts: 
\begin{equation}
    \Psi_{{\bf R}ns}({\bf k},{\bf r}) = z_{{\bf R}}({\bf k})w_n({\bf r})\chi_s.
\label{eq:wf}
\end{equation}
Here, $z_{{\bf R}}({\bf k})=\langle {\bf R}|{\bf k}\rangle$ are the plane waves of the periodic part, $w_n({\bf r})=\langle {\bf r}_n|n\rangle$ are the Wannier states with spins $\chi_s=\langle s_z|s\rangle$. All states are assumed to form complete bases in the usual way. (We drop the subscript $n$ in ${\bf r}_n$ for simplicity and insert it back when the meaning is obscured.)\footnote{The wavefunction, in its present form, is localized at ${\bf k}$ with a maximum uncertainty given by the BZ grid size of $\Delta k_x\sim (2\pi/L)$, and at ${\bf r}$ with a maximum uncertainly within a unit cell of lattice constant $\Delta x \sim a$ ($a$ is the lattice constant), along the $x$-direction. This appears to violate the uncertain relation $\Delta p\Delta x\sim 2\pi \hbar(a/L)$. But we know that the $\Psi_{{\bf R}ns}({\bf k},{\bf r})$ are periodic functions in the lattice, and hence are defined modulo ${\bf R}$ which makes $\Delta x \sim L$ and the uncertainty relation is saturated. Other possible states include wavepacket state, which can be implemented by modifying $z_{{\bf R}}({\bf k})$ from its plane wave solution to, say, a Gaussian state.}

The spinors in the orbital and spin space are defined as ${\bf W} = 1/\sqrt{N}(w_1~ w_2~ ... ~w_N)^T$ $\forall {\bf r}$, and ${\bf X}=1/\sqrt{2}(\chi_{\uparrow}~\chi_{\downarrow})^T$. The  Bloch spinor can be arranged in various ways. It is convenient to split the $L^d$-dimensional vector as sets of nearest-neighbor $({\bf Z}_{1})$, next nearest neighbor $({\bf Z}_{2})$ sites with respect to some center unit cell: ${\bf Z}({\bf k})$=${\bf Z}_{1}({\bf k})\oplus{\bf Z}_{2}({\bf k})\oplus...$, where ${\bf Z}_{a}({\bf k}):=1/\sqrt{d_a}\big(z_{{\bf R}_1}({\bf k})~...~z_{{\bf R}_{d_a}}({\bf k})\big)^T$ is the vector for the $a^{\rm th}$ nearest neighbor with $d_a$ number of sites. For short-range interactions, the Bloch spinor can be truncated up to a few nearest neighbors. Therefore, we have a single particle spinor defined at each $({\bf k},{\bf r})$ as 
\begin{eqnarray}
    \bm{\Psi}({\bf k},{\bf r})={\bf Z}({\bf k})\otimes {\bf W}({\bf r})\otimes {\bf X}.
    \label{eq:1bodywf}
\end{eqnarray}
In the Wannier spinor ${\bf W}({\bf r})$, ${\bf r}\equiv\{{\bf r}_n\}$ is assumed to be the set of Wyckoff positions of all the Wannier orbitals within a UC. A product state is a good starting point if we do not have any information on the many-body ground state; otherwise, appropriate unitary rotations can be employed if the symmetry and quantum statistics of the excitations are known (see Appendix~\ref{App:Irreps}). A key advantage of this phase space approach is that lattice symmetry and topological obstructions are encoded in the Bloch basis $z_{\bf R}({\bf k})$, enabling $w({\bf r})$ to be simple tight-binding or atomic orbitals.\footnote{For example, owing to a given site-symmetry group of the Wannier functions, one can rotate to the irreps basis of the Wannier states by suitable unitary rotation to ${\bf W}({\bf r})$ \cite{Sakuma2013, Kang2018}. Similarly, sometimes it's easier to work in the angular momentum basis for the Bloch phase ${\bf Z}({\bf k})$, as discussed in Appendix~\ref{App:Irreps}. Furthermore, due to atomic (or Rashba-type) spin-orbit coupling, the orbital (or momentum) and spin states are entangled, and one needs to go to the total angular momentum (or helicity) basis for the single-particle basis by calculating the Clebsch-Gordon coefficients. In both cases, it is easy to implement any symmetry of the theory, say $g$, as $\mathcal{U}_g=U_{g,Z}\otimes \mathcal{U}_{g,W}\otimes \mathcal{U}_{g,X}$, where $U_{g,Z}$ is the $\mathcal{N}\times \mathcal{N}$ unitary (or anti-unitary) representation of the symmetry $g$ in the ${\bf Z}$ spinor, and so on. The single particle state transforms as ${\bm \Psi}({\bf k},{\bf r})\rightarrow  \mathcal{U}_g{\bm \Psi}({{g\bf k}},g{\bf r}) = \left(U_{g,Z}{\bf Z}({g{\bf k}})\right)\otimes \left(U_{g,W}{\bf W}(g{\bf r})\right)\otimes \left(U_{g,X}{\bf X}\right)$.}\cite{Sakuma2013,Kang2018}

\subsection{Many body Wannier fields and Density Matrices}\label{Sec:2WannFields}

\begin{figure}[!htbp]
\centering
\includegraphics[width=0.45\textwidth]{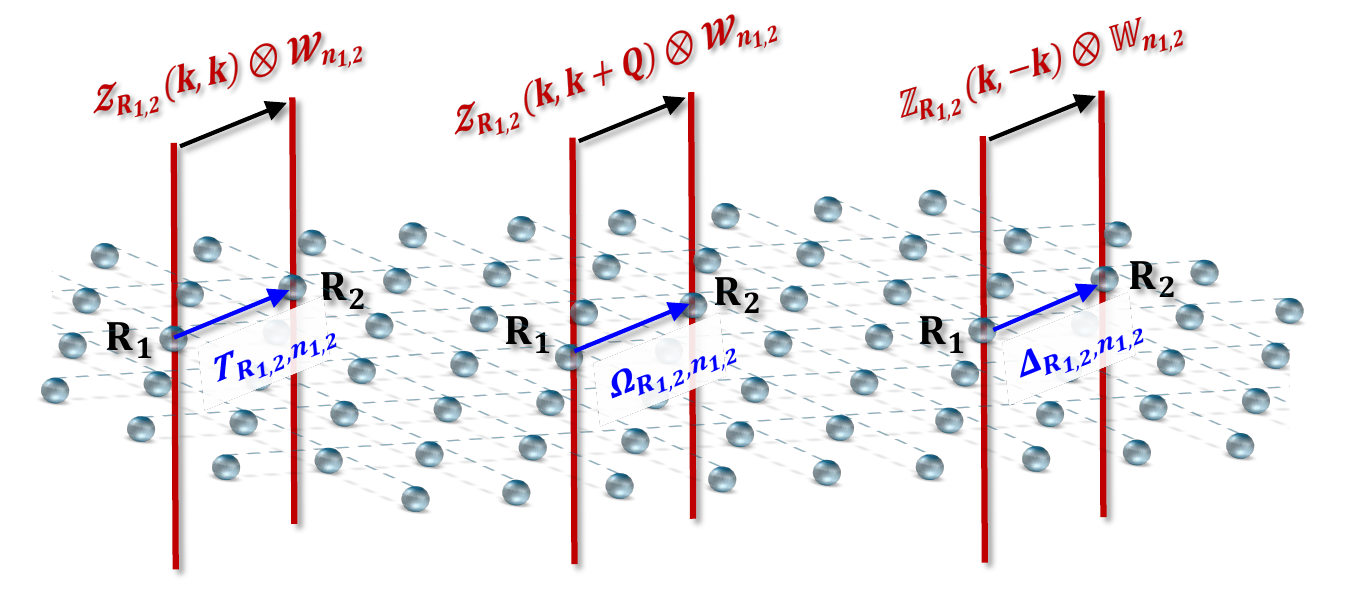}
\caption{\label{fig:twobody_blochbundle} We illustrate how the transfer matrices (${\pmb \Psi}$ and ${\pmb \varrho}$ defined in Eq. \eqref{eq:2bodywf} and \eqref{eq:DensityMatrix} ) and their matrix-elements (${\bf T}$, ${\bm \Omega}$, ${\bm \Delta}$) are defined on the fiber and base manifold, respectively. $T_{{\bf R}_{1,2},{\bm n}_{1,2}}$ gives quantum tunneling between two sites ${\bf R}_{1}$ and ${\bf R}_{2}$ through states ${\bm n}_{1}$ and ${\bm n}_{2}$. Momentum conservation in the tunneling process is embedded in the Bloch phase $\mathcal{Z}_{{\bf R}_{1,2}}({\bf k},{\bf k})$. Similarly, ${\bm \Omega}$ gives a density wave order parameter that absorbs a finite momentum $\mathcal{Z}_{{\bf R}_{1,2}}({\bf k},{\bf k}+{\bf Q})$, while ${\bm \Delta}$ corresponds to a Cooper pair field with zero-center-of-mass momentum in $\mathbb{Z}_{{\bf R}_{1,2}}({\bf k},-{\bf k})$.}
\end{figure}

A many-body ground state can be viewed as the vacuum or condensate of low-energy excitations.
A typical many-body state contains three types of correlations: quantum corrections owing to quantum statistics, `classical correlation' driven by Hamiltonian-induced interactions, and `geometric correlation' induced by Hilbert space constraints. Without loss of generality, we can start with a product state of the many-body ground state and incorporate all three correlations as variational parameters: 
\begin{eqnarray}
{\pmb {\Psi}}({\bf k}_{1,2},{\bf r}_{1,2})
&=&\mathsf{P}_{\theta}\left[{\bm \Psi}({{\bf k}_1},{\bf r}_1)\otimes {\bm \Psi}({{\bf k}_2},{\bf r}_2)\right],\nonumber\\
&=&\mathbb{{Z}}({\bf k}_{1,2})\otimes\mathbb{W}({\bf r}_{1,2})\otimes\mathbb{X}.
\label{eq:2bodywf}
\end{eqnarray}
For simplicity, we define a composite index $a_{1,2}=(a_1,a_2)$. Here, we partition the pair density fields in the three channels as $\mathbb{Z}({\bf k}_{1,2})$=$\mathsf{P}_{\theta(Z)}\left[{\bf Z}({{\bf k}_1})\otimes {\bf Z}({{\bf k}_2})\right]$, $\mathbb{W}({\bf r}_{1,2})$=$\mathsf{P}_{\theta(W)}\left[{\bf W}({\bf r}_1)\otimes {\bf W}({\bf r}_2)\right]$, and $\mathbb{X}$=$\mathsf{P}_{\theta({X})}\left[{\bf X}\otimes {\bf X}\right]$. $\mathsf{P}_{\theta}$ is the exchange parity such that the total wavefunction is an eigenstate of the exchange parity operator with eigenvalue $e^{i\theta}$. It is split among its quantum numbers as $\mathsf{P}=\mathsf{P}_{\theta(Z)}\otimes \mathsf{P}_{\theta(W)}\otimes\mathsf{P}_{\theta(X)}$, such that individual partitions can assume arbitrary value as long as $\theta(Z)+\theta(W) +\theta(X)=\pi$ (for fermion) is maintained. \footnote{Since each quantum number does not necessarily have to be the eigenvalue of their fermion parities, even and odd parity mixings lead to exotic superconducting or entangled states. For example, in non-unitary superconductors, such mixing between singlet and triplet governs novel phase transition \cite{Ramires2022}. Furthermore, the Bloch phase factors can have geometric phases such as the Ahronov-Bohm phase or Berry phase on non-trivial topological geometry, which makes the SC states more interesting. Similarly, generalization of Eq.~\eqref{eq:2bodywf} to anyonic cases will be easier.  One can, in principle, treat $\theta$ as a variational parameter for exotic many-body systems such as for fractional quantum Hall state or in flat band Chern insulators where anyon, Majorana or parafermmion excitations are expected. }\cite{Ramires2022} We can read these pair fields as Cooper pairs in superconductors or Bell pairs in spin liquid or projective symmetry group constructions. The projection $\mathsf{P}_{\theta}$  onto the pair states introduces a quantum correlation between the excitations.

An interaction term in the Hamiltonian introduces correlation between particles, which we, loosely speaking, call `classical correlation'. It can be included via a projection operator $e^{{\bf J}_{{\bf R}_{12}}({\bf k}_{1,2},{\bf r}_{1,2})}$ where ${\bf J}$ is a rank-(1,1) Jastrow factor tensor. The specific form of ${\bf J}$ is Hamiltonian dependent. 

Finally, we discuss the correlation property that arises from geometric effects. A low-energy theory serves as an effective description confined to a specific subspace of the full Hilbert space (e.g. a few bands near the Fermi level, a flat band, the lowest Landau level in the fractional quantum Hall state, singlet pair for superconductivity or a spin liquid phase.) To ensure completeness and orthogonality for the orbitals defined in a subspace, we invoke a projection operator $\mathcal{P}({\bf k})$ to the low-energy subspace of ${\bm n}$ states of our interest. This projection introduces correlations (or, equivalently, imposes constraints on the correlation function), which we call geometry-induced correlations.\cite{Yogendra2024} 

For a pure state, a projection operator is similar to a reduced-density matrix. A density matrix field or transfer matrix in the phase space can be defined as 
\begin{eqnarray}
    {\pmb \varrho}({{\bf k}_{1,2}},{\bf r}_{1,2})&=&\Psi({{\bf k}_2},{\bf r}_2)\otimes\Psi^{\dagger}({{\bf k}_1},{\bf r}_1),\nonumber\\
    &=&\bm{\mathcal{Z}}({\bf k}_{1,2})\otimes\bm{\mathcal{W}}({\bf r}_{1,2})\otimes\bm{\mathcal{X}},
    \label{eq:DensityMatrix}
\end{eqnarray}
where $\bm{\mathcal{Z}}({\bf k}_{1,2})  =  {\bf Z}({\bf k}_2)\otimes{\bf Z}^{\dagger}({\bf k}_1)$, $\bm{\mathcal{W}}({\bf r}_{1,2}) = {\bf W}({\bf r}_2)\otimes {\bf W}^{\dagger}({\bf r}_1)$, and $\bm{\mathcal{X}} = {\bf X}\otimes {\bf X}^{\dagger}$.

We show below that we can represent any operator in many-body states in terms of these two `transfer matrices': the pair field ${\pmb {\Psi}}({\bf k}_{1,2},{\bf r}_{1,2})$, and the density matrix ${\pmb \varrho}({\bf k}_{1,2},{\bf r}_{1,2})$, whose matrix-element in the Bloch and orbital states are visualized in Fig.~\eqref{fig:twobody_blochbundle}.

\subsection{Operators in the phase space}\label{Sec:Operator}
Next, we represent the one-body operator ${\bf T}({\bm r})$, and two-body operator ${\bf V}({\bm r}_{1,2})$ in the phase space in Eqs.~\eqref{eq:1bodywf}, and \eqref{eq:2bodywf}. The operators are represented as tensors in the Bloch vector space, with its each component being marices in the orbital and spin Hilbert space. Proceeding as in Eq.~\eqref{eq:func}, we obtain the tensor components of a one-body operator ${\bf T}({\bf r})$ in the Bloch phase basis as 
\begin{equation}
T({\bf k},{\bf r})={\bf Z}^{\dagger}({\bf k}){\bf T}({\bf r}){\bf Z}({\bf k})=\sum_{{\bf R}_{1,2}}T_{{\bf R}_{1,2}}({\bf r})\mathcal{Z}_{{\bf R}_{1,2}}({\bf k}), 
\label{eq:1body}
\end{equation}
where $T_{{\bf R}_{1,2}} \equiv T_{{\bf R}_{1},{\bf R}_{2}}$ is a (1,1)-tensor tight-binding hopping field. Its matrix-element in the orbital and spin basis $T_{{\bf R}_{1,2},{\bm n}_{1,2}}$ = $\int_{\bf r}~T_{{\bf R}_{1,2}}({\bf r})\mathcal{W}_{n_{1,2}}({\bf r})\mathcal{X}_{s_{1,2}}$ gives the tight-binding hopping tensor between ${\bm n}_{1}, {\bm n}_{2}$ species localized at  ${\bf R}_{1}$ and ${\bf R}_{2}$, respectively. The components of $T_{{\bf R}_{1,2},{\bm n}_{1,2}}$ are exemplified in Appendices~\ref{lab:RPA-2} and ~\ref{App:Multiband}. We have implemented translational invariance, which makes ${\bf k}$ to be the same for the incoming ${\bf Z}({\bf k})$ and departing ${\bf Z}^{\dagger}({\bf k})$ states. This makes the density operator $\mathcal{Z}_{{\bf R}_{1,2}}({\bf k},{\bf k})\equiv\mathcal{Z}_{{\bf R}_{1,2}}({\bf k})=z_{{\bf R}_{2}-{\bf R}_{1}}({\bf k})$ to be local-in-${\bf k}$, as shown in  Fig.~\ref{fig:twobody_blochbundle}. The corresponding one-body Hamiltonian becomes
\begin{eqnarray}
H_{T}  
&=&\int_{\bf k} \sum_{{\bm \kappa}_{1,2}} T_{{\bm \kappa}_{1,2}}\mathcal{Z}_{{\bf R}_{1,2}}({\bf k})|{\bm \kappa}_{1},{\bf k} \rangle \langle {\bm \kappa}_{2},{\bf k}|.
\label{eq:1bodyH}
\end{eqnarray}
We have introduced another composite index ${\bm \kappa}_i\equiv({\bf R}_i,{\bm n}_i)\equiv({\bf R}_i,n_i,s_i)$ for simplicity in presentation.  We use the notation $\int_{{\bf r}}\equiv\int_{\rm UC} d{\bf r}$ and $\int_{{\bf k}}\equiv\sum_{{\bf k}\in {\rm BZ}}$ as the integration over the phase space volume where ${\bf k}\in$BZ and ${\bf r}\in$UC integrations are assumed to be normalized with their corresponding BZ and UC volumes, respectively. 

Similarly, a two-body operator ${\bf V}({\bf r}_{1,2})$ is written in the Wannier basis as
\begin{eqnarray}
V({\bf k}_{1-4},{\bf r}_{1,2})&=&
\mathbb{Z}^{\dagger}({\bf k}_{1,2}){\bf V}({\bf r}_{1,2})\mathbb{Z}({\bf k}_{3,4})\delta_{{\bf k}_{1-4}},\nonumber\\
&=&\sum_{{\bf R}_{1-4}}\bar{V}^{{\bf R}_{2,4}}_{{\bf R}_{1,3}}({\bf r}_{1,2})\mathbb{Z}_{{\bf R}_{1,2}}^{*}({\bf k}_{1,2})\mathbb{Z}_{{\bf R}_{3,4}}({\bf k}_{3,4}).
\label{eq:2body}
\end{eqnarray}
Here $\bar{V}^{{\bf R}_{2,4}}_{{\bf R}_{1,3}}$ is a (2,2)-rank interaction tensor, whole matrix-elements in the orbital and spin basis are given by
\begin{eqnarray}
\bar{V}^{{\bm \kappa}_{2,4}}_{{\bm \kappa}_{1,3}}({\bf k}_{1-4}),
%
&=&\int_{{\bf r}_{1,2}} ~
\bar{V}^{{\bf R}_{2,4}}_{{\bf R}_{1,3}}({\bf r}_{1,2})\delta_{{\bf k}_{1-4}}({\bf r}_{1,2})\nonumber\\
&&~\times \mathbb{W}_{n_{1,2}}^{*}({\bf r}_{1,2})\mathbb{W}_{n_{3,4}}({\bf r}_{1,2})\mathbb{X}_{s_{1,2}}^{*}\mathbb{X}_{s_{3,4}}.
\label{eq:IntW}
\end{eqnarray}
$\delta_{{\bf k}_{1-4}}({\bf r}_{1,2})=\delta_{{\bf k}_{1-4}}\exp{(i\sum_{a=1}^4{\bf k}_a\cdot{\bf r}_{a})}$, with ${\bf r}_{a}$ being the center of the $a^{\rm th}$ Wannier orbital, with ${\bf r}_{3,4}={\bf r}_{1,2}$ for this case of density-density interaction. We split the interaction in the product basis as $\bar{V}^{{\bm \kappa}_{2,4}}_{{\bm \kappa}_{1,3}}$ = $\mathsf{P}_{\theta}\left[V^{{\bf R}_{2,4}}_{{\bf R}_{1,3}}\otimes V^{n_{2,4}}_{n_{1,3}}\otimes V^{s_{2,4}}_{s_{1,3}}\right]\mathsf{P}_{\theta}^{-1}$. In addition, the Jastrow factor ${\bf J}$ and low-energy projector $\mathcal{P}$ are also assumed to be applied on $\bar{V}$. We, henceforth, omit the bar symbol from the two-body terms for simplicity in notation, and all such two-body terms are assumed to include $\mathsf{P}_{\theta}$, ${\bf J}$, and $\mathcal{P}$ projections. The corresponding two-body Hamiltonian reads:
\begin{eqnarray}
H_{V} 
&=&\int_{{\bf k}_{1-4}}\sum_{{\bm \kappa}_{1-4}} {V}^{{\bm \kappa}_{2,4}}_{{\bm \kappa}_{1,3}}\mathbb{Z}^{*}_{{\bf R}_{1,2}}({\bf k}_{1,2})\mathbb{Z}_{{\bf R}_{3,4}}({\bf k}_{3,4})\nonumber\\
&&~~~\qquad\times\delta_{{\bf k}_{1-4}} |{\bm \kappa}_{1,2},{\bf k}_{1,2}\rangle\langle {\bm \kappa}_{3,4},{\bf k}_{3,4}|.
\label{eq:intH2}
\end{eqnarray}

\subsection{Density-density interaction and RPA}\label{Sec:RPA}

A common origin of interactions in materials is density-density interactions. Here, we only consider `onsite' interactions so that the Bloch basis part gives $V^{{\bf R}_{2,4}}_{{\bf R}_{1,3}}=\mathbb{I}\delta_{{\bf R}_{1-4},0}$. For the orbital part, we have two sources of interaction, namely the orbital density-density $V_{nn}^{mm}=V^{n_{2,4}}_{n_{1,3}}\delta_{n_{1},n_3}\delta_{n_{2},n_{4}}$, and the orbital exchange part, also known as pair hopping interaction, $V_{nm}^{nm}$. Similarly, for spins, we have spin density-density interaction $V_{ss}^{tt}$, and spin-flip term $V_{st}^{st}$. Their combined interactions that follow the fermion parity are recognized in the literature as follows:
 \begin{eqnarray}
&& V_{nn}^{nn}V_{ss}^{\bar{s}\bar{s}} = U_n, \quad V_{nn}^{mm}V_{ss}^{tt}=U'_{mn},\nonumber\\
&& V_{nn}^{mm}V_{st}^{st} = J_{mn},\quad   V_{nm}^{nm}V_{ss}^{\bar{s}\bar{s}}=J'_{mn}.
\label{eq:Int}
\end{eqnarray}
Here, we assume $m\ne n$, while both $s=t$ and $s\ne t$ are allowed and $\bar{s}=-s$. The interactions are often spin-independent unless spin-orbit coupling is included. Here $U$, and $U'$ are the intra- and inter-Wannier orbitals Hubbard interactions, respectively, while $J$ and $J'$ are Hund's coupling and pair hopping terms. The corresponding interacting Hamiltonian is called the Hubbard-Kanamori Hamiltonian \cite{Kanamori1963, Georges2013, Das2015}.

Such an onsite (repulsive) interaction is often inapt to give an unconventional (momentum-dependent) pairing state. However, many-body effects, such as spin and charge fluctuations, can produce short-range interactions, which can be attractive between Cooper pairs of opposite signs.\cite{KohnLuttinger,ChubukovKL,ScalapinoRMP} The lowest order renormalization of the density-density interaction arises from the random-phase approximation (RPA) for intermediate to weak coupling theory \cite{Pines1952, Bohm1953, Pines2016, Gell-Mann1957, McLACHLAN1964, Oddershede1978, Szabo1977, Das2014}. We derive the RPA-dressed interaction (${\bf \Gamma}$) in Appendix~\ref{lab:RPA-2}, and quote the result here:
\begin{eqnarray}
&&\Gamma_{{\bm \kappa}_{13}}^{{\bm \kappa}_{24}}({\bf q}) = {V}_{{\bm n}_{13}}^{{\bm n}_{24}}\nonumber\\
&&
\qquad+\frac{3}{2}\sum_{n_{5678},s_{56}}{V}_{{\bm n}_{15}}^{{\bm n}_{26}}({\Pi}_s)_{{\bf R}_{13},n_{57},s_{5}\bar{s}_5}^{{\bf R}_{24},n_{68},s_{6}\bar{s}_6}({\bf q})
{V}_{n_{73},\bar{s}_5s_{3}}^{n_{84},\bar{s}_6s_{4}} \nonumber\\
&&\qquad - \frac{1}{2}\sum_{n_{5678},s_{56}}{V}_{{\bm n}_{15}}^{{\bm n}_{26}}({\Pi}_c)_{{\bf R}_{13},n_{57},s_{55}}^{{\bf R}_{24},n_{68},s_{66}}
({\bf q}){V}_{n_{73},s_{53}}^{n_{84},s_{64}}.
\label{eq:IntRPA}
\end{eqnarray}
Here ${\bf q}$ gives the momentum transfer between the scattering states. We have occasionally expanded the compact index ${\bm n}=(n,s)$ into orbital and spin indices or kept it as a compact index ${\bm n}$ according to convenience. For example, in the third (fourth) term, we have substituted $s_7=\mp s_5$, and $s_8=\mp s_6$. The first term is the onsite interaction, the second term corresponds to the spin-flip ($\uparrow\downarrow\leftrightarrow\downarrow\uparrow$) fluctuation $-$ called the transverse spin fluctuation, and the third term corresponds to occupation density or charge fluctuation. ${\Pi}_{s/c}$ refer to the RPA susceptibilities along the spin/charge channels, respectively, which are related to the Lindhard susceptibility $\Pi$ as $\bm{\Pi}_{s/c}=\bm{\Pi}\left(\bf{I}\mp\bm{V}\bm{\Pi}\right)^{-1}$ where $\bf{I}$ is the unit matrix. The $\bm{A}$ symbol denotes that we have written the rank-(2,2) tensor ${\bm A}$ in matrix form. The details are given in Appendix~\ref{lab:RPA-2}.

\subsection{Mean fields in the phase space}\label{lab:MF}

Here, we introduce the mean-field decomposition of the interaction term in Eq.~\eqref{eq:intH2} after we replace the bare interaction ($V$) with the RPA interaction $\Gamma$ in which $z_{\bf R}({\bf k})$, $w_n({\bf r})$, and $\chi_{s}$ are read to be quasiparticle/excitation states with possibly generalized statistics embedded in the variational parameters $\theta_{Z,W,X}$. We discuss density wave order parameter in Appendix~\ref{App:MFT} and focus here on a mean-field Cooper pairing tensor field, defined as 
\begin{eqnarray}
{\Delta}_{{\bm \kappa}_{1,2}}({\bf k}_1,{\bf r}_{1})=
\int_{{\bf k}_2,{\bf r}_2}\sum_{{\bm \kappa}_{3,4}}\Gamma^{{\bm \kappa}_{2,4}}_{{\bm \kappa}_{1,3}}({\bf r}_{1,2}){\pmb \Psi}_{\bm \kappa_{3,4}}({\bf k}_2,{\bf r}_2),
\label{Eq:IntMF}
\end{eqnarray}
where ${\pmb \Psi}$ is given in Eq.~\eqref{eq:2bodywf}. We have used shorthand notation ${\pmb \Psi}({\bf k},{\bf r})\equiv{\pmb \Psi}({\bf k},-{\bf k},{\bf r},{\bf r})$, ${\Delta}({\bf k},{\bf r})\equiv{\Delta}({\bf k},-{\bf k},{\bf r})$, and $\mathbb{Z}({\bf k})\equiv\mathbb{Z}({\bf k},-{\bf k})$. Once we substitute the pair field in Eq.~\eqref{eq:intH2}, we obtain a single-particle (mean-field) Hamiltonian, which can be written as
\begin{eqnarray}
H_{\Delta}  =\int_{{\bf k}}\sum_{{\bm \kappa}_{1,2}} {\Delta}_{{\bm \kappa}_{1,2}}\mathbb{Z}^*_{{\bf R}_{1,2}}({\bf k})|{\bm \kappa}_{1},{\bf k}\rangle|{\bm \kappa}_{2},-{\bf k}\rangle. 
\label{eq:HDelta}
\end{eqnarray}
Here, we call ${\Delta}_{{\bm \kappa}_{1,2}}$ as the `Wannier pairing field' tensor between the Wannier sites ${\bf R}_1$, ${\bf R}_2$ and orbitals ${\bm n}_1$ and ${\bm n}_2$ as shown in Fig.~\ref{fig:twobody_blochbundle}. The pair field tensor is antisymmetric under the fermion parity.  

\subsection{Wannier pairing and Bogoliubov orbitals}\label{Sec:BOrbital}

The nature of excitations depends on the Bell or Cooper pair fields condensed in the ground state, and entangled or fractional excitations emerge by breaking such pairs. These excitations are often defined by a superposition of the two particles that formed the pair state. \cite{Adhikary2020fwave} For the Cooper (Bell)  pair case, such excitations are called the Bogoliubov/Majorana (spinon) states. In what follows, we express the Bogoliubov orbitals as superpositions of single-particle spinors for particles and hole states,
\begin{eqnarray}
\Phi({{\bf k}},{\bf r})&=&\left[\Psi({{\bf k}},{\bf r})\oplus \Psi^*(-{\bf k},{\bf r})\right]\nonumber\\
&=&\left[{\bf Z}({{\bf k}})\otimes{\bf W}({\bf r})\otimes{\bf X}\right]\oplus
\left[{\bf Z}(-{\bf k})\otimes{\bf W}^*({\bf r})\otimes {\bf X}^*\right]. 
\label{Eq:Bogwf}
\end{eqnarray}
Here, we have already implemented a zero center-of-mass momentum for the Cooper pair by setting $-{\bf k}$ for the second electron.  `*' stands for complex conjugation, which gives the hole states at ${\bf k}$ for the corresponding particle state at ${\bf k}$, and vice versa. The crucial advantage of our phase space approach is that we can separate the particle-hole entanglement among the spatial, orbital, and spin parts so that localization of the Bogoliubov quasiparticles is warranted.  

We express a (mean-field) one-body BdG Hamiltonian in the particle-hole spinor as 
\begin{eqnarray}
H&=&\int_{{\bf k},{\bf r}}~\Phi^{\dagger}({{\bf k}},{\bf r})\left({\bf T}({\bf r})+{\bm  \Delta}({\bf r})\right)\Phi({\bf k},{\bf r}),\nonumber\\
&=&\int_{{\bf k}}~\sum_{{\bm \kappa}_{1,2}}\Big[T_{{\bm \kappa}_{1,2}}\mathcal{Z}_{{\bf R}_{1,2}}({\bf k})
|{\bm \kappa}_{1},{\bf k}\rangle \langle {\bm \kappa}_{2},{\bf k}|\nonumber\\
&&\qquad~ +{\Delta}_{{\bm \kappa}_{1,2}}\mathbb{Z}^*_{{\bf R}_{1,2}}({{\bf k}})|{\bm \kappa}_{1},{\bf k}\rangle |{\bm \kappa}_{2},-{\bf k}\rangle\Big] \nonumber\\
\label{Eq:BdG}
\end{eqnarray}
Note that the Bloch phase parts for the density and pairing terms (with ${\bf k}=-{\bf k}$) follow the identity $\mathcal{Z}_{{\bf R}_{1,2}}({\bf k})=\mathbb{Z}^*_{{\bf R}_{1,2}}({\bf k})=z_{{\bf R}_{2}-{\bf R}_1}({\bf k})$.  Therefore, the Bloch phase part can be taken as common to the above two terms, and we can define a BdG Hamiltonian that lives on the link or bond between the Wannier sites ${\bf R}_1$ and ${\bf R}_2$ at each ${\bf k}$, see Fig.~\ref{fig:twobody_blochbundle}, as \cite{Das2011}
\begin{eqnarray}
{\bm H}_{{\bf R}_{1,2}} =\left(\begin{array}{cc}
     {\bm T}_{{\bf R}_{1,2}} & {\bm \Delta}_{{\bf R}_{1,2}}  \\
     {\bm \Delta}_{{\bf R}_{2,1}}^{\dagger} &  -{\bm T}_{{\bf R}_{2,1}}^T
\end{array}\right),
\label{Eq:BdGmatrix}
\end{eqnarray}
where ${\bm T}$ and ${\bm \Delta}$ are rank-(1-1) tensors (indexed by ${\bf R}_{1,2}$) with each component being a matrix in the orbital and spin space (index by ${\bm n}_{1,2}$). The Hermitian conjugate and transpose operations are defined in the orbital and spin basis. 

Generally, $[{\bm H}_{{\bf R}_{1,2}},{\bm H}_{{\bf R}_{3,4}}]\ne 0$, and so exact eigenstates at each point on the phase space are not possible, by construction. Therefore, to diagonalize the Hamiltonian, we first go to the momentum space, and obtain eigenvalues and eigenvectors at each ${\bf k}$ (or one can go to a fully local basis in real space), corresponding to the fully global coherence (BCS) limit [or local pairing (BEC limit)]. In our case here, we assume a global (momentum space) coherence $\Delta\ne 0$ for a given irrep of $z_{{\bf R}_{2}-{\bf R}_{1}}({\bf k})$ and zero otherwise. Eq.~\eqref{Eq:BdG} is one of the key result of our theory: we obtain a mean-field Hamiltonian that is local in phase space, living at a given ${\bf k}$-point and on the bond/link between sites in real space, as exhibited in Fig.~\ref{fig:twobody_blochbundle}. We have considered examples of multiband case and mean-field density order and SC orders and contrasted with the traditional tight-binding methods in Sec.~\ref{Sec:Flatband} and in Appendix~\ref{App:Examples}.

Finally, to obtain the Bogoliubov orbital states from the eigenvectors of Eq.~\eqref{Eq:BdG}, we return to the products basis as
\begin{eqnarray}
{\bf B}({\bf r})\oplus {\bf B}^*({\bf r})&=&[\mathcal{U}\mathcal{S}({\bf W}({\bf r})\otimes{\bf X})]\oplus\nonumber\\
&& [-\mathcal{V}\mathcal{S}^*({\bf W}^*({\bf r})\otimes{\bf X}^*)],
\label{Eq:BdG4}
\end{eqnarray}
where $\mathcal{S}$, $\mathcal{U}$, $\mathcal{V}$ are the components of the unitary operators of the eigenvectors, as defined in Appendix~\ref{App:MFT}. Since the unitary transformations mix the orbital and spins, there is no general way to disentangle them in the Bogoliubov quasiparticle basis. ${\bf B}({\bf r})$ is a spinor of the Wannier orbitals of the Bogoliubov quasiparticles.  However, one can proceed similarly to define Fourier components of the Bogoliubov orbitals in phase space as ${\bf B}({\bf k},{\bf r})={\bf Z}({\bf k})\otimes {\bf B}({\bf r})$ as in Eq.~\eqref{eq:func}. We cannot separate the spin and orbital part here since spin and orbitals are mixed in the Bogoliubov states. Under certain conditions, ${\bf B}({\bf r})$ become Wannier spinor of Majorana orbitals.\cite{Yogendra2024, Sticlet2012, Wahl2013, Schindler2020} There may arise obstructions in gauge fixing in the Wannier orbitals of Bogoliubov (or general entangled) quasiparticles, which needgauge fields to obtain orthonormalized orbital states \cite{Yogendra2024}. However, we do not pursue this endeavor in this paper.

\subsection{Self-consistent SC gap equation}\label{Sec:gapself}

Our final step is to obtain the self-consistent gap equation. We can switch from each pair of indices ${\bm \kappa}\equiv({\bf R},n,s)$ to their even/odd parity indices ${\bm \nu}\equiv(\nu_{\bf R},\nu_{n},\nu_s)$ where $\nu_i=-\pm$ runs over the even and odd combinations of the index under the constraint that the total parity is odd $\nu=-1$ (see Appendix~\ref{App:Irreps}). This aids in implementing the fermion odd parity in the total wavefunction. For example, we denote $\Delta_{\nu_{\bf R},{\nu}_{\bm n}}$  to imply that if it's even (odd) under the exchange between ${\bf R}_{1,2}$, it must be odd (even) in the band basis ${\bm n}_{1,2}$ and vice versa. Obtaining the expectation value of the right-hand side of Eq.~\eqref{Eq:IntMF} in the Bogoliubov state in Eq.~\eqref{Eq:BdG4},  we find a phase-space eigenvalue equation for the SC gap (see Appendix~\ref{App:MFT}) as
\begin{eqnarray}
 \int_{{\bf k}_2} \sum_{{\bm \nu}'}\Gamma_{{\bm \nu}}^{{\bm \nu}'}({\bf k}_{1,2})\Delta_{{\bm \nu}'}z_{\bm \nu'}({\bf k}_2) =-\lambda  {\Delta}_{{\bm \nu}}z_{\bm \nu}({\bf k}_1).
    \label{Eq:eivenval1}
\end{eqnarray}
Note that although the Bloch basis seems to be separated from the orbital part in the above formalism, the odd parity constraint entangles all the indices. 

As a variational approach, we can first find the symmetry-allowed Bloch functions $z_{\nu}({\bf k})$ in the pair state and find expectation values of ${\bm \Gamma}$ in those basis states. Then, the highest negative value of this expectation value gives us the pairing symmetry. We demonstrate that analytically in Sec.~\ref{Sec:Flatband} and numerically in Sec.~\ref{sec-3.1}.

\section{Results and Discussions}\label{sec-3}

\subsection{Single- (flat-) band superconductivity: Analytical solution}\label{Sec:Flatband}

\begin{figure}[ht]
    \centering
    \includegraphics[width=0.45\textwidth]{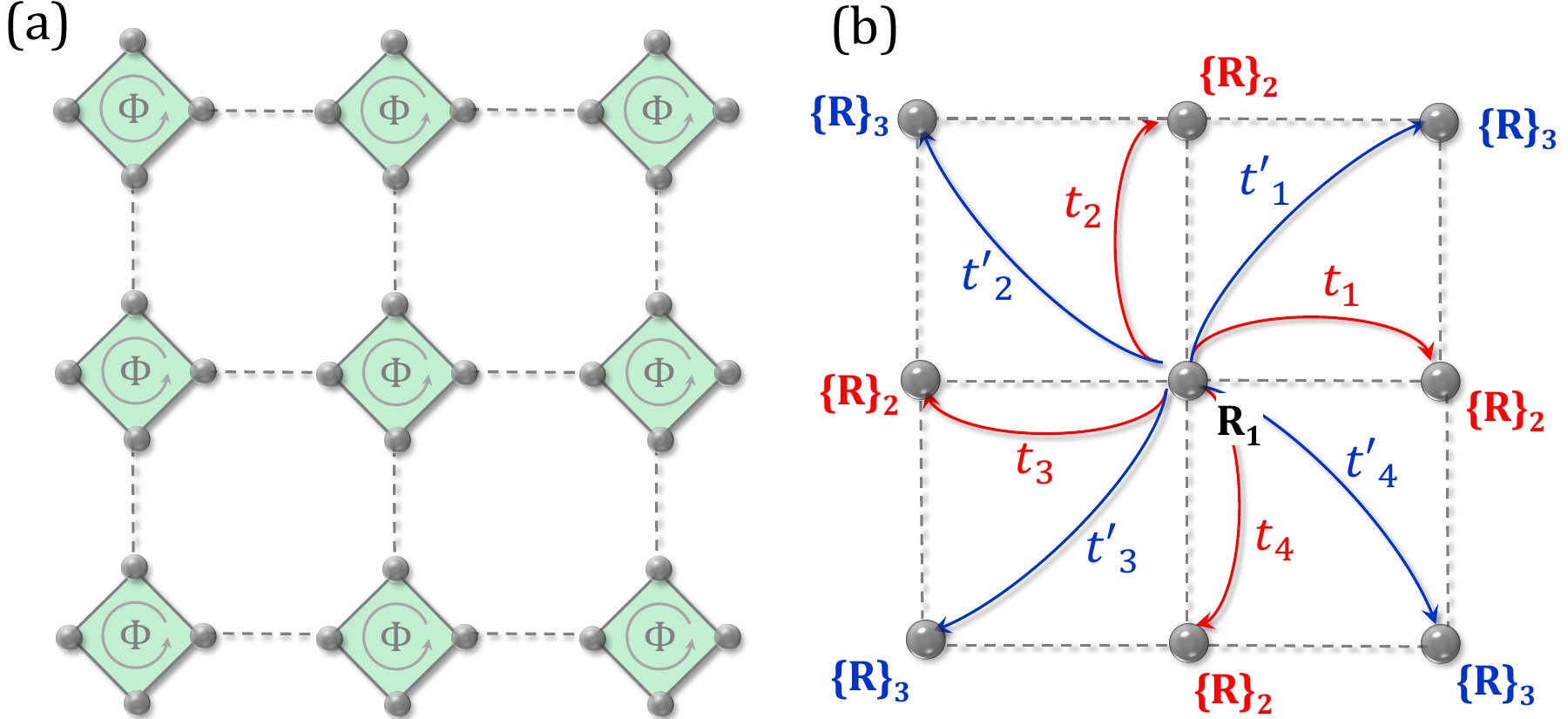}
    \caption{(a) A 2D SO lattice with four sublattices forming a diamond plaquette (green shading). Following Ref.~\cite{Pal2018} we consider an integer flux threading each diamond plaquette, which produces a flat band (see Fig.~\ref{fig:2dSOL}). (b) We consider a single Wannier orbital model for the flat band, which is localized at the diamond center. We consider first- (second-) nearest-neighbor hoppings $t_i$ ($t'_i$)  between these orbitals to draw a phase diagram of pairing symmetry as summarized in Table~\ref{tab:irreps}.}
    \label{fig:sol_new}
\end{figure}

\begin{figure*}[!t]
\centering
\includegraphics[width=0.9\textwidth]{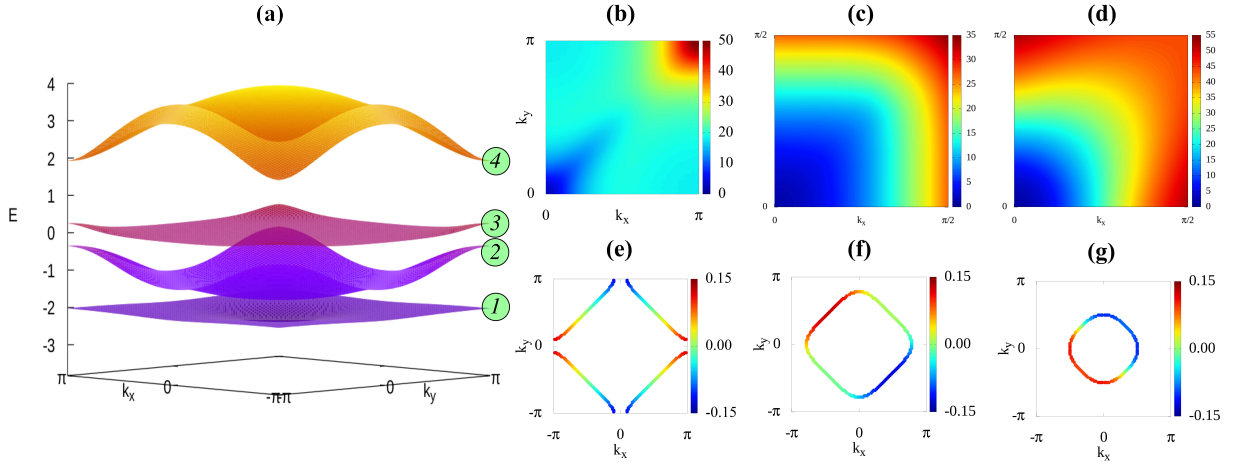}
\caption{\label{fig:2dSOL}(a) We reproduce the band structure of the tight-binding model from Ref.~\cite{Pal2018} in 2D SO lattice. The number inside the circle enumerates the bands. We consider band 3 for further analysis of the flat band.  (b)-(d) Static RPA susceptibility  ${\rm Tr}\Pi_s({\bf q})$ is plotted in the 2D BZ for three representative values of the chemical potential:  (b,e) $\mu$=0.045, (c,f) $\mu$=0.04 and (d,g) $\mu$=0.03. We plot the corresponding FS topologies for the three cases discussed in the above panels. The chemical potential is chosen to capture the metallic state on the flat band only. The gradient color gives the pairing eigenfunction $z_{\bar{\nu}}$ of Eq.~\eqref{Eq:eivenval1} for the largest eigenvalue in the three cases. }
\end{figure*}

To illustrate the formalism, we begin with a simple four-band tight-binding model on a 2D square octagon (SO) lattice, as shown in Fig.~\ref{fig:sol_new}(a).\footnote{Note that we explored several monolayer SO materials using DFT calculations and so far found no stable lattice to report here. Thus, the 2D structure studied here serves purely as a theoretical example.} Previous tight-binding calculations for a spinless orbital per sublattice on the 2D SO lattice with four sublattices have shown four dispersive metallic bands\cite{Nie2015, Bao2014, Pal2018}. Pal has shown that with a suitable insertion of magnetic flux through the diamond plaquette, as shown in Fig.~\ref{fig:2dSOL}(b) below, a topological flat band can be achieved.\cite{Pal2018}

\begin{figure*}[ht!]
\centering
\includegraphics[width=0.8\textwidth]{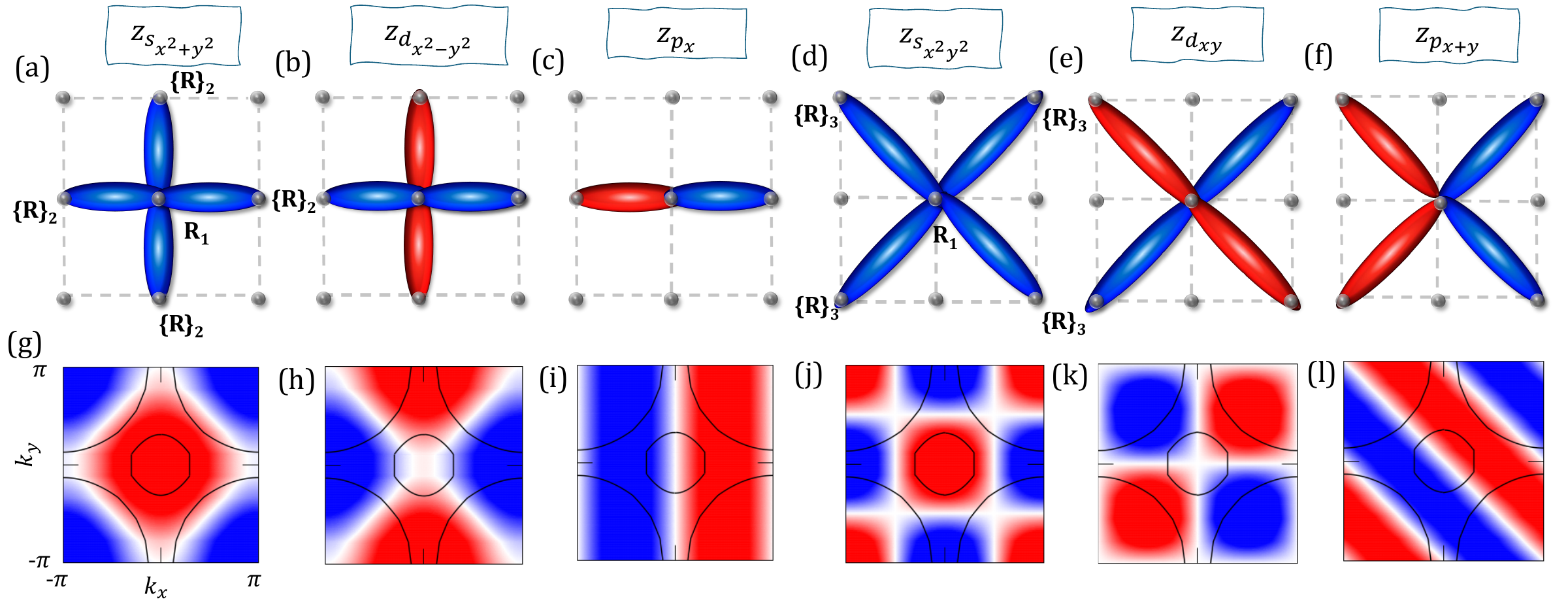}
\caption{\label{fig:z_nu} We visualize some of the Bloch basis irreps $z_{\nu}$ in real (top row) and momentum (bottom row) space of the square lattice. The red and blue colors represent the positive and negative values of the function, respectively. The solid black lines in (g-l) show schematic FS contours of two representative filling factors to demonstrate how the pairing function changes sign on the corresponding FS. A small FS centered at the $\Gamma$ point corresponds to an electron-like FS, while a large FS centered at the ${\bf k}=(\pi,\pi)$ point corresponds to a slight hole doping. }
\end{figure*}

Lattice models for gauge and matter fields are often studied by expanding the lattice into a magnetic unit cell that threads an integer flux, as demonstrated for U(1) gauge field in Ref.~\cite{Thouless1982}, and for $Z_2$ field in Ref.~\cite{Yogendra2024}. Here, we focus on single flux quantum ($\Phi=\Phi_0$) per diamond plaquette (unit cell), which ensures the magnetic unit cell coincides with the flux-free unit cell. For higher integer fluxes, the extended magnetic unit cell formalism becomes analogous to supercell models for superconductivity,\cite{supercellZhu,supercellHan} though in our phase-space model, the dimension of ${\bf Z}({\bf k}/{\bf R})$ and the one-body tensors change accordingly. Finally, among the four bands, we focus on the spinless single flat band, whose Wannier orbital is localized at the center of the diamond plaquette, yielding a dual square lattice. Since there is no Peierls phase between the diamond plaquette, the hopping between the Wannier orbitals, ${\bm T}_{{\bf R}_{1,2}}$, is purely real. Our primary focus lies in nearly flat bands, where a well-defined Fermi surface (FS) contour exists. This allows us to construct a superconductivity theory based on FS instability, and solve it analytically.

We arrange the Bloch basis in the invariant subspace of the point-group symmetry as ${\bf Z}({\bf k})={\bf Z}_{1}({\bf k})\oplus{\bf Z}_{2}({\bf k})\oplus...$. Here for the $i^{\rm th}$ - nearest neighbors with $d_i$ number of sites, we define a $d_i$-dimensional vector as ${\bf Z}_{i}:=\big(z_{1}~...~z_{d_i}\big)^T$, $\forall {\bf k}$. For the 2D SO lattice belonging to the $\mathsf{D}_{4h}$ group, the irreducible representations (irreps) are $\mathsf{A}\otimes\mathsf{B}\otimes \mathsf{E}$. We show that each nearest neighbor vector belongs to one of the irreps.

\begin{table*}[ht!]
\caption{\label{tab:irreps} The irreps from Eq.~\eqref{Eq:irrepsiband} are presented with corresponding ${\bf k}$ space representation, parity and tight-binding hoppings and required FS topology to produce pairing states in a repulsive interaction scenario. $t_{\bf R}$ ($t'_{\bf R}$) are the nearest-neighbor-hopping elements from ${\bf R}=0$ to four of the $\{{\bf R}\}_1$ ($\{{\bf R}\}_2$) Wannier sites.  The results are the same for $\Delta_{\nu}$ in terms of nearest neighbors $\Delta_{\bf R}$.}
\begin{tabular}{llllll}
 Basis ($z_{\nu}$)& ${\bf k}$-rep& $\mathsf{G}$ irrep & Parity & $t_{\nu}$ & FS  for pairing\\ 
 \hline
$z_{s_{x^2+y^2}}$& $2(\cos{k_x} +\cos{k_y})$ & $\mathsf{A}_1$ & $+1$ & $(t_1+t_2+t_3+t_4)/4$ & No pairing \\
     $z_{d_{x^2-y^2}}$& $2(\cos{k_x} -\cos{k_y})$ &  $\mathsf{B}_1$& $+1$& $(t_1-t_2+t_3-t_4)/4$ & All FS topologies\\
     $z_{p_x/p_y}$& $i\sqrt{2}\sin{k_{x/y}}$&  $\mathsf{E}$& $-1$& $\sqrt{2}(t_{1,2}-t_{3,4})$ & Small (e-/h-) FSs\\
    $z_{s_{x^2y^2}}$& $2\cos{k_x} \cos{k_y}$& $\mathsf{A}_1$& $+1$& $(t'_1+t'_2+t'_3+t'_4)/4$ & Near half-filling\\
     $z_{d_{xy}}$& $2\sin{k_x}\sin{k_y}$&  $\mathsf{B}_2$& $+1$& $(t'_1-t'_2+t'_3-t'_4)/4$ & For small (e-/h-) FSs\\
     $z_{p_{x\pm y}}$& $i\sqrt{2}\sin{(k_{x}\pm k_y)}$&  $\mathsf{E}$& $-1$& $\sqrt{2}(t'_{1,2}-t'_{3,4})$ & For small (e-/h-) FSs\\
      \hline
\end{tabular}
\end{table*}

On a square lattice, we have $d_i=4$ $\forall i$, as shown in Fig.~\ref{fig:sol_new}. For the first-nearest-neighbor ($i=1$), we have $\{{\bf R}\}_1=(\pm 1,0), (0,\pm 1)$, for the second-nearest-neighbor ($i=2$)  $\{{\bf R}\}_2=(\pm 1,\pm 1)$, and so on. The invariant subspaces can be written in the irreps as
\begin{eqnarray}
{\bf Z}_1({\bf k})&=&(z_{s_{x^2+y^2}}, z_{d_{x^2-y^2}}, z_{p_x}, z_{p_y}),\nonumber\\
{\bf Z}_2({\bf k})&=&(z_{s_{x^2y^2}}, z_{d_{xy}}, z_{p_{x+y}}, z_{p_{x-y}}),\nonumber\\
{\bf Z}_3({\bf k})&=&{\bf Z}_1(2{k}_x,2{ k}_y),~{\bf Z}_6({\bf k})={\bf Z}_2(2{k}_x,2{ k}_y),\nonumber\\
{\bf Z}_4({\bf k})&=&{\bf Z}_2(2{k}_x,{ k}_y),~{\bf Z}_5({\bf k})={\bf Z}_2({k}_x,2{ k}_y),
\label{Eq:irrepsiband}
\end{eqnarray}
and so on. The basis functions are given in Table~\ref{tab:irreps}, visualized in Fig.~\ref{fig:z_nu}, and derived in Appendix~\ref{App:Irreps}.\footnote{Note that the Wannier states in the $z_{\nu}$ basis can be a realization of the compact Wannier orbital \cite{Bergman2008} $w_{\nu}({\bf r})=\int_{\bf k}z_{\nu}({\bf k})u({\bf k},{\bf r})e^{i{\bf k}\cdot{\bf r}}$.}\cite{Bergman2008} (Note that $z_{p_{x/y}}$ are not the irreps of $\mathsf{D}_{4h}$ group, rather their complex combinations $z_{p_{\pm}}=z_{p_x}\pm i z_{p_y}$ are. The same goes for $z_{p_{x\pm y}}$.) Since different irreps do not mix, we can write the BdG equation Eq.~\eqref{Eq:BdG} in the irreps basis of $z_{\nu}$ as
\begin{eqnarray}
H&=&\int_{{\bf k}}~\sum_{{\nu}}H_{\nu}z_{\nu}({\bf k})
|{\nu},{\bf k}\rangle \langle {\nu},{\bf k}|,
\label{Eq:BdG1band}
\end{eqnarray}
where the chemical potential term is included for $\nu=0$. $H_{\nu}$ is a $2\times 2$ matrix in the particle-hole basis given by
\begin{eqnarray}
{H}_{\nu} =\left(\begin{array}{cc}
     t_{\nu} & {\Delta}_{\nu}  \\
     {\Delta}_{\nu}^{*} &  -{t}_{\nu}
\end{array}\right).
\label{Eq:BdGiband}
\end{eqnarray}
($t_{\nu}$ and $\Delta_{\nu}$ are obtained from $T_{{\bf R}_{1,2}}$ and $\Delta_{{\bf R}_{1,2}}$ with the same basis transformation map employed on ${\bf Z}_{{d}_i}=U_i {\bf Z}_{\nu}$, as given in Table~\ref{tab:irreps}.)  $t_{\nu}$ ( $\Delta_{\nu}$) corresponds to the hopping (pairing) between a Wannier state at ${\bf R}_1=0$ and the $\nu^{\rm th}-$ compact orbital. In the SC ground state, only a single (or degenerate) irrep $\bar{\nu}$ is present $-$ say, $\Delta_{\bar{\nu}}\ne 0$, while other irreps vanish. In this case, the $\bar{\nu}$ irrep (a compact particle-hole state) fractionalizes into two Bogoliubov orbitals within each unit cell, whereas the remaining orbitals (${\nu\ne \bar{\nu}}$) remain unchanged. The Fermi momenta ${\bf k}_F$ and the pairing interaction $\Gamma({\bf k}_{1,2})$ are determined by all irreps, ensuring that the self-consistent solution  $\Delta_{\bar{\nu}}\ne 0$ is realized. A detailed discussion of this mechanism is provided in Appendix~\ref{App:MFT}.

We focus on specific points in the tight-binding parameter space where $t_{\nu=\bar{\nu}}\ne \epsilon $ and $t_{\nu\ne\bar{\nu}}=0$, with $\epsilon$ kept small to ensure a nearly flat band. At these points, only one irrep $z_{\nu}$ contributes, simplifying the analytical solution of the SC gap equation in Eq.~\eqref{Eq:eivenval1}. In the irreps space, the gap equation takes a reduced form: 
\begin{eqnarray}
 \int_{{\bf k}_2} \Gamma_{\bar{\nu}}^{\bar{\nu}}({\bf k}_{1,2})z_{\bar{\nu}}({\bf k}_2)=-\lambda  z_{\bar{\nu}}({\bf k}_1).
    \label{Eq:eivenval}
\end{eqnarray}
$\Delta_{\bar{\nu}}$ drops out from both sides. Here the RPA interaction $\Gamma_{\bar{\nu}}^{\bar{\nu}}$ is diagonal in the $\bar{\nu}$ irrep, but not diagonal between ${\bf k}_{1,2}$.  The ${\bf k}$ integration is reduced to the Fermi momenta ${\bf k}_F$, which is determined by the contour $|z_{\bar{\nu}}({\bf k}_F)|=t_{\bar{\nu}}/\mu$ where $\mu$ is the chemical potential.\footnote{Since $z_{\nu}$ are linearly independent Bloch states at each ${\bf k}$, a flat band solution $\xi({\bf k})=0, \forall {\bf k}$ demands $t_{\nu}=0, \forall\nu$. Alternatively, if one or more $t_{\nu}$ turns out to be degenerate, a linear superposition state $z_{\mu}({\bf k})=\sum_{\nu}' C_{\nu,\mu}({\bf k})z_{\nu}({\bf k})$, while summation runs over the degenerate states, is also valid for all values of ${\bf C}({\bf k})$. If there exists a function ${\bf C}({\bf k})$ such that $z_{\mu}({\bf k})=0, \forall{\bf k}$ we have a flat band.} Accordingly, $\Gamma_{\bar{\nu}}^{\bar{\nu}}({\bf k}_{1,2})$ has a strong peak  owing to FS nesting and/or RPA instability at some nesting vector ${\bf Q}={\bf k}_{2F}-{\bf k}_{1F}$. Then a finite pairing strength $\lambda>0$ exists if $z_{\nu}({\bf k}_{1F})$ and  $z_{\bar{\nu}}({\bf k}_{1F}+{\bf Q})$ change sign across the FS nesting vector. Otherwise, a SC solution does not exist for this repulsive interaction $\Gamma>0$. \footnote{Note that when other $t_{\nu}\ne 0$ terms contribute, the ${\bf k}_F$ topology changes and, hence, the nesting profile and the nesting strength change, while the main criterion for sign reversal in the SC irrep $z_{\bar{\nu}}$ remains the same. }

We denote $t_i$ and $t'_i$ as the first and second nearest neighbor hoppings. We consider several representative special points in ($t_i, t'_i$) where only one $t_{\bar{\nu}}$ is finite, see Table~\ref{tab:irreps}. 

(i) $t_i=\epsilon$, and $t'_i=0$. In this case $t_{s_{x^2+y^2}}=\epsilon$, and the rest are zero. This makes only $\Delta_{s_{x^2+y^2}}$ pairing possible. For the single band case with either electron or hole-like FSs, where the FS is centered at $\Gamma=(0,0)$ or M=$(\pi,\pi)$, the FS does not change sign.  As a result, Eq. ~\eqref{Eq:eivenval} does not have any non-trivial solution except $\lambda=0$. However, a finite $\lambda$ solution can arise for a two- (or higher) band model with both electron and hole-like pockets, as observed in Fe-pnictide superconductors.\cite{Chubukov2012, Mazin2008, Das2012JoP, Das2015SR}.

(ii) Next, consider $t_1=-t_2=t_3=-t_4=\epsilon$, which gives the only non-zero irrep is $t_{d_{x^2-y^2}}=\epsilon$. In this case, any single band FS will have a $\lambda>0$ solution. This gives a $\Delta_{d_{x^2-y^2}}$  spin singlet pairing. 

(iii) For $t_1=t_2=-t_3=-t_4$, we have $t_{x/y}=\pm 2\sqrt{2}\epsilon$, while the rest is zero. Here, a single band pairing is possible only for an electron like FS, but not for any hole-doped case. Since the pairing is odd under spatial parity, the spin sector is a triplet $\mathbb{X}_-$ in a single band case. 

Next, we examine next-nearest-neighbor pairing channels, which are promoted by the next nearest neighbor hopping $t'_i$. Achieving $t^{'}_i\ne 0$ while $t_i=0$ in challenging unless unless the system is in the flat-band limit. Consequently, next-nearest-neighbor pairing does not arise in isolation but coexists with nearest-neighbor pairing (as in graphene \cite{Black-Schaffer_2014,Ray2019}). We can similarly discuss special cases: (iv) For $t'_{s_{{x^2y^2}}}=\epsilon$, an $s_{x^2y^2}$ pairing emerges near half-filling, where the FS is large enough to support sign changes. In multiband systems,  this pairing can occur if different FSs exhibit opposite sign of $z_{s_{{x^2y^2}}}$. (v) When $t'_{d_{xy}}=\epsilon$ (with all other hoppings being zero),  a $d_{xy}$ pairing occurs near half-filling. (vi) Finally, in rare cases where $t'_{p_{x\pm y}}=\epsilon$, an exotic $p_{x\pm y}$ pairing state may form for small electron- and hole- like FSs, though not near half-filling. All these scenarios are summarized in Table~\ref{tab:irreps}. 

\subsection{Bogoliubov orbitals}

Finally, we evaluate the Wannier-like states of the Bogoliubov quasiparticles following the framework established in Eq.~\ref{Eq:BdG4}. Starting from the non-local Bogoliubov transformation for a single-band case, ${U}_{{\bf R}_1}{H}_{{\bf R}_{1,2}}U_{{\bf R}_2}^{\dagger}=-D_{{\bf R}_{1,2}}$, we set ${\bf R}_1=0$, and transform to the irreps space $\nu$, yielding ${U}{H}_{\nu}U_{\nu}^{\dagger}=-D_{\nu}$. Here ${H}_{\nu}$ is defined in Eq.~\eqref{Eq:BdGiband}, while $D_{\nu}$ represents the diagonal matrix with components $\pm E_{\nu}$. The Bogoliubov transformation takes the form $U_{{\nu}}=(I\cos\theta_{{\nu}}+\sigma_y\sin\theta_{{\nu}})$ with  ${\rm tan}2\theta_{{\nu}}=-\Delta_{{\nu}}/t_{{\nu}}$ \cite{tinkham2004}.  By substituting $U_{\nu}$ in Eq.~\eqref{Eq:BdG4}, we find that the Bogoliubov orbital localizes as a compact orbital in the pairing irrep $\bar{\nu}$.  

\subsection{Tight-binding model and numerical solution}

To validate our analytical results, we numerically compute the pairing symmetry for the full four-band model\cite{Pal2018}, by tuning the chemical potential to the flat band. The corresponding band structure is reproduced in Fig.~\ref{fig:2dSOL}(a). We examine three representative FS topologies: hole doping in Fig.~\ref{fig:2dSOL}(e), near half-filling in Fig.~\ref{fig:2dSOL}(f), and electron doping in Fig.~\ref{fig:2dSOL}(g). The corresponding RPA susceptibility ${\rm Tr}\Pi_s({\bf q})$ in the dominant spin channel is plotted in the upper panels of Fig. \ref{fig:2dSOL}(b-d). By solving the multiband eigenvalue equation in Eq.~\eqref{Eq:eivenval1}, we determine the pairing eigenvalues and eigenfunctions, with the dominant symmetry $z_{\bar{\nu}}$ plotted as a color gradient on the FS in Figs.~\ref{fig:2dSOL}(e-g).

For the hole-doped case with a large FS, the FS nesting near ${\bf q}_1=(\pi,\pi)$ dominates, leading to a sign-reversing $z_{d_{x^2-y^2}}$ pairing symmetry, as shown in Fig.~\ref{fig:z_nu}(b). In contrast, the electron-doped regime exhibits a squarish FS near the flat band, shifting the nesting wavevector to ${\bf q}_2=(\pi,0)/(0,\pi)$ [Fig.~\ref{fig:2dSOL}(d)]. This favors a $z_{p_{x/y}}$ pairing state, as confirmed numerically in Fig.~\ref{fig:2dSOL}(g). 

At the critical point near half-filling, where  ${\bf q}_1$ and ${\bf q}_2$ nestings are degenerate, an intriguing question arises: Do the $z_{d_{x^2-y^2}}$ and $z_{p_{x/y}}$ pairings coexist or compete? Since they belong to different irreps and different fermion parity sectors,  their coexistence requires lifting spin degeneracy (by spin-orbit coupling) or lattice symmetry breaking.\cite{Sigrist1991} Alternatively, a quantum liquid-like phase involving disordered mixtures of the two pairing symmetries may emerge, as proposed in ReHf$_2$ \cite{Mandal2022}, or a first-order phase transition separates the two pairings, as seen in Sr$_{0.1}$Bi$_2$Se$_3$ \cite{Neha2019}. However, it is worth noting that exact degeneracy between these nesting instabilities occurs only in the ideal flat-band limit;  any finite dispersion lifts this degeneracy, favoring one pairing channel over the other.

\subsection{First-principles calculations}\label{sec-3.1}

\begin{figure}[t]
\centering
\includegraphics[width=0.45\textwidth]{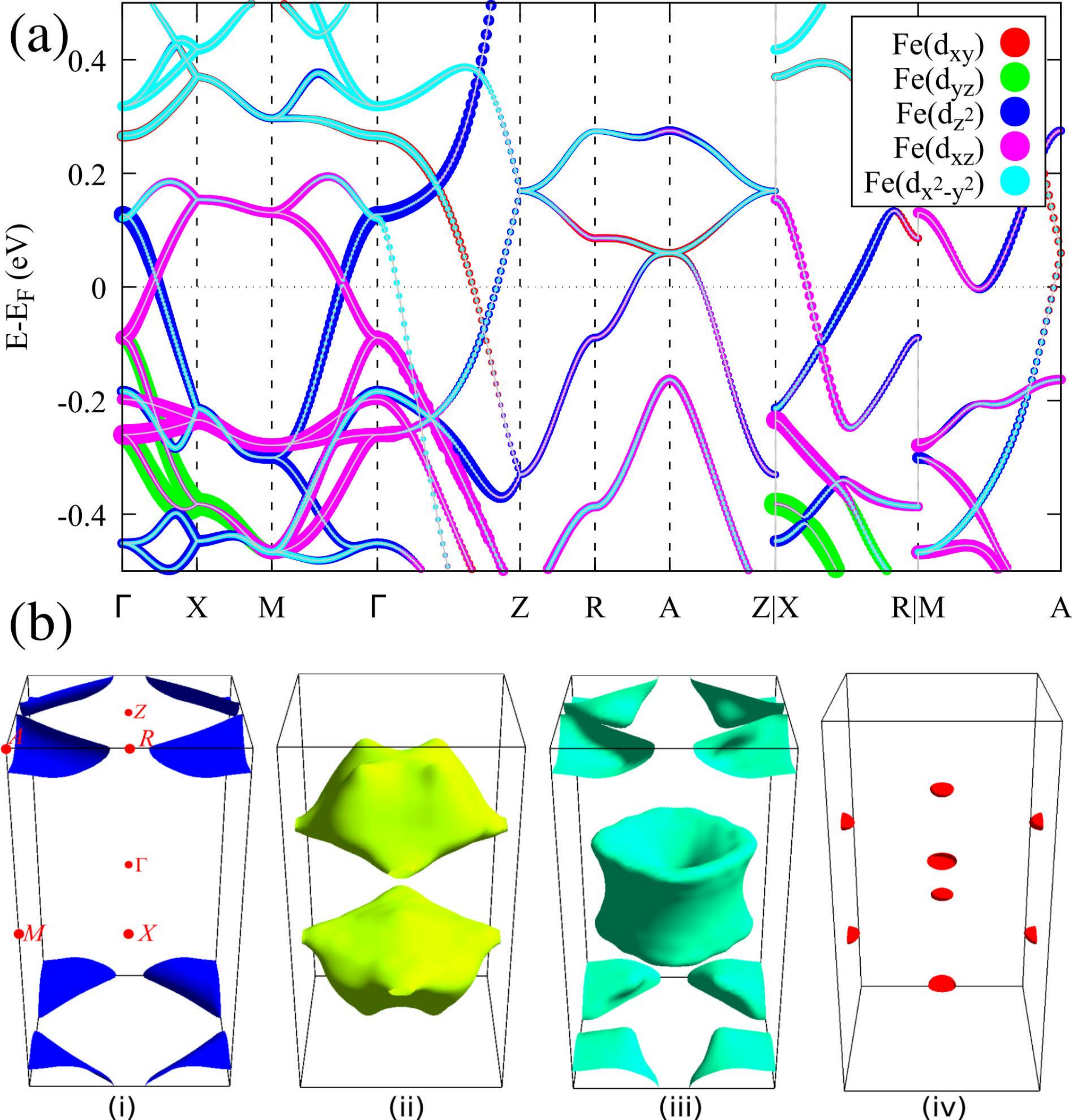}
\caption{\label{fig:bnd-fs}(a) Calculated band structure and projected orbitals weight of Fe-$d$ orbitals are shown here. The contribution from other orbitals of Fe and other atoms is negligibly small near the Fermi level. (b) The FSs of Lu$_2$Fe$_3$Si$_5$ are composed of four bands, which we label as band 1,2,3 and 4. Band 1 has a negligibly small FS. Here the colors have no meaning except to distinguish different bands.}
\end{figure}%

Next, we extend our analysis to a multiband Wannier orbital description of  Lu$_2$Fe$_3$Si$_5$. Figure\ \ref{fig:unitcell} displays the tetragonal structure (space group $P4/mnc$, No. 128), with detailed symmetry analysis and DFT calculations provided in Appendix~\ref{sec-dft_result}. The calculated band structure, shown in Fig.~\ref{fig:bnd-fs}(a), reveals dominant $d_{z^2}$ and $d_{xz}/d_{yz}$ orbital weights near the Fermi level, in contrasts to weaker contributions from basal plane $d$ orbitals. This orbital selectivity, combined with significant FS anisotropy along $k_z$ underscores the materials's strong three-dimensional character, see Fig.~\ref{fig:bnd-fs}(b). As analyzed in Sec.~\ref{sec-3} [particularly for the Hamiltonian in Eq.\eqref{Eq:BdG1band}], the enhanced hopping along the $z$-direction is expected to favor pairing symmetry along the same $k_z$ direction.

\begin{figure}[!ht]
\centering
\includegraphics[width=0.45\textwidth]{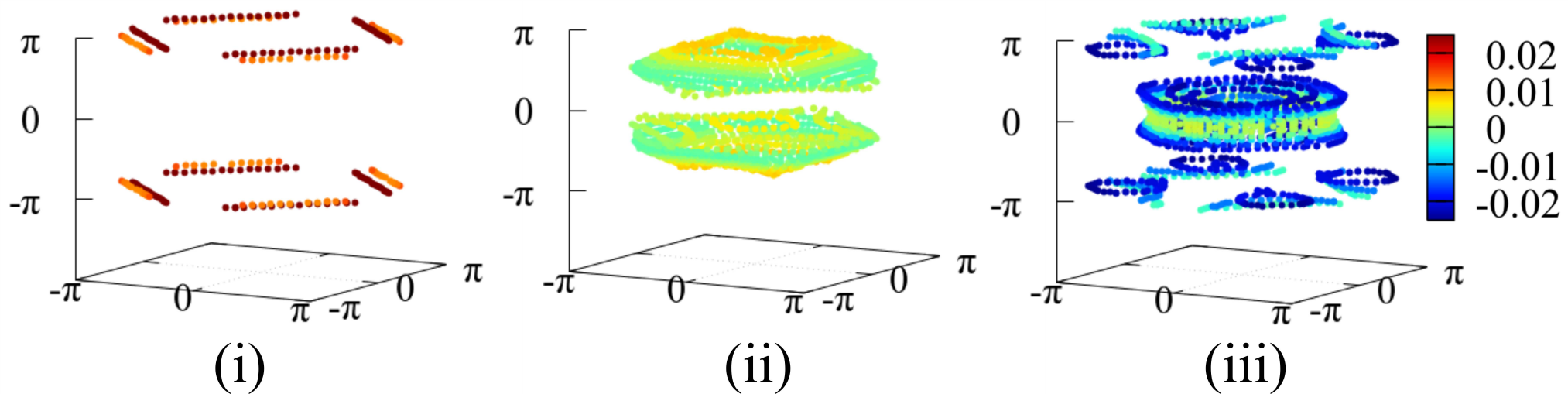}
\caption{\label{fig:pair} Pairing eigenfunctions for the largest eigenvalues for the solution of Eq.\eqref{Eq:eivenval1} are shown on the three FS. The gradient color maps show that the gap function changes sign between bands 1 and 3 with very weak momentum anisotropy, which we attribute to $s^{\pm}$ pairing symmetry. The colormap for band 2 suggests that it is a $s_{z^2}$ pairing symmetry.}
\end{figure}

We construct Wannier orbitals for the three bands using the Wannier90 program \cite{Wannier90}, with the results displayed in Fig. \ref{fig:wf_Fs}. As expected from the strong $k_z$ anisotropy and basal plane isotropy, the Wannier orbitals exhibit similar $k_z$-dependent anisotropy. These orbitals predominantly show $s_{z^2}$ and $d_{z^2}$ orbital symmetries. 

Given the system's multiband nature, an analytical treatment of pairing symmetry following Sec.~\ref{sec-3} becomes cumbersome. Instead, we numerically evaluate the eigenvalues of the interaction vertex $\Gamma$ in Eq.~\eqref{Eq:eivenval1}. This is equivalent to calculating the expectation values of $\Gamma$ for symmetry-allowed pairing channels $z_{\nu}$. We consider onsite ($s$-wave) and first nearest-neighbor pairing irreps $s_{x^2+y^2}$, $d_{x^2-y^2}$ $s_{z^2}$, $p_{x/y/z}$ (see Appendix~\ref{App:Irreps}). Our results reveal two degenerate eigenvalues $\lambda_s$ between bands 1 and 3, and $\lambda_{s_{z^2}}$ for band 2, with both pairing states being in the spin singlet channels (see Fig.~\ref{fig:pair}). 

The emergence of these unconventional pairing symmetries can be understood through the principle that strong FS nesting at a specific wavevector stabilizes a pairing symmetry that changes sign on the FS regions connected by this nesting vector. Bands 1 and 2 exhibit significant inter-band nesting, with minimal momentum anisotropy, favoring a pairing gap function that changes sign between these bands. The lack of anisotropy in their FS results in an isotropic $s$-wave for each band, but with opposite sign between them (an $s^{\pm}$ wave state), analogous to the iron-pnictide superconductors \cite{Chubukov2012, Mazin2008, Das2012JoP, Das2015SR}. 

In contrast, band 3 displays a pronounced anisotropy along the $k_z$ direction, with its Wannier state possessing $z_{s_{z^2}}$ symmetry. This anisotropy leads to intra-band FS nesting between the Fermi pockets located at $k_z=0$ and $k_z=\pm \pi$, naturally stabilizing a pairing that changes sign between these $k_z$ planes. The $z_{s_{z^2}}$ irreps of the Wannier state consistently stabilizes this axial pairing symmetry, aligning with the nesting profile. This axial pairing symmetry, driven by the same-symmetric Wannier irrep $z_{s_{z^2}}$, was predicted and subsequently confirmed in infinite-layer nickelate superconductors.\cite{Adhikary2020}. 

Based on the DFT results, we provide a three band phase space BdG model with the above pairing state in the irreps of the Bloch vectors in Eq.~\eqref{Eq:multigapApp} in Appendix~\ref{App:Examples}.

Our finding of band resolved two $s$-wave pairings is consistent with the penetration depth data fitted with $s+s$ pairing symmetry in  Lu$_2$Fe$_3$Si$_5$ compound.\cite{Biswas2011} The coexistence of two sign-reversal pairing gaps leads to multiple magnetic resonance peaks arising from  distinct Cooper pair fluctuations\cite{Das2011PRL} and can be detected by inelastic neutron scattering measurements.

\section{Summary and Outlook}\label{sec-4}
We devoted significant parts of the paper to developing a formalism for constructing a lattice model that can be applied to non-interacting band structure, many-body ground states, and systems coupled to a bath. Unlike conventional approaches that rely on localized real-space orbitals, our framework begins by describing states, operators, and fields in phase space, with final results projected onto either real or momentum space as needed.Crucially, rather than treating real and momentum spaces as a unified manifold, we work within momentum space while embedding real-space information via the Bloch vector space at each momentum point. Orbital and spin degrees of freedom are incorporated as product states with the Bloch vector, while quantum statistics, topology, and entanglement emerge naturally from the Hamiltonian.

This approach offers several advantages: 

(i) {\it Built-in localization and uncertainty}: The localization and uncertainty of states in real and momentum space are inherently fixed before adding correlation, entanglement, and topology. 

(ii) {\it Unified treatment of global and local properties: } The method bypasses any obstruction between  global and local properties defined in momentum and real space, as they are included via unitary rotation and projection operators on the phase space. 

{\it Applications.} For demonstrations and applications of the framework, we focused on superconductivity in SO materials. First, we analyzed a single flat-band model, where the method enabled an analytical solution for unconventional pairing symmetries within a spin-fluctuation mechanism. A central insight is that while the SC ground state exhibits global coherence across all bands in momentum space, the pairing symmetry is determined by compact orbitals aligned with the irreps of the lattice symmetry.

We then extended the analysis to the multiband SO material Lu$_2$Fe$_3$Si$_5$, combining DFT-derived band structures and Wannier orbitals. We found that the interplay between multibands and three dimensionality produces Wannier orbitals of corresponding symmetries. In fact, consistent with the analytical solutions, we find here as well that the obtained pairing symmetries $-$ $s^{\pm}$ wave across two bands and a nodal $s_{z^2}$ symmetry for a third band $-$ are consistent with their Wannier orbital irreps. 

{\it Future prospects.} While our focus here has been on unconventional superconductivity, the formalism is kept general and adaptable to other exotic many-body states. The method can be viewed as a variational approach with variational parameters including single quasiparticles excitations for a Fermi liquid state, density wave orders for symmetry-breaking phases, or even exotic Bell pairs or Cooper pairs or entangled pairs for spin liquid or fractional quantum Hall states. This flexibility opens new possibilities, such as: (i) Assigning quantum statistics and braiding phases to orbitals and Bloch vectors as variational parameters. (ii) Modeling fractional quantum Hall or fractional Chern states by attaching vortices in Bloch vector space, rather than in orbital states. (iii) Stabilizing liquid states of compact orbitals in flat-band degenerate manifolds or spin-singlet liquids in quantum spin-liquid metals. These and other possibilities remain open for future exploration.

\begin{acknowledgments}
R.S. acknowledges the Science and Engineering Research Board (SERB), Government of India, for providing the NPDF fellowship with grant number PDF/2021/000546. T.D. acknowledges funding from Core Research Grant (CRG) of S.E.R.B. (CRG/2022/003412), under I.R.H.P.A Grant (IPA/2020/000034) and benefited from the computational resources (SERC) from the Indian Institute of Science.
\end{acknowledgments}

\appendix

\section{Density-Density interaction and Random Phase Approximation (RPA)}\label{lab:RPA-2}

The components of the two types of transfer matrices, namely the density fields (Eqs.~\eqref{eq:2bodywf}, and \eqref{eq:DensityMatrix}) and pairing fields (Eq) are written explicitly as
\begin{eqnarray}
\mathcal{Z}_{{\bf R}_{1,2}}({\bf k}_{1,2})  &=&  z_{{\bf R}_1}^*({\bf k}_1)z_{{\bf R}_2}({\bf k}_2),\nonumber\\
\mathcal{W}_{n_{1,2}}({\bf r}_{1,2}) &=& w_{n_1}^*({\bf r}_1)w_{n_2}({\bf r}_2),\nonumber\\
\mathcal{X}_{s_{1,2}} &=& \chi_{s_1}^*\chi_{s_2}.
\label{eq:densityfields}\\
\mathbb{Z}_{{\bf R}_{1,2}}({\bf k}_{1,2})&=&\mathsf{P}_{\theta({\bf R}_{12}})[z_{{\bf R}_1}({\bf k}_1)z_{{\bf R}_2}({\bf k}_2)],  \nonumber\\
\mathbb{W}_{n_{1,2}}({\bf r}_{12}) &=& \mathsf{P}_{\theta(n_{12})}[w_{n_1}({\bf r}_1)w_{n_2}({\bf r}_2)],\nonumber\\
\mathbb{X}_{s_{1,2}} &=& \mathsf{P}_{\theta(s_{1,2)}}[\chi_{s_1}\chi_{s_2}].
\label{eq:pairfields}
\end{eqnarray}
Recall that the composite index ${\bm n}_{i}=(n_i,s_i)$ combines Wannier orbital and spin indices. In the basis of density fields, we obtain the matrix-elements of the one-body operator ${\bf T}({\bf r})$ as follows (refer to Eq.~\eqref{eq:1bodyH}): 
\begin{eqnarray}
H_{T}  &=& 
\int_{{\bf k},{\bf r}} {\Psi}^{\dagger}({\bf k},{\bf r}){\bf T}({\bf r}){\Psi}({\bf k},{\bf r})~|{\bf k}\rangle \langle {\bf k}|\nonumber\\
&=&\int_{{\bf k},{\bf r}}\sum_{{\bm n}_{1,2}} {T}({\bf k},{\bf r})\mathcal{W}_{n_{1,2}}({\bf r})\mathcal{X}_{s_{1,2}} |{\bm n}_1,{\bf k} \rangle \langle {\bm n}_2,{\bf k}|,\nonumber\\
&=&\int_{\bf k} \sum_{{\bm \kappa}_{1,2}} T_{{\bm \kappa}_{1,2}}\mathcal{Z}_{{\bf R}_{1,2}}({\bf k}) |{\bm \kappa}_1,{\bf k} \rangle \langle {\bm \kappa}_2,{\bf k}|.\nonumber\\
&=&\int_{\bf k} \sum_{{\bm n}_{1,2}} {T}_{{\bm n}_{1,2}}({\bf k}) |{\bm n}_1,{\bf k} \rangle \langle {\bm n}_2,{\bf k}|.
\label{eq:1bodyHApp}
\end{eqnarray}
We define $T_{{\bm \kappa}_{1,2}}$ = $ \int_{\bf r}~T_{{\bm \kappa}_{1,2}}({\bf r})\mathcal{W}_{n_{1,2}}({\bf r})\mathcal{X}_{s_{1,2}}$ as tight-binding hopping tensor,  and ${T}_{{\bm n}_{1,2}}({\bf k})$ = $\sum_{{\bf R}_{1,2}}T_{{\bm \kappa}_{1,2}}\mathcal{Z}_{{\bf R}_{1,2}}({\bf k})$ is the corresponding dispersion.  (Note that the other phase factor $e^{i{\bf k}\cdot{\bf r}}$ drops out here.) Proceeding similarly, for the two-body operators are expressed in the pair fields as (refer to Eq.~\eqref{eq:intH2App}) 
\begin{widetext}
\begin{eqnarray}
H_{V} &=& 
\int_{{\bf k}_{1-4},{\bf r}_{1,2}}{\pmb {\Psi}}^{\dagger}({\bf k}_{1,2},{\bf r}_{1,2}){\bf V}({\bf r}_{1,2}){\pmb {\Psi}}({\bf k}_{3,4},{\bf r}_{1,2})\delta_{{\bf k}_{1-4}}({\bf r}_{1,2})~|{\bf k}_{1,2}\rangle \langle {\bf k}_{3,4}|, \nonumber\\
&=&\int_{{\bf k}_{1-4},{\bf r}_{1,2}} ~
V({\bf k}_{1-4},{\bf r}_{1,2})\delta_{{\bf k}_{1-4}}({\bf r}_{1,2})\mathbb{W}_{n_{1,2}}^{*}({\bf r}_{1,2})\mathbb{W}_{n_{3,4}}({\bf r}_{1,2})\mathbb{X}_{s_{1,2}}^{*}\mathbb{X}_{s_{3,4}}|{\bm n}_{1,2},{\bf k}_{1,2}\rangle\langle{\bm n}_{3,4},{\bf k}_{3,4}|,\nonumber\\
&=&\int_{{\bf k}_{1-4}}\sum_{{\bm \kappa}_{1-4}} \bar{V}^{{\bm \kappa}_{2,4}}_{{\bm \kappa}_{1,3}}\mathbb{Z}^{*}_{{\bf R}_{1,2}}({\bf k}_{1,2})\mathbb{Z}_{{\bf R}_{3,4}}({\bf k}_{3,4})\delta_{{\bf k}_{1-4}} |{\bm \kappa}_{1,2},{\bf k}_{1,2}\rangle\langle {\bm \kappa}_{3,4},{\bf k}_{3,4}|.\nonumber\\
&=&\int_{{\bf k}_{1-4}}\sum_{{\bm n}_{1-4}} \bar{V}^{{\bm n}_{2,4}}_{{\bm n}_{1,3}}({\bf k}_{1-4})  |{\bm n}_{1,2},{\bf k}_{1,2}\rangle\langle {\bm n}_{3,4},{\bf k}_{3,4}|.
\label{eq:intH2App}
\end{eqnarray}
\end{widetext}
Note that Eq.~\eqref{eq:intH2App} can also be expressed in terms of the density matrices in Eq.~\eqref{eq:densityfields}, however,  it is easier to implement the exchange parity symmetry in terms of the pair fields.

Invoking translational invariance, we set ${\bf R}_{3}={\bf R}_{4}=0$ for the initial states. In addition, we often consider onsite interactions, i.e. ${\bf R}_{1}={\bf R}_{2}=0$. This means $V^{{\bf R}_{2,4}}_{{\bf R}_{1,3}} =\mathbb{I}\delta_{{\bf R}_{1-4},0}$. (The corresponding ${\bf R}_{1,2}\ne 0$ components give the so-called extended Hubbard model, which we ignore here.)  We can write the remaining orbital and spin parts in the eigenbasis of exchange parity as:
\begin{eqnarray}
V_{\theta(W)} &=& \frac{1}{2}\left(V_{n_{1,3}}^{n_{2,4}}+e^{i\theta(W)} V_{n_{2,4}}^{n_{1,3}}\right),~\nonumber\\ 
V_{\theta(X)} &=&\frac{1}{2}\left(V_{s_{1,3}}^{s_{2,4}}+e^{i\theta(X)} V_{s_{2,4}}^{s_{1,3}}\right).
\label{eq:GammaParity}
\end{eqnarray}
Typically, for the fermion case of present interest, in the absence of any gauge fields, we have $\theta=0,\pi$ as $\mathsf{P}_{\theta}^2=1$. Then we have two allowed choices: orbital symmetric, spin antisymmetric: $V_{W+}V_{X-}$, and orbital antisymmetric and spin symmetric: $V_{W-}V_{X+}$ parts. We identify these interactions in terms of known interactions in Sec.~\ref{Sec:RPA}. For systems with spin-orbit coupling and other entanglement properties, we implement the unitary rotation (\textit{c.f.} Chebsch-Gordon coefficients) in the interaction term.

The explicit expression for the onsite interactions, given in Eq.~\eqref{eq:Int}, can be written from Eq.~\eqref{eq:IntW} in terms of the pairing and density fields defined in Eqs.~\eqref{eq:densityfields} and \eqref{eq:pairfields} as
\begin{widetext}
\begin{eqnarray}
U_n&=&V_{nn}^{nn}V_{ss}^{\bar{s}\bar{s}}= 
\int_{{\rm r}_{1,2}} ~{V}^{00,nn,\bar{s}\bar{s}}_{00,nn,ss}({\bf r}_{1,2}) \mathcal{W}_{nn}({\bf r}_1)\mathcal{W}_{nn}({\bf r}_2)\mathcal{X}_{ss}\mathcal{X}_{\bar{s}\bar{s}},\nonumber\\
U'_{mn}&=&V_{nn}^{mm}V_{ss}^{tt}=\int_{{\rm r}_{1,2}}~{V}^{00,mm,tt}_{00,nn,ss}({\bf r}_{1,2}) \mathcal{W}_{nn}({\bf r}_1)\mathcal{W}_{mm}({\bf r}_2)
\mathcal{X}_{ss}\mathcal{X}_{tt},\nonumber\\
J_{mn}&=&V_{nn}^{mm}V_{st}^{st}=\int_{{\rm r}_{1,2}}~{V}^{00,mm,st}_{00,nn,st}({\bf r}_{1,2})\mathcal{W}_{mm}({\bf r}_1)\mathcal{W}_{nn}({\bf r}_2)
\mathcal{X}_{st}\mathcal{X}_{st},\nonumber\\
J'_{mn}&=&V_{nm}^{nm}V_{ss}^{\bar{s}\bar{s}}=\int_{{\rm r}_{1,2}}~{V}^{00,nm,\bar{s}\bar{s}}_{00,nm,ss}({\bf r}_{1,2}) \mathbb{W}^{*}_{nn}({\bf r}_1)\mathbb{W}_{mm}({\bf r}_2)
\mathbb{X}^{*}_{s\bar{s}}\mathbb{X}_{s\bar{s}}
\label{eq:onsite_ints}
\end{eqnarray}
\end{widetext}
It is convenient to express the abstract states and the matrix elements of the operators in the second quantized form. Introducing the creation operator in the Wannier basis as $|{\bf R}ns\rangle$=$c_{{\bf R}ns}^{\dagger}|G\rangle$, where $|G\rangle$ is the many-body ground state serving as a vacuum to the low-energy excitations. Note that here, the orbital's Wyckoff position ${\bf r}$ does not enter into the abstract states as they are included in the wavefunctions $w_{n}({\bf r})$. The corresponding field operator is defined as $\psi_{ns}^{\dagger}({\bf r})=\sum_{{\bf R}}w_{n}({\bf r})c_{{\bf R}ns}^{\dagger}$. The local densities are defined as $\mathsf{n}_{{\bf R}ns}=c_{{\bf R}ns}^{\dagger}c_{{\bf R}ns}$, we obtain the charge density as $\mathsf{n}_{{\bf R}n}=\frac{1}{2}\sum_{s}\mathsf{n}_{{\bf R}ns}$, spin operator ${\bf S}_{{\bf R}n}=\frac{1}{2}\sum_{st}c_{{\bf R}ns}^{\dagger}{\bm \sigma}_{st}c_{{\bf R}nt}$, and pair operator $\zeta_{{\bf R}n}=\mathsf{P}_{\theta}\sum_{s}c_{{\bf R}ns}c_{{\bf R}n\bar{s}}$. We write Eq.~\eqref{eq:intH2} by substituting Eqs.~\eqref{eq:onsite_ints} as
\begin{eqnarray} 
H_{V}&=&\sum_{{\bf R}ns}U_{n} \mathsf{n}_{{\bf R}ns}\mathsf{n}_{{\bf R}n\bar{s}}
+\sum_{{\bf R},n<m}\Big[(U'-J/2)_{mn}\mathsf{n}_{{\bf R}m}\mathsf{n}_{{\bf R}n} \nonumber\\ 
&& ~~~
-2J_{mn}{\bf S}_{{\bf R}m}\cdot{\bf S}_{{\bf R}n}
+ J'_{mn}\zeta_{{\bf R}m}^{\dagger}\zeta_{{\bf R}n}\Big].
\label{eq:Hubbard}
\end{eqnarray}
This is the famous Hubbard-Kanamori interaction Hamiltonian in a multiband system \cite{Kanamori1963}. To derive the RPA interaction as many-body renormalized interaction due to density-density fluctuations, we express Eq.~\eqref{eq:Hubbard} in terms of generalized density-matrix and non-local interaction form:
\begin{eqnarray} 
H_{V}(\mathsf{F}(t))&=&\sum_{{\bm \kappa}_{1-4}}{V}_{{\bm \kappa}_{1,3}}^{{\bm \kappa}_{2,4}} {\pmb \varrho}_{{\bm \kappa}_{1,2}}^{\dagger}{\pmb \varrho}_{{\bm \kappa}_{3,4}} \nonumber\\
&&
+ \sum_{{\bm \kappa}_{1,2}}\left(\mathsf{F}_{{\bm \kappa}_{1,2}}(t){\pmb \varrho}_{{\bm \kappa}_{1,2}}(t)+{\rm h.c.}\right),
\label{eq:HubbardRPA1}
\end{eqnarray}
(For local interaction ${\bf R}_{1,4}=0$, ${\bf V}$ is identified with $U$, $U'$, $J$, $J'$ defined in Eq.~\eqref{eq:onsite_ints}. The diagonal elements in the orbital indices of the density matrix ${\pmb \varrho}_{{\bm \kappa}_{1,2}}\delta_{{\bm \kappa}_{1,2}}$ correspond to the densities ($\mathsf{n}_{{\bm \kappa}_1}$, $S^{xyz}_{{\bm \kappa}_1}$, and $\zeta_{{\bm \kappa}_1}$ ) in Eq.~\eqref{eq:Hubbard}. We include $\mathsf{F}_{{\bm \kappa}_{1,2}}(t)$ as an external source term that causes a (local) fluctuation to the corresponding density matrix ${\pmb \varrho}_{{\bm \kappa}_{1,2}}$, and it then spreads over the entire lattice due to the interaction term $\Gamma$ and the one-body hopping term ($H_T$). The $\mathsf{F}_{{\bm \kappa}_{1,2}}(t)$ term drops out from the final result within the linear response theory. 

 The induced local density to the perturbation $\mathsf{F}_{{\bm \kappa}_{1,2}}(t)$ is ${\pmb \varrho}^{\rm (ind)}_{{\bm \kappa}_{1,2}}(t)$=${\pmb \varrho}_{{\bm \kappa}_{1,2}}(\mathsf{F}(t))-{\pmb \varrho}_{{\bm \kappa}_{1,2}}(\mathsf{F}=0)$. Within RPA, ${\pmb \varrho}^{\rm (ind)}_{{\bm \kappa}_{1,2}}(t)$ takes a \textit{mean} value, i.e., it is a $c$-number. Substituting this in the interaction term, we obtain a non-interacting term ${{\pmb \varrho}}_{{\bm \kappa}_{1,2}}^{\dagger}{{\pmb \varrho}}_{{\bm \kappa}_{3,4}}\rightarrow -({\pmb \varrho}^{\dagger}_{{\bm \kappa}_{1,2}}{{\pmb \varrho}}^{\rm (ind)}_{{\bm \kappa}_{3,4}} + {\rm h.c.}) + |{\pmb \varrho}^{\rm (ind)}|^2 + \mathcal{O}({\pmb \varrho}^2)$, where $|{\pmb \varrho}^{\rm (ind)}|^2$ is a number that shifts the energy, and we neglect the quadratic term in the density with respect to the mean value. Substituting this mean-field expansion in Eq.~\eqref{eq:HubbardRPA1}, we define an effective force experienced by the density as $\mathsf{F}^{\rm (tot)}_{{\bm \kappa}_{1,2}}(t)=\mathsf{F}_{{\bm \kappa}_{1,2}}(t) + \sum_{{\bm \kappa}_{3,4}}{V}_{{\bm n}_{1,3}}^{{\bm n}_{2,4}}{{\pmb \varrho}}^{\rm (ind)}_{{\bm \kappa}_{3,4}}(t)$. Then, using linear response theory, we define as
\begin{eqnarray}
 {\pmb \varrho}^{\rm (ind)}_{{\bm \kappa}_{1,2}}(t)&=&-i\int_{t'}\sum_{{\bm \kappa}_{3,4}}{\Pi^{(0)}}^{{\bm \kappa}_{2,4}}_{{\bm \kappa}_{1,3}}(t-t')\mathsf{F}_{{\bm \kappa}_{3,4}}(t'),\nonumber\\
 &=&-i\int_{t'}\sum_{{\bm \kappa}_{3,4}}{\Pi}_{{\bm \kappa}_{2,4}}^{{\bm \kappa}_{1,3}}(t-t')\mathsf{F}^{\rm (tot)}_{{\bm \kappa}_{3,4}}(t').  
\end{eqnarray}
$\Pi$, $\Pi^{\rm(0)}$ are the RPA and non-interacting density-density correlation functions, which can now be related to each other as ${\bm \Pi}_{s/c}= {\bm \Pi}^{(0)}\left[{\bf I}\mp {\bf V}{\bm \Pi}^{(0)}\right]^{-1}$, where ${\bm \Pi}$, ${\bf V}$ denote square matrices defined by the (2,2)- tensor. Subscript `$s$' corresponds to the spin fluctuation ${\pmb \varrho}\equiv S^{\pm}=S_x\pm iS_y$ term, and `$c$' corresponds to ${\pmb \varrho}\equiv \mathsf{n}, S_z, \zeta$ density matrices. The bare density-density correlator is defined in the Wannier basis as $({\Pi^{(0)}})^{{\bm \kappa}_{2,4}}_{{\bm \kappa}_{1,3}}(t-t)=-i\theta(t-t')\langle[{\pmb \varrho}_{{\bm \kappa}_{1,3}}(t),{\pmb \varrho}_{{\bm \kappa}_{2,4}}(t')]\rangle_0$, where $\theta$ is the step function for retarded response, $[.,.]$ corresponds to a commutator, and $<\cdots>_0$ gives the expectation value in the ground state of the non-interacting Hamiltonian.  The Fourier transformation of the susceptibility to the momentum space is given by 
$[\Pi^{(0)}({\bf q},\omega)]_{{\bm n}_{1,3}}^{{\bm n}_{2,4}}=\sum_{{\bf R}_{1,2}}\int_t e^{i\omega t}({\Pi^{(0)}})^{{\bm \kappa}_{2,4}}_{{\bm \kappa}_{1,3}}(t-t) z_{{\bf R}_1-{\bf R}_2}({\bf q})$.

In the absence of SOC or superconductivity or magnetism in the non-interacting Hamiltonian, the only term in the bare susceptibility that contributes is the particle-hole susceptibility, i.e. the so-called Lindhard functions defined for the multiband case as
\begin{eqnarray}
&&\left[\Pi^{(0)}(\textbf{q},\omega) \right]_{{\bm n}_{13}}^{{\bm n}_{24}} = -\sum_{\textbf{k},{\bm \alpha}_{12}} \dfrac{f(\xi_{{\bm \alpha}_1}(\textbf{k}+\textbf{q}))-f(\xi_{{\bm \alpha}_2}(\textbf{k}))}{\omega+\xi_{{\bm \alpha}_1}(\textbf{k}+\textbf{q})-\xi_{{\bm \alpha}_2}(\textbf{k})+i\epsilon}\nonumber\\
&&
\times\mathcal{S}^{\dagger}_{{\bm n}_1,{\bm \alpha}_1}({\bf k}+{\bf q})\mathcal{S}^{\dagger}_{{\bm n}_2,{\bm \alpha}_2}({\bf k})
    \mathcal{S}_{{\bm n}_3,{\bm \alpha}_1}({\bf k}+{\bf q})\mathcal{S}_{{\bm n}_4,{\bm \alpha}_2}({\bf k}).
\label{eq-chi0}
\end{eqnarray}
We denote $\xi_{{\bm \alpha}_i}({\bf k})$ as the ${{\bm \alpha}_i}^{\rm th}$-eigenvalue of $H_T({\bf k})$, and $\mathcal{S}({\bf k})$ as the unitary matrix consisting of the corresponding eigenvectors defined in Eq.~\eqref{eq:unitality}. Finally, we obtain the effective (many-body) interaction term within the RPA, which is obtained by summing over the bubble (for density fluctuations) and ladder (for pair density fluctuations) diagrams as
\begin{widetext}
\begin{eqnarray}
\Gamma_{{\bm \kappa}_{13}}^{{\bm \kappa}_{24}} &=& {V}_{{\bm n}_{13}}^{{\bm n}_{24}} 
 - \frac{1}{2}\sum_{{\bm n}_{5-8}}{\rm sgn}[s_5s_6]\mathcal{D}_{s_5}\delta_{s_5,\bar{s}_7}\delta_{s_6,\bar{s}_8} {V}_{{\bm n}_{15}}^{{\bm n}_{26}}(\Pi_s)_{{\bf R}_{13},{\bm n}_{57}}^{{\bf R}_{24},{\bm n}_{68}}{V}_{{\bm n}_{73}}^{{\bm n}_{84}}
 - \frac{1}{2}\sum_{{\bm n}_{5-8}}\delta_{s_5,s_7}\delta_{s_6,s_8}{V}_{{\bm n}_{15}}^{{\bm n}_{26}}(\Pi_c)_{{\bf R}_{13},{\bm n}_{57}}^{{\bf R}_{24},{\bm n}_{68}}{V}_{{\bm n}_{73}}^{{\bm n}_{84}}, \nonumber\\
\end{eqnarray}
\end{widetext}
where $\mathcal{D}_{s}=2s(s+1)$ is the spin degeneracy factor. ${\rm sgn}[s_5s_6]$ arise from the Fermion parity.  

Finally, the interaction vertex in the band basis $\bm{\alpha}$ is obtained with a rotation by the $\mathcal{S}_{{\bf R}}$ matrix:
\begin{eqnarray}
{\Gamma}_{{\bf R}_{1,3},{\bm \alpha}_{1,3}}^{{\bf R}_{2,4},{\bm \alpha}_{2,4}} = \sum_{{\bm n}_{1-4}} \Gamma_{{\bm \kappa}_{1,3}}^{{\bm \kappa}_{2,4}}\mathcal{S}^{\dagger}_{{\bm \kappa}_1,{\bm \alpha}_1}\mathcal{S}^{\dagger}_{{\bm \kappa}_2,{\bm \alpha}_2}\mathcal{S}_{{\bm \kappa}_3,{\bm \alpha}_3}\mathcal{S}_{{\bm \kappa}_4,{\bm \alpha}_4}.
    \label{eq:Intband}
\end{eqnarray}
To solve the eigenvalue equation in Eq.~\eqref{Eq:eivenval1}, we construct a superoperator ${\bm \Gamma}$ and supervector $|\Delta\rangle$ by vectorizing each pair of indices \cite{Ray2019}. Typically, the SC gap function is short-ranged, and hence, we can truncate the Bloch phase basis up to, say, $N_R$ nearest neighbors. We consider $N_{\alpha}^F$ number of Fermi momenta (in $d$ dimension)  for the ${\alpha}^{\rm th}$  energy band. Then we have a total of $N_F=\sum_{\alpha} N_{\alpha}^F$. With $s=1/2$ spins, we have a  $N_{\rm tot}=2N_RN_F$ dimensional vector $|\Delta\rangle$ and $N_{\rm tot}\times N_{\rm tot}$ dimensional matrix ${\bm \Gamma}$. We solve for the largest eigenvalue $\lambda>0$ of ${\bm \Gamma}$ and the corresponding eigenvector $|\Delta\rangle$ gives the pairing symmetry of the theory. We express the eigenvector in matrix form between different bands, spins, and Wannier sites, and this is the final result.

 \section{Details of phase-space mean-field theory}\label{App:MFT}
In Sec.~\ref{lab:MF}, we discussed how to decomposition of the interaction term in Eq.~\eqref{eq:intH2} in terms of a Cooper pair field ${\bm \Delta}({\bf r})$. We generalize this analysis for a local density wave order parameter field in the particle-hole channel ${\bm \Omega}({\bf r})$, as well as develop the full mean-field theory for both the density wave order and the SC order parameters. Both order parameters are defined to be `local' in the phase space, i.e., to be local in ${\bf k}$, but non-local in real space, living on the bond/link, as discussed in Fig.~\ref{fig:twobody_blochbundle}. Implementing,  the momentum conservation in the interaction vertex, we define a density wave order as 
\begin{eqnarray}
&&\Omega_{{\bm \kappa}_{1,3}}({\bf k}_1,{\bf k}_1+{\bf Q},{\bf r}_1)=
\int_{{\bf k}_2,{\bf r}_2}\sum_{{\bm \kappa}_{2,4}}~\nonumber\\
&&\qquad \qquad\qquad {\Gamma}^{{\bm \kappa}_{2,4}}_{{\bm \kappa}_{1,3}}({\bf r}_{1,2}){\pmb \varrho}_{{\bm \kappa}_{2,4}}({\bf k}_2,{\bf k}_2-{\bf Q},{\bf r}_2). 
\label{Eq:IntMFM}
\end{eqnarray}
Similarly, an exchange field would be defined as $\Omega_{{\bm \kappa}_{1,4}}({\bf k},{\bf k}+{\bf Q},{\bf r}_{1,2})$. From Eqs.~\eqref{Eq:IntMFM}, and \eqref{Eq:IntMF}, we can define local density and pair fields using Eq.~\eqref{eq:1body} as 
\begin{eqnarray}
 \Omega({\bf r})&=&\int_{\bf k} \Psi({\bf k},{\bf r})\Omega({\bf k},{\bf k}+{\bf Q},{\bf r})\Psi^{\dagger}({\bf k}+{\bf Q},{\bf r}),\nonumber\\
 &=&\int_{\bf k} \sum_{{\bm \kappa}_{1,3}}\Omega_{{\bm \kappa}_{1,3}}({\bf k},{\bf k}+{\bf Q},{\bf r}){\pmb \varrho}_{{\bm \kappa}_{1,3}}({\bf k},{\bf k}+{\bf Q},{\bf r})\nonumber\\
 \Delta({\bf r})&=&\int_{\bf k} \Psi({\bf k},{\bf r})\Delta({\bf k},{\bf r})\Psi(-{\bf k},{\bf r}),\nonumber\\
 &=&\int_{\bf k} \sum_{{\bm \kappa}_{1,2}} \Delta_{{\bm \kappa}_{1,2}}({\bf k},{\bf r}){\pmb \Psi}_{{\bm \kappa}_{2,4}}({\bf k},-{\bf k},{\bf r})
 .
 \end{eqnarray}
Similar to the Nambu spinor ${\bm \Phi}$ defined in the SC state in \eqref{Eq:Bogwf} , we define a density wave spinor for the density wave order ground state as 
\begin{eqnarray}
{\bm \zeta}({{\bf k}},{\bf Q},{\bf r})&=&\left[\Psi({{\bf k}},{\bf r})\oplus \Psi({\bf k}+{\bf Q},{\bf r})\right]\nonumber\\
&=&\left[{\bf Z}({{\bf k}})\otimes{\bf W}({\bf r})\otimes{\bf X}\right]\oplus
\left[{\bf Z}({\bf k}+{\bf Q})\otimes{\bf W}({\bf r})\otimes {\bf X}\right]. 
\label{Eq:Bogwf}
\end{eqnarray}
Note that ${\bf Q}$ is a fixed wavevector. Replicating the mean-field equation \eqref{Eq:BdG}, we now have an equivalent a (mean-field) one-body density wave Hamiltonian as 
\begin{widetext}
\begin{eqnarray}
H_{\Omega}&=&\int_{{\bf k},{\bf r}}~{\bm \zeta}^{\dagger}({\bf k},{\bf Q},{\bf r})\left({\bf T}({\bf r})+{\bm  \Omega}({\bf r})\right){\bm \zeta}({\bf k},{\bf Q},{\bf r}),\nonumber\\
&=&\int_{{\bf k}}~\sum_{{\bm \kappa}_{1,2}}\Big[T_{{\bm \kappa}_{1,2}}\mathcal{Z}_{{\bf R}_{1,2}}({\bf k},{\bf k})
|{\bm \kappa}_{1},{\bf k}\rangle \langle{\bm \kappa}_{2},{\bf k}| +T_{{\bm \kappa}_{1,2}}\mathcal{Z}_{{\bf R}_{1,2}}({\bf k}+{\bf Q},{\bf k}+{\bf Q})
|{\bm \kappa}_{1},{\bf k}+{\bf Q}\rangle \langle {\bm \kappa}_{2},{\bf k}+{\bf Q}|\nonumber\\
&&\quad\qquad  +{\Omega}_{{\bm \kappa}_{1,2}}\mathcal{Z}_{{\bf R}_{1,2}}({\bf k},{\bf k}+{\bf Q})|{\bm \kappa}_{1},{\bf k}\rangle \langle {\bm \kappa}_{2},{\bf k}+{\bf Q}|+{\rm H.c.}\Big].
\label{Eq:DensityH}
\end{eqnarray}
\end{widetext}
We have $\mathcal{Z}_{{\bf R}_{1,2}}({\bf k},{\bf k})=z_{{\bf R}_{2}-{\bf R}_{1}}({\bf k})$, and $\mathcal{Z}_{{\bf R}_{1,2}}({\bf k}+{\bf Q},{\bf k}+{\bf Q})=z_{{\bf R}_{2}-{\bf R}_{1}}({\bf k}+{\bf Q})=z_{{\bf R}_{2}-{\bf R}_{1}}({\bf k})z_{{\bf R}_{2}-{\bf R}_{1}}({\bf Q})$, and $\mathcal{Z}_{{\bf R}_{1,2}}({\bf k},{\bf k}+{\bf Q})=z_{{\bf R}_{2}-{\bf R}_{1}}({\bf k})z_{{\bf R}_{2}}({\bf Q})$. Due to translational invariance, we can write $H^{\Omega}$ to be local in both momentum ${\bf k}$ and real space on the link ${\bf R}_2-{\bf R}_1$ as
\begin{eqnarray}
H^{\Omega}
&=&\int_{\bf k}~\sum_{{\bm \kappa}_{1,2}}{\bm H}^{\Omega}_{{\bm \kappa}_{1,2}}z_{{\bf R}_{2}-{\bf R}_{1}}({\bf k})|{\bm \kappa}_{1},{\bf k}\rangle \langle {\bm \kappa}_{2},{\bf k}|,
\label{Eq:DensityH2}
\end{eqnarray}
where ${\bm H}^{\Omega}_{{\bm \kappa}_{1,2}}$ is a $2\times 2$ matrix at each ${\bf R}_{1,2}$ and ${\bm n}_{1,2}$, and is defined to be 
\begin{eqnarray}
{\bm H}^{\Omega}_{{\bf R}_{1,2}} =\left(\begin{array}{cc}
     {\bm T}_{{\bf R}_{1,2}} & {\bm \Omega}_{{\bf R}_{1,2}}z_{{\bf R}_{2}}({\bf Q}) \\
     {\bm \Omega}_{{\bf R}_{1,2}}^{\dagger}z^*_{{\bf R}_{2}}({\bf Q}) &  {\bm T}_{{\bf R}_{1,2}}z_{{\bf R}_{2}-{\bf R}_{1}}({\bf Q})
\end{array}\right).
\label{Eq:DensityWavematrix}
\end{eqnarray}
In the above equation, ${\bm T}_{{\bf R}_{1,2}}$, and ${\bm \Omega}_{{\bf R}_{1,2}}$ are matrices in the orbital and spin basis. We diagonalize mean-field Hamiltonians (Eq.~\eqref{Eq:DensityWavematrix}) and BdG Hamiltonian (Eq.~\eqref{Eq:BdGmatrix}) in two steps. In both cases, we first employ a (local) similarity transformation $\mathcal{S}_{\bf R}$ that diagonalizes the orbital and spin parts of $\bm{T}$. Note that $\mathcal{S}_{\bf R}$ is a local rank-1 tensor which is a unitary matrix  in the orbital and spin basis, and follows a `unitality' condition between the Wannier sites as
\begin{eqnarray}
\sum_{{\bf R}_{1,2}}\mathcal{S}^{\dagger}_{{\bf R}_1}\mathcal{S}_{{\bf R}_2}\mathcal{Z}_{{\bf R}_{1,2}}({\bf k}_{1,2})=\mathbb{I}\delta_{{\bf k}_{1,2}}.
\label{eq:unitality}
\end{eqnarray}
This gives the hopping and the pairing interaction in the band basis (denoted by ${\bm \alpha}$) as
\begin{eqnarray}
t_{{\bf R}_{1,2},{\bm \alpha}_1}&=&\sum_{{\bm n}_{1,2}}{T}_{{\bm \kappa}_{1,2}}\mathcal{S}^{\dagger}_{{\bm \kappa}_2,{\bm \alpha}_1}\mathcal{S}_{{\bf \kappa}_1,{\bm \alpha}_1},\label{eq:hoppingband}\nonumber\\
\Omega_{{\bf R}_{1,2},{\bm \alpha}_{1,2}}&=&\sum_{{\bm n}_{1,2}}{\Omega}_{{\bm \kappa}_{1,2}}\mathcal{S}^{\dagger}_{{\bm \kappa}_2,{\bm \alpha}_2}\mathcal{S}_{{\bf \kappa}_1,{\bm \alpha}_1},\label{eq:Omegaband}\nonumber\\
\Delta_{{\bf R}_{1,2},{\bm \alpha}_{1,2}}&=&\sum_{{\bm n}_{1,2}}\Delta_{{\bm \kappa}_{1,2}}\mathcal{S}_{{\bm \kappa}_2,{\bm \alpha}_2}\mathcal{S}_{{\bf \kappa}_1,{\bm \alpha}_1}. 
\label{eq:gapband}
\end{eqnarray}
(We have kept the same notation ${\bm \Omega}$, and ${\bm \Delta}$ in both orbital and band basis for simplicity in notation.) ${\bm \alpha}_i$ index denotes the band basis, in which the orbital and spin can be mixed if there is a spin-orbit coupling term. Since ${\bm T}$ and ${\bm \Omega}$ or ${\bm T}$ and ${\bm \Delta}$ do not necessarily commute, ${\bm \Omega}$ and ${\bm \Delta}$ are not necessarily diagonal in the band basis.  This local transformation faces gauge obstruction if the band structure or the pairing state is topologically non-trivial. There an additional parallel transport/Wilson line is to be included between the ${\bf R}_{1,2}$ sites with a flux conservation constraint \cite{Yogendra2024}.

The diagonalization of the Hamiltonian in the momentum space, which is assumed to give a have coherent ground state, mixes these irreps. We assume that the density wave and SC phases do not coexist. So we individually diagonalize the density wave Hamiltonian and BdG Hamiltonian in Eqs.~\eqref{Eq:MFTDWirreps}, and \eqref{Eq:MFTSCirreps} by the Bogoliubov transformation as defined by the unitality matrix at each tensor component ${\bf R}$: 
\begin{eqnarray}
U^{\Omega}_{{\bf R}} =\left(\begin{array}{cc}
     \mathcal{F}_{{\bf R}} & -\mathcal{G}_{{\bf R}}  \\
    \mathcal{G}^{\dagger}_{{\bf R}} &  \mathcal{F}^{\dagger}_{{\bf R}}
\end{array}\right),
~~
U_{{\bf R}} =\left(\begin{array}{cc}
     \mathcal{U}_{{\bf R}} & -\mathcal{V}_{{\bf R}}  \\
    \mathcal{V}^{\dagger}_{{\bf R}} &  \mathcal{U}^{\dagger}_{{\bf R}}
\end{array}\right).
\label{Eq:BdG3}
\end{eqnarray}
$\mathcal{F}$, $\mathcal{G}$,  $\mathcal{U}$ and $\mathcal{V}$ are rank-1 tensors in the Bloch phase basis, with each component being a matrix on the band basis. The non-local Bogoliubov transformation for the SC Hamiltonian is defined as ${U}_{{\bf R}_2}\mathcal{S}_{{\bf R}_2}{\bm H}_{{\bf R}_{1,2}}S_{{\bf R}_1}^{\dagger}U_{{\bf R}_1}^{\dagger}$=${\bm D}_{{\bf R}_{1,2}}$ where ${\bm D}_{{\bf R}_{12}}$ is a diagonal matrix consistent of the Bogoliubov quasiparticle energies. The `unitality' condition of $U_{\bf R}$, as for $\mathcal{S}_{\bf R}$ in Eq.~\eqref{eq:unitality}, yields
\begin{eqnarray}
\sum_{{\bf R}_{1,2}}(\mathcal{U}^{\dagger}_{{\bf R}_1}\mathcal{U}_{{\bf R}_2}+\mathcal{V}^{\dagger}_{{\bf R}_1}\mathcal{V}_{{\bf R}_2})\mathcal{Z}_{{\bf R}_{1,2}}({\bf k}_{1,2})&=&\mathbb{I}\delta_{{\bf k}_{1,2}},\nonumber\\
\sum_{{\bf R}_{1,2}}(\mathcal{U}_{{\bf R}_1}\mathcal{V}^{\dagger}_{{\bf R}_2}-\mathcal{U}_{{\bf R}_2}\mathcal{V}_{{\bf R}_1}^{\dagger})\mathcal{Z}_{{\bf R}_{1,2}}({\bf k}_{1,2})&=&0.
\label{eq:UVunitality}
\end{eqnarray}
The above formalism is similar for the density wave case and hence not repeated here.

To obtain the self-consistent gap equation, we apply the $S_{\bf R}$ rotation on both sides of Eqs.~\eqref{Eq:IntMF}, and ~\eqref{eq:gapband}, followed by the $U_{\bf R}$ rotation.  Following Eqs.~\eqref{eq:GammaParity}, and ~\eqref{Eq:oddwf}, the interaction $\Gamma$ and gap function $\Delta$ are decoupled in the even and odd fermion parity channels for the space, orbital, and spin parts. This gives the self-consistent gap equation as
\begin{eqnarray}
 \Delta_{{\bm \nu}}z_{\nu_{\bf R}}({\bf k}_1) &=&\int_{{\bf k}_2}\sum_{{\bm \nu}'} \Gamma_{\bm \nu}^{{\bm \nu}'}({\bf k}_{1,2}) 
    \left\langle (\mathcal{U}\mathcal{V})_{{\bm \nu}'}z_{\nu'_{\bf R}}({\bf k}_2)\right\rangle_{\rm BCS}.
    \label{Eq:gapeq}
\end{eqnarray}
Here we have used $\mathbb{Z}_{\nu_{\bf R}}({\bf k},-{\bf k})=z_{\nu_{\bf R}}({\bf k})$. We recall that the indices $\nu_i=\pm$ stand for even and odd parity basis for each quantum number $i$. We denote $\bar{\nu}_i$  as the opposite parity to $\nu_i$. For example, we denote $\Delta_{\nu_{\bf R},\bar{\nu}_{\bm \alpha}}$  to imply that if it is even (odd) under the exchange between ${\bf R}_{1,2}$, it must be odd (even) in the band basis ${\bm \alpha}_{1,2}$ and vice versa. Note that although the Bloch basis seems to be separated from the orbital part in the above formalism, the odd parity constraint entangles all the indices.  At temperature $T\rightarrow 0$, the expectation value on the wavefunction can be expanded in a Taylor series in terms of $\Delta$, and restricting ourselves to the linear term in $\Delta$, we define $\left\langle (\mathcal{U}\mathcal{V})_{{\bm \nu}'}z_{\nu'_{\bf R}}({\bf k}_2))\right\rangle_{\rm BCS}\approx -\lambda^{-1}{\Delta}_{{\bm \nu}'}z_{\nu'_{\bf R}}({\bf k}_2) +\mathcal{O}(({\Delta}/T)^2)$, where $T$ represents the 
bandwidth. Here $\lambda$ is a SC coupling constant, and its dimension is $[{\rm Energy}]$. A minus sign is chosen as we expect that a solution exists for an attractive potential $\Gamma$. This converts Eq.~\eqref{Eq:gapeq} into an eigenvalue equation as given in Eq.~\eqref{Eq:eivenval1}. As a variational approach, we can first find the symmetry-allowed Bloch functions $\mathbb{Z}$ in the pair state and find expectation values of ${\bm \Gamma}$ in these basis states. Then, the highest negative value of this expectation value gives us the pairing symmetry. 

\section{Example for Implementation of Fermion Parity}\label{App:Irreps}
The implementation of the fermion parity is an important distinction of the pairing field $\Delta$ compared to the density wave field. Without losing generality, we can continue to implement the parity in the product basis for the three quantum numbers and move to any entangled or irreps basis with appropriate unitary transformations after incorporating the fermion parity.\cite{Zhou2008, Venderbos2018, Brydon2016} In the total state, we can also arrange the even and odd parity states as
\begin{eqnarray}
\mathbb{Z}\otimes\mathbb{W}\otimes\mathbb{X}
    &&\xrightarrow{{U_{\theta(Z)}\otimes U_{\theta(W)}\otimes U_{\theta(X)}}}
    \left(\begin{array}{c}
         \mathbb{Z}_{+}  \\
         \mathbb{Z}_{-} 
    \end{array}\right)\otimes
    \left(\begin{array}{c}
         \mathbb{W}_{+}  \\
         \mathbb{W}_{-} 
    \end{array}\right)\otimes
    \left(\begin{array}{c}
         \mathbb{X}_{+}  \\
         \mathbb{X}_{-} 
    \end{array}\right)\nonumber\\
    && ~~ 
    \rightarrow\left(\begin{array}{c}
         \mathbb{Z}_{+}\otimes \mathbb{W}_{+}\otimes\mathbb{X}_{-}  \\
         \mathbb{Z}_{+}\otimes \mathbb{W}_{-}\otimes\mathbb{X}_{+}\\
         \mathbb{Z}_{-}\otimes \mathbb{W}_{+}\otimes\mathbb{X}_{+}\\
         \mathbb{Z}_{-}\otimes \mathbb{W}_{-}\otimes\mathbb{X}_{-}
    \end{array}\right)\oplus( {\rm even~parity}).
    \label{Eq:oddwf}
\end{eqnarray}
On the right-hand side, we have rearranged the odd and even parity states for presentation purposes. We introduce a projection operator $\mathsf{P}_{\theta}$ to eliminate the even parity state. These are the four odd-parity states allowed for superconductivity \cite{Sigrist1991, Das2012, Linder2019, Brydon2016, Suh2023}. By substituting Eq.~\eqref{Eq:oddwf} in Eq.~\eqref{Eq:IntMF} we obtain the mean-field expression for the SC gap functions in the product basis of orbital, spin, and Bloch phase. 

We demonstrate the above construction for two orbitals with spin-1/2 case on a tetragonal lattice ($\mathsf{D}_4$ group). For spins, the even and odd states are simply the triplets and singlet, respectively. This transformation is simply done with a basis (unnormalized) transformation on $\mathbb{X}$ as
\begin{eqnarray}
\left(\begin{array}{c}
\uparrow\uparrow\\
\downarrow\downarrow\\
\uparrow\downarrow\\
\downarrow\uparrow\\
\end{array}\right)
=\frac{1}{2}\left(\begin{array}{cccc}
2 & 0 & 0 & 0\\
0 & 2 & 0 & 0\\
0 & 0 & 1 & 1\\
0 & 0 & 1 & -1\\
\end{array}\right)
\left(\begin{array}{c}
\uparrow\uparrow\\
\downarrow\downarrow\\
\uparrow\downarrow + \downarrow\uparrow\\
\uparrow\downarrow - \downarrow\uparrow\\
\end{array}\right),
\end{eqnarray}
which gives the corresponding unitary transformation $U_{\theta(X)}$ in Eq.~\eqref{Eq:oddwf}. 

For two orbitals, we similarly define interorbital terms into even and odd parity states $\mathbb{W}_{\pm} = 1/\sqrt{2}(w_1w_2\pm w_2w_1)$, and intraorbital states $w_1w_1$ $w_2w_2$ are even parity. This transformation is done with a similar unitary transformation $U_{\theta(W)}$. These four states are also the irreps of the $\mathsf{D}_4$ symmetry. 

For the Bloch basis $\mathbb{Z}$, the fermion parity becomes the same as the spatial inversion for zero-momentum Cooper pairs. We can write $\mathbb{Z}$ as a polynomial of either (even) $\cos{({\bf k}\cdot {\bm \delta})}$ or (odd) $\sin{({\bf k}\cdot {\bm \delta})}$, where ${\bm \delta}={\bf R}_1-{\bf R}_2$. This is easier to demonstrate with an example. Consider a $D=3$ - dimensional simple tetragonal lattice. The onsite pairing at $\delta_0=(0,0,0)$ gives a trivial $s$-wave superconductivity. For the nearest neighbor term, we have $d=6$ with their distances being $\delta_{1,2}=(\pm 1,0,0)$, $\delta_{3,4}=(0,\pm 1,0)$, $\delta_{5,6}=(0,0,\pm 1)$, for the lattice constants $a=b=c=1$. So we have a six-dimensional Bloch spinor ${\bf Z}_{6}({\bf k})$=$ (e^{ik_x}~e^{-ik_x}~e^{ik_y}~e^{-ik_y}~e^{ik_z}~e^{-ik_z})^T$. Setting one electron at the origin, i.e., ${\bf R}_1=0$, we express the two-particle Bloch spinor in terms of single particle Bloch spinor as $\mathbb{Z}({\bf k})=z_0{\bf Z}_{6}({\bf k})$, where $z_0=1$. More generally we can move to the irreps of the point group symmetry for ${\bf Z}_6$, which for the $\mathsf{D}_4$ group are trivial to find. We choose (unnormalized) bases as 
\begin{eqnarray}
|s_{x^2+y^2}\rangle &=& \left(\begin{array}{cccccc}
     1 & 1 & 1 & 1 & 0 & 0 
\end{array}\right)^T,\nonumber\\
|s_{z^2}\rangle &=& \left(\begin{array}{cccccc}
     0 & 0 & 0 & 0 & 1 & 1 
\end{array}\right)^T,\nonumber\\
|p_x\rangle &=& \left(\begin{array}{cccccc}
     1 & -1  & 0 & 0 & 0 & 0 
\end{array}\right)^T,\nonumber\\
|p_y\rangle &=& \left(\begin{array}{cccccc}
     0 & 0  & 1 & -1 & 0 & 0 
\end{array}\right)^T,\nonumber\\
|p_z\rangle &=& \left(\begin{array}{cccccc}
     0 & 0 & 0 & 0 & 1 & -1 
\end{array}\right)^T,\nonumber\\
|d_{x^2-y^2}\rangle &=& \left(\begin{array}{cccccc}
     1& 1& -1 & -1 & 0 & 0
\end{array}\right)^T.
\label{Eq:irreps_3D}
\end{eqnarray}
These bases constitute the unitary transformation $U_{\theta(Z)}$ in Eq.~\eqref{Eq:oddwf}. We make two observations here. The basis states are not unique, and one needs to pay attention to the symmetry of the lattice. Secondly, the $t_{2g}$ irrep states are not included as they arise from the second nearest neighbor pairings (as shown for square lattice in Eq.~\eqref{Eq:irrepsiband}. We write the corresponding Bloch wavefunctions as (up to a normalization) 
\begin{eqnarray}
z_{s_{x^2+y^2}}({\bf k}) &=& \langle s_{x^2+y^2}|\mathbb{Z}({\bf k})\rangle = 2(\cos {k_x}+\cos{k_y}),\nonumber\\
z_{s_{z^2}}({\bf k}) &=& \langle s_{z^2}|\mathbb{Z}({\bf k})\rangle = 2\cos {k_z},\nonumber\\
z_{p_{x/y/z}}({\bf k}) &=& \langle p_{x/y/z}|\mathbb{Z}({\bf k})\rangle = i\sqrt{2}\sin {k_{x/y/z}},\nonumber\\
z_{d_{x^2-y^2}}({\bf k}) &=& \langle d_{x^2-y^2}|\mathbb{Z}({\bf k})\rangle = 2(\cos {k_x}-\cos{k_y}).
\label{eq:Blochsquare}
\end{eqnarray}
Note that these functions are the eigenstates of the fermion parity $\mathsf{P}_{\theta(Z)}$ being $+1$ for $s$- and $d$- waves and $-1$ for $p$-wave states. 

To illustrate how Eq.~\eqref{Eq:oddwf} operates , we consider a few representaive examples. In the single orbital case ($\mathbb{W}=1$), the spin channels allow for singlet or triplet pairings. For nearest-neighbor pairing on a $\mathsf{D}_4$ lattice with the Bloch wavefunctions described above,  the $s_{x^2+y^2}$, $s_{z^2}$, and $d_{x^2-y^2}$ wave states with $\mathsf{P}_{\theta(Z)}=+1$ will exclusively occur in the spin singlet channel. Conversely,  $p_{x/y/z}$ wave pairing will be triplet. The scenario becomes more versatile for two or more orbitals, say, two orbitals $w_{1}$ and $w_2$. In this case, the $s_{x^2+y^2}$, $s_{z^2}$, and $d_{x^2-y^2}$ pairing symmetries manifests as spin singlets in the intra-orbital channels ($w_1w_1$, $w_2w_2$) or in the bonding inter-orbital channel ($w_1w_2+w_2w_1$). Alternatively, they can also appear in the spin triplet channel in the anti-binding inter-orbital channel ($w_1w_2-w_2w_1)$. The spin-channel association is simply reversed for the $p_{x/y/z}$ pairing case.

\section{Examples and comparison with tight-binding model}\label{App:Examples}

\begin{figure}[t]
\centering
\includegraphics[width=.4\textwidth]{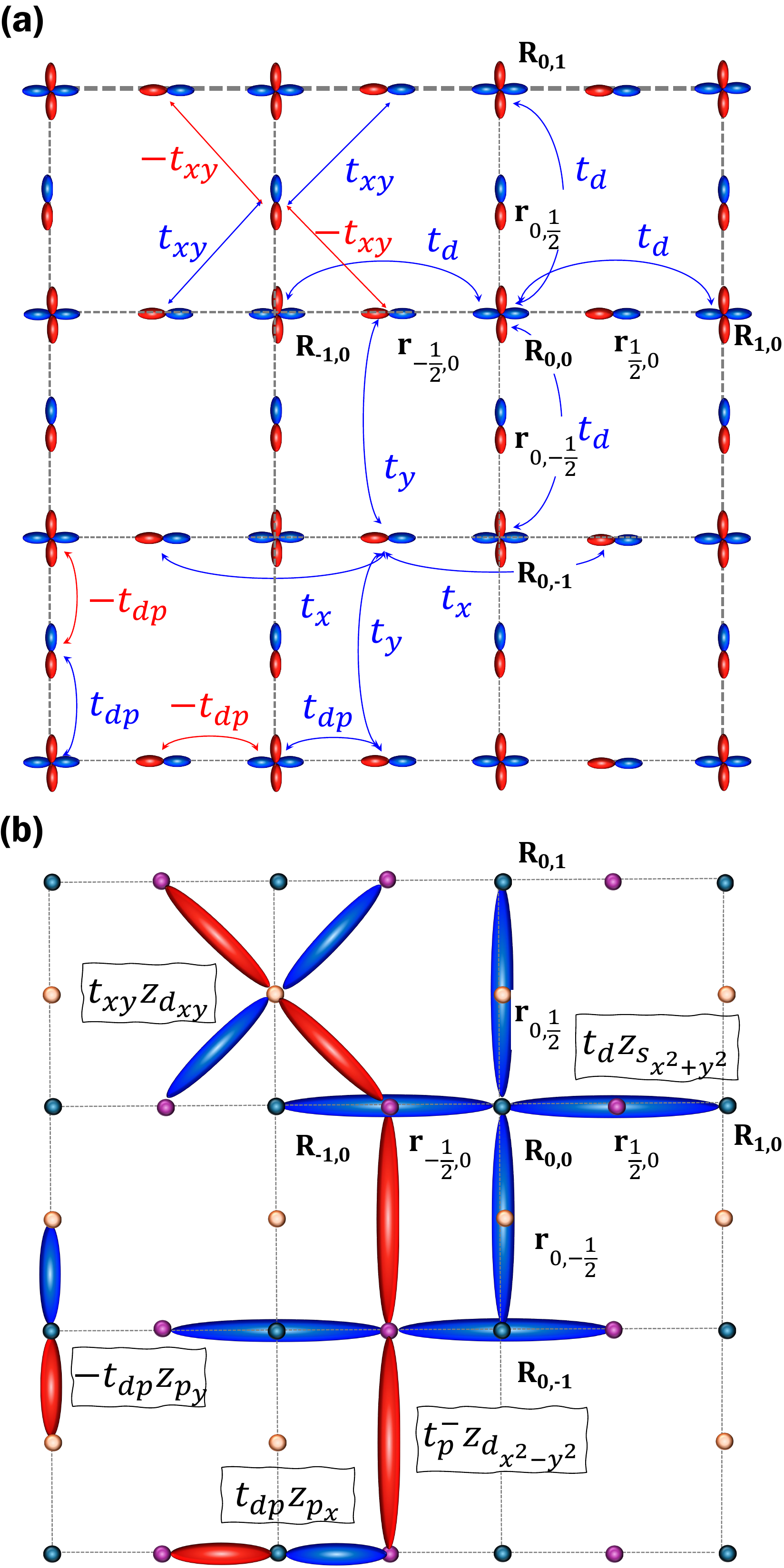}
\caption{\label{fig:threeband} We visualize the difference between a quintessential tight-binding model in (a) and our phase space model in (b) for three band model given in Appendix~\ref{App:Multiband}. (a) We show the tight-binding hopping parameters for the case of a square lattice with $d_{x^2-y^2}$ orbital sitting at the corner and $p_{x/y}$ orbitals lying on the bonds. Blue and red colored arrows distinguish between the $\pm$ signs of the hopping terms. (b) The same dispersions are now expressed in terms of the irreps $z_{\nu}$ multiplied by a hopping matrix $T_{\nu}$. Note that in case (a), the orbitals $w({\bf r})$ contain the angular momentum symmetry, which dictates the sign of the hopping integrals, while in case (b), the orbitals are $\delta({\bf r})$-functions and the corresponding angular momentum symmetry is transported to the Bloch phase $z_{\nu}$.
}
\end{figure}

It is worthwhile pointing out the momentum space representation of these quantities: The band dispersion and the gap function can be retrieved in the momentum space as usual by 
\begin{eqnarray}
\xi_{{\bm \alpha}_{1}}({\bf k})&=&\sum_{{\bf R}_{1,2}}t_{{\bf R}_{1,2},{\bm \alpha}_1}z_{{\bf R}_{2}-{\bf R}_{1}}({\bf k}),\\
\xi_{{\bm \alpha}_{1}}({\bf k}+{\bf Q})&=&\sum_{{\bf R}_{1,2}}t_{{\bf R}_{1,2},{\bm \alpha}_1}z_{{\bf R}_{2}-{\bf R}_{1}}({\bf Q})z_{{\bf R}_{2}-{\bf R}_{1}}({\bf k}),\\
\Omega_{{\bm \alpha}_{1,2}}({\bf k},{\bf Q})&=&\sum_{{\bf R}_{1,2}}\Omega_{{\bf R}_{1,2},{\bm \alpha}_{1,2}}z_{{\bf R}_{2}}({\bf Q})z_{{\bf R}_{2}-{\bf R}_{1}}({\bf k}),\\
\Delta_{{\bm \alpha}_{1,2}}({\bf k})&=&\sum_{{\bf R}_{1,2}}\Delta_{{\bf R}_{1,2},{\bm \alpha}_{1,2}}z_{{\bf R}_{2}-{\bf R}_{1}}({\bf k}),
\end{eqnarray}
$\mathcal{S}({\bf k})=\sum_{\bf R}\mathcal{S}_{{\bf R}} z_{{\bf R}}({\bf k})$. Given the fact that all one-body terms are not expanded in the same Bloch basis of $z_{{\bf R}_{2}-{\bf R}_{1}}({\bf k})$, we can analyze the mean-field excitation spectrum locally in both momentum and real spaces.  As demonstrated in Sec.~\ref{Sec:Flatband}, it is easier to work in the irreducible representation (irreps) of the corresponding point group such as $z_{{\bf R}_{2}-{\bf R}_{1}}({\bf k})\rightarrow z_{\nu}$ where $\nu$ are the irreps. This allows us to write the mean-field Hamiltonians (Eqs.~\eqref{Eq:DensityWavematrix}, and \eqref{Eq:DensityWavematrix})  as in Eq.~\eqref{Eq:BdG1band}. Because different irreps do not mix, we can simply write the mean-field Hamiltonians in the irreps basis. We consider two examples here.

\subsection{Multi-band non-interaction Hamiltonian}\label{App:Multiband}
We consider a case of three spinless orbitals in a square lattice, and compare the Hamiltonian formalism of the traditional Wannier orbital model versus our phase space orbital model in Fig.~\ref{fig:threeband}. We consider a $d_{x^2-y^2}$ orbital is at the site center ${\bf r}_1=(0,0)$ and $p_{x,y}$ orbitals sitting at the Wyckoff position ${\bf r}_{2,3}=(0.5,0),(0,0.5)$. This situation represents a single layer cuprate superconductor.\cite{cuprate3band} In the traditional approach, we associate the symmetry of the wavefunction in the ${\bf r}-$ dependence of the  Wannier orbitals, in which, we have $w_d(r_1,\theta_1,\phi_1)=f_d(r_1)\frac{1}{\sqrt{2}}(Y_{2}^{-2}+Y_{2}^{2})(\theta_1,\phi_1)$, and  $w_{i}(r_{i},\theta_{i},\phi_{i})=f_{i}(r_{i})\frac{1}{\sqrt{2}}(Y_{1}^{-1}+(-1)^i Y_{1}^{1})(\theta_{i},\phi_{i})$ for $i=2,3$ for $p_{x/y}$ orbitals. $f_i(r_{i})$ are some radial function such that the overlap between the two orbitals is optimized to obtain maximal localization. We assume the global U(1) symmetry is present, which means there is only one chemical potential $\mu$. The charge transfer gap between $d$ and $p$ orbital is $\delta$. We restrict ourselves to only first nearest neighbor hoppings as shown in Fig.~\ref{fig:threeband}(a). Then the corresponding dispersions are $\xi_{dd}({\bf k})=t_dz_{s_{x^2+y^2}}({\bf k})-\mu$, $\xi_{xx/yy}({\bf k})=t_{x/y}z_{s_{x^2}}({\bf k})+t_{y/x}z_{s_{y^2}}({\bf k})-\delta-\mu$, $\xi_{d{x/dy}}({\bf k})=t_{dp}z_{p_{x/y}}({\bf k}/2)$, and $\xi_{xy}({\bf k})=t_{xy}z_{d_{xy}}({\bf k}/2)$. Here $z_{s_{x^2}/s_{y^2}}=2\cos{k_{x/y}}$, while the remaining Bloch phases are given in Eqs.~\eqref{Eq:irrepsiband} and \eqref{eq:Blochsquare}. The corresponding $3\times 3$  one-body Hamiltonian in the Wannier spinor $\Psi({\bf k})=(w_d~ w_x~w_y)^T$ read as
 \begin{eqnarray}
       {\bm H}_{T} ({\bf k})&=&\left(\begin{array}{ccc}
     \xi_{dd}({\bf k}) & \xi_{dx}({\bf k}) & \xi_{dy}({\bf k}) \\
      \xi_{dx}^{*}({\bf k}) & \xi_{xx}({\bf k}) & \xi_{xy}({\bf k})\\
      \xi_{dy}^{*}({\bf k}) & \xi_{xy}^*({\bf k}) & \xi_{yy}({\bf k})\\
\end{array}\right).
\label{Eq:WannierH}
\end{eqnarray}

In our phase space approach, the Wannier orbitals can be taken simply as delta function $w_{i}({\bf r}_{i})=\delta({\bf r}_{i})$, $\forall i$, and a spinor $\Psi({\bf k})=(w_1({\bf r}_1)~ w_2({\bf r}_2)~w_3({\bf r}_1))^T$ is sitting at each unit cell. The angular symmetry of the orbitals is now transported into the irreps of the Bloch phases $z_{\nu}$ associated with hopping tensor $(T_{\nu_{\bf R}})_{n_1,n_2}$, where $n_1,n_2=1,2,3$ are the orbital indices. Notice that $z_{s_{x^2}}$, $z_{s_{y^2}}$, $z_{p_{x}}$, $z_{p_{y}}$ are not the irreps of square lattice, rather their combinations $z_{s_{x^2+y^2}}=z_{s_{x^2}}+z_{s_{y^2}}$, $z_{d_{x^2-y^2}}=z_{s_{x^2}}-z_{s_{y^2}}$, and $z_{p_{\pm}}=z_{p_{x}}\pm i z_{p_{y}}$ are the corresponding irreps. As shown in Fig.~\ref{fig:threeband}(b), the same Hamiltonian Eq.~\eqref{Eq:WannierH} in now expanded in the basis of $z_{\nu}$ with coefficient $T_{\nu}$ as $3\times 3$ matrices given by
\begin{widetext}
 \begin{eqnarray}
       {\bm H}_{T} ({\bf k})&=&\left(\begin{array}{ccc}
      -\mu  & 0 & 0\\
      0 & -\mu-\delta & 0\\
      0 & 0 & -\mu-\delta
\end{array}\right)z_{0}({\bf k})
+ \left(\begin{array}{ccc}
      t_d  & 0 & 0 \\
      0 & t_p^+ & 0\\
      0 & 0 & t_p^+\\
\end{array}\right)z_{s_{x^2+y^2}}({\bf k})
+ \left(\begin{array}{ccc}
      0  & 0 & 0 \\
      0 & t_p^- & 0\\
      0 & 0 & -t_p^-\\
\end{array}\right)z_{d_{x^2-y^2}}({\bf k})\nonumber\\
      && + \frac{1}{2}\left(\begin{array}{ccc}
      0 & t_{dp} & 0\\
      t_{dp} & 0 & 0 \\
      0 & 0 &0
\end{array}\right)z_{p_{+}}({\bf k}/2) 
      + \frac{1}{2}\left(\begin{array}{ccc}
      0 & 0& it_{dp}\\
      0 & 0 & 0 \\
      -it_{dp} & 0 &0
\end{array}\right)z_{p_{-}}({\bf k}/2) 
 + \left(\begin{array}{ccc}
      0 & 0 & 0\\
      0 & 0 & t_{xy} \\
      0 & t_{xy} &0
\end{array}\right)z_{d_{xy}}({\bf k}/2) .
\label{Eq:PhaseSpaceMultiband}
\end{eqnarray} 
\end{widetext}
Here we define $t_p^{\pm}=(t_x\pm t_y)/2$. $z_0({\bf k})=1$. As we emphasized, the form of the Hamiltonian does not change, and only the basis representation changes here.

\subsection{Single band mean-field Hamiltonians}\label{App:Singleband}

Next, we consider the examples of single band mean-field Hamiltonian in the density wave and superconducting states and represent them in the Bloch basis. For a single orbital case, we can  expand Eqs.~\eqref{Eq:DensityWavematrix}, \eqref{Eq:BdGiband} as follows in the Bloch phases as 
\begin{widetext}
    \begin{eqnarray}
        {\bm H}^{\Omega} ({\bf k})&=&\left(\begin{array}{cc}
     t_{\bar{\nu}} & \Omega_{\bar{\nu}}z_{\bar{\nu}}({\bf Q}) \\
      \Omega^{\dagger}_{\bar{\nu}}z^{\dagger}_{\bar{\nu}}({\bf Q}) & t_{\bar{\nu}} z_{\bar{\nu}}({\bf Q})
\end{array}\right)z_{\bar{\nu}}({\bf k})
+ \sum_{\nu\ne\bar{\nu}} \left(\begin{array}{cc}
     t_{{\nu}} & 0\\
      0      & t_{{\nu}} z_{\nu}({\bf Q})\end{array}\right)z_{\nu}({\bf k}),
      \label{Eq:MFTDWirreps}\\
       {\bm H} ({\bf k})&=&\left(\begin{array}{cc}
     t_{\bar{\nu}} & \Delta_{\bar{\nu}} \\
      \Omega^{\dagger}_{\bar{\nu}}z & -t_{\bar{\nu}} 
\end{array}\right)z_{\bar{\nu}}({\bf k})
+ \sum_{\nu\ne\bar{\nu}} \left(\begin{array}{cc}
     t_{{\nu}} & 0\\
      0      & -t_{{\nu}} \end{array}\right)z_{\nu}({\bf k}) .
      \label{Eq:MFTSCirreps}
    \end{eqnarray}
\end{widetext}
In the above we assume that the density wave order and SC pair only occur in one irrep $\bar{\nu}$ basis, for simplicity. We are not yet done. We have to pay attention to the corresponding spin basis, and this turns out to be more more crucial for the pairing case due to fermion parity.  If we have a spin density wave, we consider the corresponding spin density matrix $\mathcal{X}_{\uparrow\uparrow}$, $\mathcal{X}_{\downarrow\downarrow}$ for ferromagnet, or $\frac{1}{2}(\mathcal{X}_{\uparrow\uparrow}-\mathcal{X}_{\downarrow\downarrow})$ for antiferromagnet or altermagnet, or $\frac{1}{2}(\mathcal{X}_{\uparrow\downarrow}-\mathcal{X}_{\downarrow\uparrow})$ for spin-orbit density wave or spin spiral phases (note that this is not a two-particle pair state. but only a density matrix in the spin sector).  For the case of altermagnet, spin-orbit density wave, or spin-momentum locking, we have to worry about the mixing between the Bloch irreps and spin irreps, which we ignore here. 

This is, however, crucial for the SC case. As a simple example, we consider a single orbital case without any spin orbit coupling on a  2D square lattice with upto the first nearest neighbor terms, i.e.,  ${\bf Z}_1({\bf k})=(z_{s_{x^2+y^2}}, z_{d_{x^2-y^2}}, z_{p_x}, z_{p_y})$ from Eq.~\eqref{Eq:irrepsiband}, in addition to the onsite term $z_0({\bf k})=1$. Here  $z_0$, $z_{s_{x^2+y^2}}$, $z_{d_{x^2-y^2}}$ phases are coupled to spin singlet, while $z_{p_x}$, $z_{p_y}$ are coupled with spin triplet pairing (see Appendix~\ref{App:Irreps} for further discussions). 

The important lesson of Eqs.~\eqref{Eq:MFTDWirreps}, and \eqref{Eq:MFTSCirreps} is that in this product state expansion, only the irrep Bloch phase $z_{\bar{\nu}}$ that forms an order state becomes a superposition of two electrons with different wavevectors for the density wave case or particle-hole states in the SC state, respectively. The rest of the Bloch irreps remain same to the previous non-interacting case. 

\subsection{A multigap BdG equation}\label{App:multigapBdG}
In the main text, the DFT-derived numerical simulation predicted a coexistence of $s^{\pm}$ pairing symmetries in bands 1 and 2 and $s_{z^2}-$ symmetry in band 3. A $s$-wave pairing corresponds to onsite pairing, belonging to the $z_0$ irreps, while the $s_{z^2}$ symmetry arises from the nearest-neighbor-pairing along the $z-$direction, which belongs to the $z_{s_{z^2}}$ irrep, as presented in Eq.~\eqref{Eq:irreps_3D}. Both pairing states occur in the spin singlet channel, i.e., $\mathbb{X}_-$. Since $s$ and $s_{z^2}$ irreps are even under spatial inversion, therefore, the orbital part must be symmetric, i.e, $\mathbb{W}_+$.  Based on these inputs, we can construct a phase space multiband BdG model for three bands as follows. 
\begin{widetext}
 \begin{eqnarray}
       {\bm H} ({\bf k})&=&\left(\begin{array}{cccccc}
      -\mu  & 0 & 0 & \Delta_s & 0 & 0 \\
      0 & -\mu-\delta_1 & 0 &0& -\Delta_s & 0 \\
      0 & 0 & -\mu-\delta_2& 0 & 0 & 0 \\
      \Delta_s & 0 & 0 & \mu  & 0 & 0 \\
      0 & -\Delta_s & 0 & 0 & \mu+\delta_1  & 0 \\
      0 & 0 & 0 & 0 & 0&\mu+\delta_2 \\
\end{array}\right)z_{0}({\bf k})\nonumber\\
&& ~~~+ \left(\begin{array}{cccccc}
      0  & 0 & 0 & 0 & 0 & 0 \\
      0 & 0 & 0 &0& 0 & 0 \\
      0 & 0 & t_{s_{z^2}} & 0 & 0 & \Delta_{s_{z^2}} \\
      0 & 0 & 0 &  0  & 0 & 0 \\
      0 & 0 & 0 & 0 & 0  & 0 \\
      0 & 0 & \Delta_{s_{z^2}} & 0 & 0 &  -t_{s_{z^2}}\\
\end{array}\right)z_{s_{z^2}}({\bf k})
+ \sum_{\nu\ne 0,s_{z^2}}\left(\begin{array}{cc}
     {\bf T}_{\nu} & {\bf 0}_{3\times 3}\\
     {\bf 0}_{3\times 3} & -{\bf T}^T_{\nu}
\end{array}\right)z_{\nu}({\bf k}).
\label{Eq:multigapApp}
\end{eqnarray} 
\end{widetext}
Here $\mu$ is the chemical potential, $\delta_{1,2}$ are the charge transfer gaps, $t_{s_{z^2}}$ is the intra-orbital tight-binding hopping for the Wannier orbital $w_{3}$ sitting at the lattice sites that are connected by the $z_{s_{z^2}}({\bf k})$ along the $z$ direction. $\Delta_s$ and $\Delta_{s_{z^2}}$ are the pairing amplitude for the corresponding pairing channels. $\pm\Delta_s$ is set for bands 1 and 2 due to $s^{\pm}$ pairing state. There is no pairing in the other irreps, giving only $3\times 3$ hopping integrals ${\bf T}_{\nu}$ for all other irreps. 

\section{First-principles electronic structure calculations}\label{sec-dft_result}

\begin{figure}[t]
\centering
9\includegraphics[width=.45\textwidth]{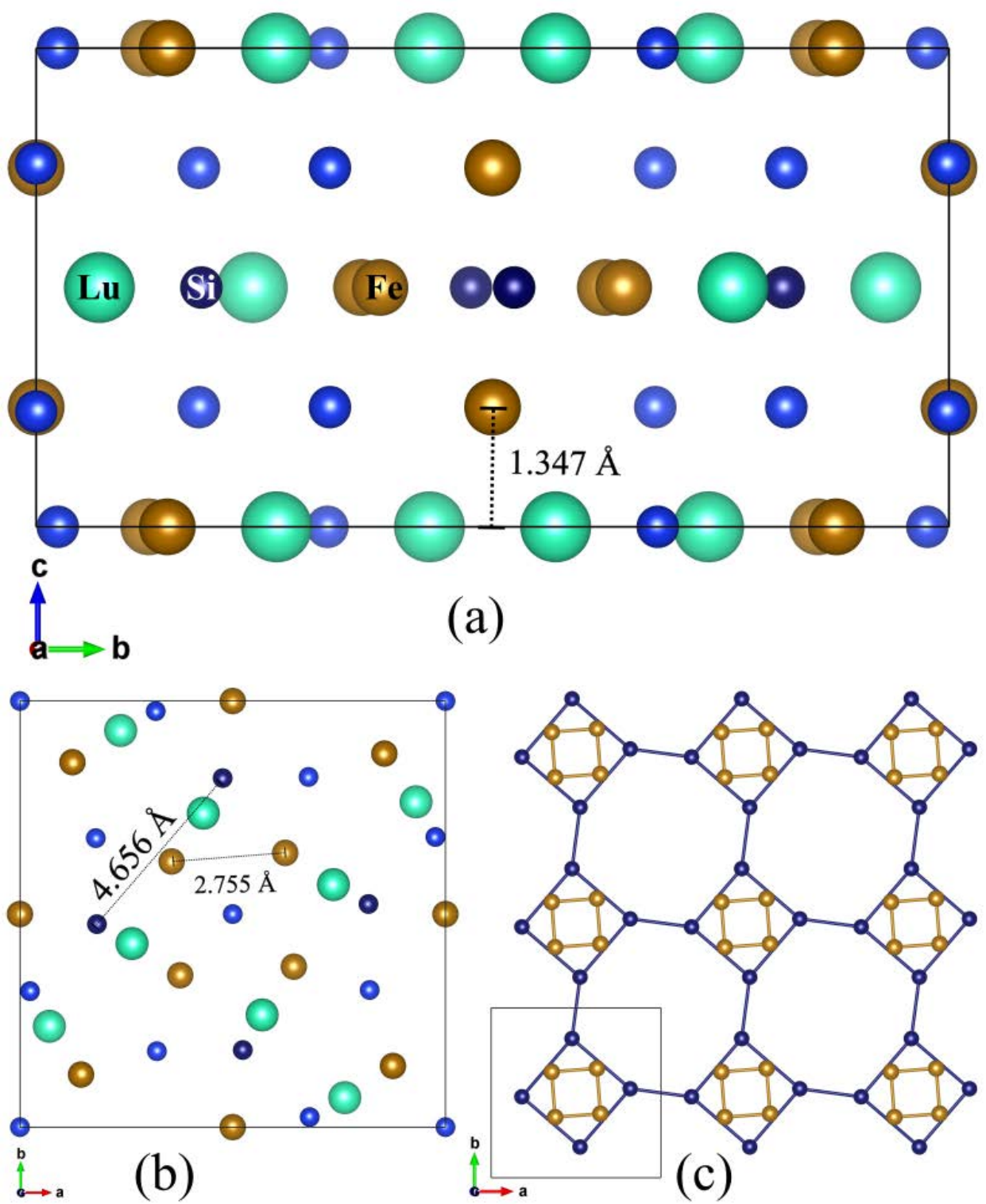}
\caption{\label{fig:unitcell} (a,b) Side (top) view of the Lu$_2$Fe$_3$Si$_5$ unit-cell. The Si atoms that form the square-octagon (SO) lattice are highlighted by dark blue color. The two Si atoms of the square-octagon lattice are separated by 4.656 \AA. (c) A 2D view of the SO lattice formed by Si atoms is shown here. Each square of Si holds a square of Fe atoms. The distance between these Fe atoms is 2.755 \AA.
}
\end{figure}

We used first-principles electronic structure calculations to examine the band structure and the Fermi surface of the system to understand the origins of multiband superconductivity in Lu$_2$Fe$_3$Si$_5$. We perform density functional theory (DFT) simulation using the Vienna Ab initio Simulation Package (VASP) \cite{vasp1,vasp2}. For the exchange-correlation between electrons, we employed the generalized gradient approximation (GGA), parametrized by Perdew$-$Burke$-$Ernzerhof (PBE) \cite{pbe}. We employed the projector$-$augmented wave (PAW) approach to characterize the interaction between the core and valence electrons \cite{paw1,paw2}. The plane$-$wave energy cutoff is set at 400 eV. The Brillouin zone (BZ) is sampled using the Monkhorst$-$Pack technique \cite{Monkhorst1976} with a grid spacing of 0.01 for all calculations, which yields equivalent $29\times29\times29$ and $30\times30\times3$ k-point meshes for  Rhombohedral and Hexagonal unitcells, respectively. The structural relaxation was carried out until the forces acting on each atoms were less than 0.0001 eV$/$\AA. The convergence threshold for energy in the electronic self$-$consistent cycle was set to $10^{-8}$ eV. The simplified method suggested by Dudarev et al. \cite{Dudarev1998} was used to make the PBE$+$U calculations, which simply considers the difference between $U$ and $J$ ($U_\textrm{eff}=U-J$). The value of $U_\textrm{eff}$ for Fe 3$d$ was set at 4.0 eV \cite{Guo2022}, while it is zero for the other atoms. We also performed the spin$-$orbit coupling (SOC) calculations.

The lattice parameters of the relaxed structure are $a$ = $b$ = 10.2898 \AA\ and $c$ = 5.3884 \AA. This structure  consisting of two-dimensional iron squares perpendicular to the c axis as shown in Fig.\ \ref{fig:unitcell}(a). Each iron plane is separated by a distance of 1.3471 Å and contains silicon atoms. The central silicon plane forms a square-octagon lattice, and within each silicon squares have iron square. The distance between silicon atoms in the squares is 4.656 Å and between irons are 2.755 Å as shown in Fig.\ \ref{fig:unitcell}(b). Figure\ \ref{fig:unitcell} shows the $3\times3\times1$ sheet containing Si and Fe atoms from the central layer. Si atoms forms the square-octagon lattice. All the crystal structures shown herein were created using VESTA software \cite{Momma2011}.

The band structure of Lu$_2$Fe$_3$Si$_5$ is dominated by Fe 3$d$ electrons and exhibits a 3D configuration with electron-like and hole-like sheets. Figure \ref{fig:bnd-fs}(a) shows the band diagram of Lu$_2$Fe$_3$Si$_5$ with the weights of the various Fe 3$d$ orbitals. We also calculated the spin-polarized band structure, which was found to be spin-degenerate, suggesting the non-magnetic nature of the Fe atoms in this system. Therefore, we did not apply spin-orbit coupling. We found that four bands (122, 123, 124, and 125) cross the Fermi level, among which bands 122 and 123 are electron-like, and the remaining ones are hole-like. Both hole-like bands are degenerate at the high-symmetry point $A$ of the BZ, where they form a dome. This portion of the band diagram is dominated by the Fe 3$d_{z^2}$ orbital with some contributions from the $d_{xz}$ and $d_{x^2-y^2}$ orbitals near the Fermi surface. Both electron-like bands cross the Fermi surface along $\Gamma-Z$ and $X-R$, with contributions from the Fe 3$d_{x^2-y^2}$, $d_{xy}$, $d_{z^2}$, and $d_{xz}$ orbitals. The interactions between different orbitals and their contributions to the electronic structure can result in the emergence of multiple Fermi surface sheets and potentially lead to complex multiband superconductivity.

The Fermi surface of Lu$_2$Fe$_3$Si$_5$ is composed of several bands originating from mainly three distinct branches labeled as i-iii in Figure \ref{fig:bnd-fs}(b). Each of these bands possesses specific features that play a crucial role in the superconducting state of the compound. This Fermi surface exhibits both hole-like and electron-like characteristics in its topology. Notably, FS pockets i and iii exhibit a hole-like nature, while FS sheet ii is electron-like. In FS pockets i and iii, there are pockets centered at the corner ($A$) of the Brillouin zone (BZ) with a quasi-2D structure. Conversely, the electron-like Fermi surface sheet ii takes on a three-dimensional form, resembling a roughly spherical shape positioned along the $\Gamma-Z$ path. Moreover, FS sheet iii also displays z 3D pocket that takes on a cylindrical form centered around the $\Gamma$ point. These Fermi surface sheets confirm the fulfillment of conditions for multi-band superconductivity in Lu$_2$Fe$_3$Si$_5$. This finding is in good agreement with previous works \cite{Winiarski2013, Samsel-Czekaa2012, Nakajima2008}.

\begin{figure}[!ht]
\centering
\includegraphics[width=0.45\textwidth]{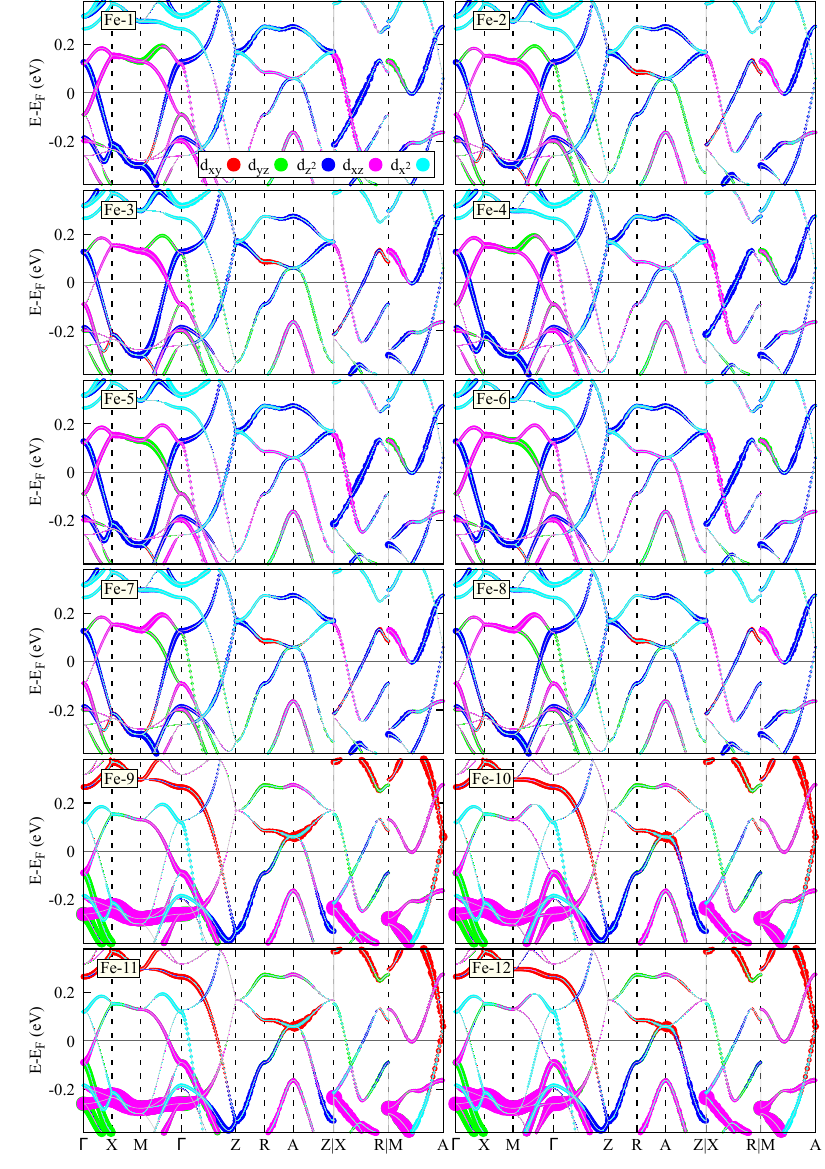}
\caption{\label{fig:bnd-fe} Projected bands of all Fe atoms are depicted, with the weight of $d$ orbitals varying according to the symbol size. The colors represent different $d$ orbitals. The vertical dashed lines indicate the high-symmetry points of the BZ, and the gray horizontal line represents the offset from the Fermi energy to zero. The legend is displayed in the top-left sub-figure. Fe-i represents the label of Fe atoms.}
\end{figure}

In Figure \ref{fig:bnd-fe}, we present the projected band structure of all Fe atoms in the unit cell, highlighting the weight of Fe 3$d$ orbitals using the symbol size. Upon inspecting Fig. \ref{fig:bnd-fe}, we observe that for Fe-1 and Fe-4, the $d_{xz}$ and $d_{z^2}$ orbitals contribute predominantly near the Fermi level. For Fe-2 and Fe-3, the $d_{xz}$, $d_{z^2}$, and $d_{yz}$ orbitals exhibit significant contributions. Similarly, for Fe-5 to Fe-8, the $d_{xz}$, $d_{z^2}$, and $d_{yz}$ orbitals play a major role, with a larger weight of $d_{yz}$ observed for Fe-7 and Fe-8. In the case of Fe-9 to Fe-10, the $d_{xz}$, $d_{x^2}$, $d_{xy}$, and $d_{yz}$ orbitals predominantly contribute, while Fe-10 and Fe-12 exhibit a higher weight of $d_{xz}$ compared to others.  Recall the positions of Fe atoms in the unit-cell, Fe-5 to Fe-8 form the central layer along the $c$ axis, while Fe-1 to Fe-4 constitute the outer layers. Fe-9 to Fe-12 are sandwiched between the outer and central layers of Fe. The central layer of Fe, which forms a square, has a contribution from the $d_{z^2}$ orbital, whereas the sandwiched layer, which does not form a square, has no contribution from the $d_{z^2}$ orbital but exhibits the $d_{x^2}$ orbital.

\begin{figure}[!ht]
    \centering
    \includegraphics[width=0.49\textwidth]{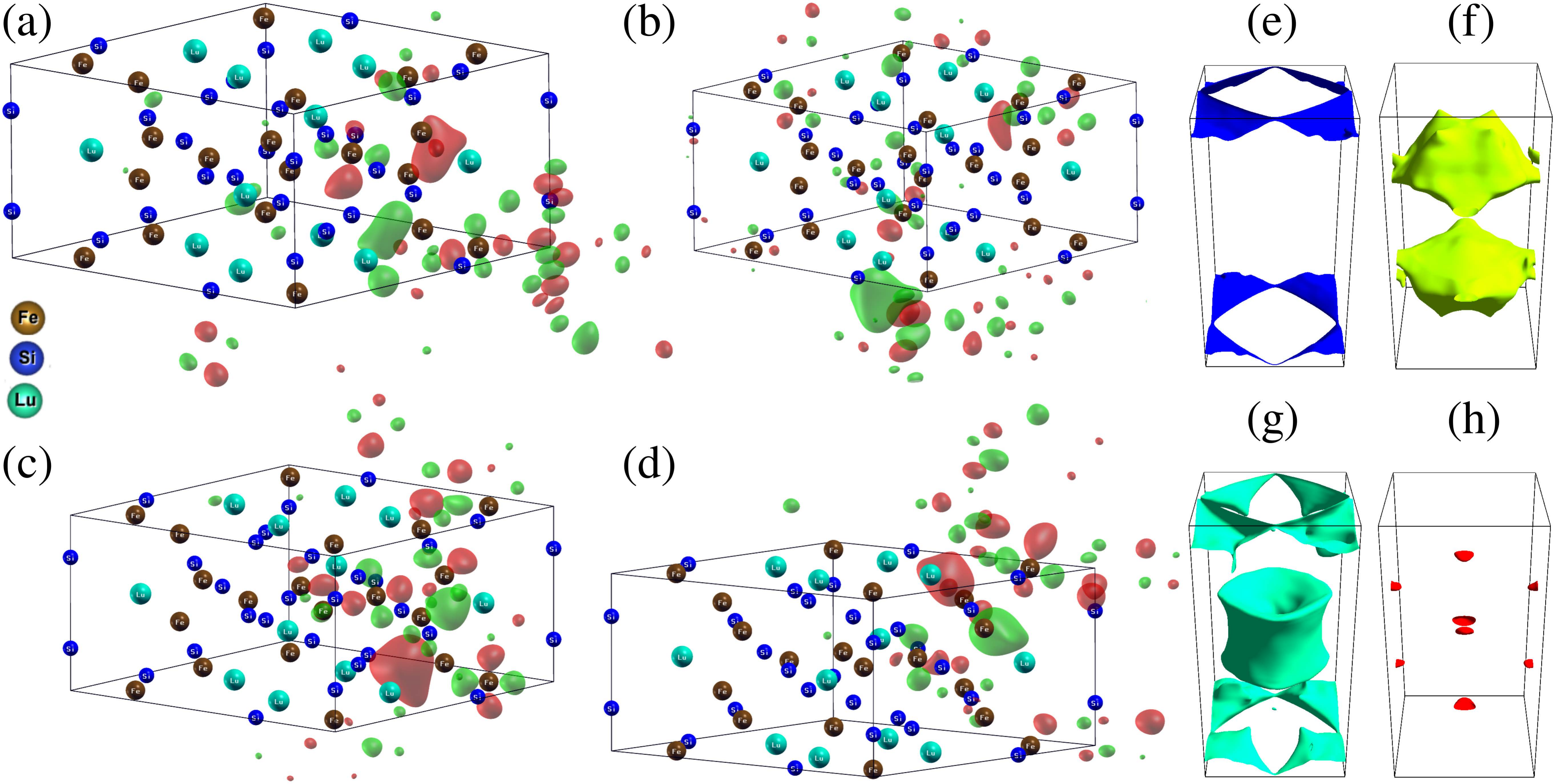}
    \caption{(a-d) Wannier functions calculated using random projections via the selected columns of the density matrix (SCDM) method \cite{Damle2015, Damle2018}. (e-h) The Fermi surface reproduced from the Wannier functions.}
    \label{fig:wf_Fs}
\end{figure}

Figure~\ref{fig:wf_Fs} gives the Wannier orbitals and the FS topology obtained in the Wannier orbital method.

\bibliography{Lu2Fe3Si5.bib}

\begin{thebibliography}{96}%
\makeatletter
\providecommand \@ifxundefined [1]{%
 \@ifx{#1\undefined}
}%
\providecommand \@ifnum [1]{%
 \ifnum #1\expandafter \@firstoftwo
 \else \expandafter \@secondoftwo
 \fi
}%
\providecommand \@ifx [1]{%
 \ifx #1\expandafter \@firstoftwo
 \else \expandafter \@secondoftwo
 \fi
}%
\providecommand \natexlab [1]{#1}%
\providecommand \enquote  [1]{``#1''}%
\providecommand \bibnamefont  [1]{#1}%
\providecommand \bibfnamefont [1]{#1}%
\providecommand \citenamefont [1]{#1}%
\providecommand \href@noop [0]{\@secondoftwo}%
\providecommand \href [0]{\begingroup \@sanitize@url \@href}%
\providecommand \@href[1]{\@@startlink{#1}\@@href}%
\providecommand \@@href[1]{\endgroup#1\@@endlink}%
\providecommand \@sanitize@url [0]{\catcode `\\12\catcode `\$12\catcode `\&12\catcode `\#12\catcode `\^12\catcode `\_12\catcode `\%12\relax}%
\providecommand \@@startlink[1]{}%
\providecommand \@@endlink[0]{}%
\providecommand \url  [0]{\begingroup\@sanitize@url \@url }%
\providecommand \@url [1]{\endgroup\@href {#1}{\urlprefix }}%
\providecommand \urlprefix  [0]{URL }%
\providecommand \Eprint [0]{\href }%
\providecommand \doibase [0]{https://doi.org/}%
\providecommand \selectlanguage [0]{\@gobble}%
\providecommand \bibinfo  [0]{\@secondoftwo}%
\providecommand \bibfield  [0]{\@secondoftwo}%
\providecommand \translation [1]{[#1]}%
\providecommand \BibitemOpen [0]{}%
\providecommand \bibitemStop [0]{}%
\providecommand \bibitemNoStop [0]{.\EOS\space}%
\providecommand \EOS [0]{\spacefactor3000\relax}%
\providecommand \BibitemShut  [1]{\csname bibitem#1\endcsname}%
\let\auto@bib@innerbib\@empty
\bibitem [{\citenamefont {Marzari}\ \emph {et~al.}(2012)\citenamefont {Marzari}, \citenamefont {Mostofi}, \citenamefont {Yates}, \citenamefont {Souza},\ and\ \citenamefont {Vanderbilt}}]{Marzari2012}%
  \BibitemOpen
  \bibfield  {author} {\bibinfo {author} {\bibfnamefont {N.}~\bibnamefont {Marzari}}, \bibinfo {author} {\bibfnamefont {A.~A.}\ \bibnamefont {Mostofi}}, \bibinfo {author} {\bibfnamefont {J.~R.}\ \bibnamefont {Yates}}, \bibinfo {author} {\bibfnamefont {I.}~\bibnamefont {Souza}},\ and\ \bibinfo {author} {\bibfnamefont {D.}~\bibnamefont {Vanderbilt}},\ }\bibfield  {title} {\bibinfo {title} {Maximally localized wannier functions: Theory and applications},\ }\href {https://doi.org/10.1103/RevModPhys.84.1419} {\bibfield  {journal} {\bibinfo  {journal} {Rev. Mod. Phys.}\ }\textbf {\bibinfo {volume} {84}},\ \bibinfo {pages} {1419} (\bibinfo {year} {2012})}\BibitemShut {NoStop}%
\bibitem [{\citenamefont {Marzari}\ and\ \citenamefont {Vanderbilt}(1997)}]{Marzari1997}%
  \BibitemOpen
  \bibfield  {author} {\bibinfo {author} {\bibfnamefont {N.}~\bibnamefont {Marzari}}\ and\ \bibinfo {author} {\bibfnamefont {D.}~\bibnamefont {Vanderbilt}},\ }\bibfield  {title} {\bibinfo {title} {Maximally localized generalized wannier functions for composite energy bands},\ }\href {https://doi.org/10.1103/PhysRevB.56.12847} {\bibfield  {journal} {\bibinfo  {journal} {Phys. Rev. B}\ }\textbf {\bibinfo {volume} {56}},\ \bibinfo {pages} {12847} (\bibinfo {year} {1997})}\BibitemShut {NoStop}%
\bibitem [{\citenamefont {Souza}\ \emph {et~al.}(2001)\citenamefont {Souza}, \citenamefont {Marzari},\ and\ \citenamefont {Vanderbilt}}]{Souza2001}%
  \BibitemOpen
  \bibfield  {author} {\bibinfo {author} {\bibfnamefont {I.}~\bibnamefont {Souza}}, \bibinfo {author} {\bibfnamefont {N.}~\bibnamefont {Marzari}},\ and\ \bibinfo {author} {\bibfnamefont {D.}~\bibnamefont {Vanderbilt}},\ }\bibfield  {title} {\bibinfo {title} {Maximally localized wannier functions for entangled energy bands},\ }\href {https://doi.org/10.1103/PhysRevB.65.035109} {\bibfield  {journal} {\bibinfo  {journal} {Phys. Rev. B}\ }\textbf {\bibinfo {volume} {65}},\ \bibinfo {pages} {035109} (\bibinfo {year} {2001})}\BibitemShut {NoStop}%
\bibitem [{\citenamefont {Soluyanov}\ and\ \citenamefont {Vanderbilt}(2011)}]{Soluyanov2011}%
  \BibitemOpen
  \bibfield  {author} {\bibinfo {author} {\bibfnamefont {A.~A.}\ \bibnamefont {Soluyanov}}\ and\ \bibinfo {author} {\bibfnamefont {D.}~\bibnamefont {Vanderbilt}},\ }\bibfield  {title} {\bibinfo {title} {Wannier representation of z2 topological insulators},\ }\href {https://doi.org/10.1103/PhysRevB.83.035108} {\bibfield  {journal} {\bibinfo  {journal} {Phys. Rev. B}\ }\textbf {\bibinfo {volume} {83}},\ \bibinfo {pages} {035108} (\bibinfo {year} {2011})}\BibitemShut {NoStop}%
\bibitem [{\citenamefont {Wu}\ \emph {et~al.}(2012)\citenamefont {Wu}, \citenamefont {Regnault},\ and\ \citenamefont {Bernevig}}]{Wu2012}%
  \BibitemOpen
  \bibfield  {author} {\bibinfo {author} {\bibfnamefont {Y.-L.}\ \bibnamefont {Wu}}, \bibinfo {author} {\bibfnamefont {N.}~\bibnamefont {Regnault}},\ and\ \bibinfo {author} {\bibfnamefont {B.~A.}\ \bibnamefont {Bernevig}},\ }\bibfield  {title} {\bibinfo {title} {Gauge-fixed wannier wave functions for fractional topological insulators},\ }\href {https://doi.org/10.1103/PhysRevB.86.085129} {\bibfield  {journal} {\bibinfo  {journal} {Phys. Rev. B}\ }\textbf {\bibinfo {volume} {86}},\ \bibinfo {pages} {085129} (\bibinfo {year} {2012})}\BibitemShut {NoStop}%
\bibitem [{\citenamefont {Winkler}\ \emph {et~al.}(2016)\citenamefont {Winkler}, \citenamefont {Soluyanov},\ and\ \citenamefont {Troyer}}]{Winkler2016}%
  \BibitemOpen
  \bibfield  {author} {\bibinfo {author} {\bibfnamefont {G.~W.}\ \bibnamefont {Winkler}}, \bibinfo {author} {\bibfnamefont {A.~A.}\ \bibnamefont {Soluyanov}},\ and\ \bibinfo {author} {\bibfnamefont {M.}~\bibnamefont {Troyer}},\ }\bibfield  {title} {\bibinfo {title} {Smooth gauge and wannier functions for topological band structures in arbitrary dimensions},\ }\href {https://doi.org/10.1103/PhysRevB.93.035453} {\bibfield  {journal} {\bibinfo  {journal} {Phys. Rev. B}\ }\textbf {\bibinfo {volume} {93}},\ \bibinfo {pages} {035453} (\bibinfo {year} {2016})}\BibitemShut {NoStop}%
\bibitem [{\citenamefont {Gresch}\ \emph {et~al.}(2017)\citenamefont {Gresch}, \citenamefont {Autès}, \citenamefont {Yazyev}, \citenamefont {Troyer}, \citenamefont {Vanderbilt}, \citenamefont {Bernevig},\ and\ \citenamefont {Soluyanov}}]{Gresch2017}%
  \BibitemOpen
  \bibfield  {author} {\bibinfo {author} {\bibfnamefont {D.}~\bibnamefont {Gresch}}, \bibinfo {author} {\bibfnamefont {G.}~\bibnamefont {Autès}}, \bibinfo {author} {\bibfnamefont {O.~V.}\ \bibnamefont {Yazyev}}, \bibinfo {author} {\bibfnamefont {M.}~\bibnamefont {Troyer}}, \bibinfo {author} {\bibfnamefont {D.}~\bibnamefont {Vanderbilt}}, \bibinfo {author} {\bibfnamefont {B.~A.}\ \bibnamefont {Bernevig}},\ and\ \bibinfo {author} {\bibfnamefont {A.~A.}\ \bibnamefont {Soluyanov}},\ }\bibfield  {title} {\bibinfo {title} {Z2pack: Numerical implementation of hybrid wannier centers for identifying topological materials},\ }\href {https://doi.org/10.1103/PhysRevB.95.075146} {\bibfield  {journal} {\bibinfo  {journal} {Phys. Rev. B}\ }\textbf {\bibinfo {volume} {95}},\ \bibinfo {pages} {075146} (\bibinfo {year} {2017})}\BibitemShut {NoStop}%
\bibitem [{\citenamefont {Cornean}\ and\ \citenamefont {Monaco}(2017)}]{Cornean2017}%
  \BibitemOpen
  \bibfield  {author} {\bibinfo {author} {\bibfnamefont {H.~D.}\ \bibnamefont {Cornean}}\ and\ \bibinfo {author} {\bibfnamefont {D.}~\bibnamefont {Monaco}},\ }\bibfield  {title} {\bibinfo {title} {On the construction of wannier functions in topological insulators: the 3d case},\ }\href {https://doi.org/10.1007/s00023-017-0621-y} {\bibfield  {journal} {\bibinfo  {journal} {Annales Henri Poincaré}\ }\textbf {\bibinfo {volume} {18}},\ \bibinfo {pages} {3863} (\bibinfo {year} {2017})}\BibitemShut {NoStop}%
\bibitem [{\citenamefont {Po}\ \emph {et~al.}(2018)\citenamefont {Po}, \citenamefont {Watanabe},\ and\ \citenamefont {Vishwanath}}]{Po2018}%
  \BibitemOpen
  \bibfield  {author} {\bibinfo {author} {\bibfnamefont {H.~C.}\ \bibnamefont {Po}}, \bibinfo {author} {\bibfnamefont {H.}~\bibnamefont {Watanabe}},\ and\ \bibinfo {author} {\bibfnamefont {A.}~\bibnamefont {Vishwanath}},\ }\bibfield  {title} {\bibinfo {title} {Fragile topology and wannier obstructions},\ }\href {https://doi.org/10.1103/PhysRevLett.121.126402} {\bibfield  {journal} {\bibinfo  {journal} {Phys. Rev. Lett.}\ }\textbf {\bibinfo {volume} {121}},\ \bibinfo {pages} {126402} (\bibinfo {year} {2018})}\BibitemShut {NoStop}%
\bibitem [{\citenamefont {Anisimov}\ \emph {et~al.}(2005)\citenamefont {Anisimov}, \citenamefont {Kondakov}, \citenamefont {Kozhevnikov}, \citenamefont {Nekrasov}, \citenamefont {Pchelkina}, \citenamefont {Allen}, \citenamefont {Mo}, \citenamefont {Kim}, \citenamefont {Metcalf}, \citenamefont {Suga}, \citenamefont {Sekiyama}, \citenamefont {Keller}, \citenamefont {Leonov}, \citenamefont {Ren},\ and\ \citenamefont {Vollhardt}}]{Anisimov2005}%
  \BibitemOpen
  \bibfield  {author} {\bibinfo {author} {\bibfnamefont {V.~I.}\ \bibnamefont {Anisimov}}, \bibinfo {author} {\bibfnamefont {D.~E.}\ \bibnamefont {Kondakov}}, \bibinfo {author} {\bibfnamefont {A.~V.}\ \bibnamefont {Kozhevnikov}}, \bibinfo {author} {\bibfnamefont {I.~A.}\ \bibnamefont {Nekrasov}}, \bibinfo {author} {\bibfnamefont {Z.~V.}\ \bibnamefont {Pchelkina}}, \bibinfo {author} {\bibfnamefont {J.~W.}\ \bibnamefont {Allen}}, \bibinfo {author} {\bibfnamefont {S.-K.}\ \bibnamefont {Mo}}, \bibinfo {author} {\bibfnamefont {H.-D.}\ \bibnamefont {Kim}}, \bibinfo {author} {\bibfnamefont {P.}~\bibnamefont {Metcalf}}, \bibinfo {author} {\bibfnamefont {S.}~\bibnamefont {Suga}}, \bibinfo {author} {\bibfnamefont {A.}~\bibnamefont {Sekiyama}}, \bibinfo {author} {\bibfnamefont {G.}~\bibnamefont {Keller}}, \bibinfo {author} {\bibfnamefont {I.}~\bibnamefont {Leonov}}, \bibinfo {author} {\bibfnamefont {X.}~\bibnamefont {Ren}},\ and\ \bibinfo {author} {\bibfnamefont {D.}~\bibnamefont {Vollhardt}},\ }\bibfield  {title}
  {\bibinfo {title} {Full orbital calculation scheme for materials with strongly correlated electrons},\ }\href {https://doi.org/10.1103/PhysRevB.71.125119} {\bibfield  {journal} {\bibinfo  {journal} {Phys. Rev. B}\ }\textbf {\bibinfo {volume} {71}},\ \bibinfo {pages} {125119} (\bibinfo {year} {2005})}\BibitemShut {NoStop}%
\bibitem [{\citenamefont {Lechermann}\ \emph {et~al.}(2006)\citenamefont {Lechermann}, \citenamefont {Georges}, \citenamefont {Poteryaev}, \citenamefont {Biermann}, \citenamefont {Posternak}, \citenamefont {Yamasaki},\ and\ \citenamefont {Andersen}}]{Lechermann2006}%
  \BibitemOpen
  \bibfield  {author} {\bibinfo {author} {\bibfnamefont {F.}~\bibnamefont {Lechermann}}, \bibinfo {author} {\bibfnamefont {A.}~\bibnamefont {Georges}}, \bibinfo {author} {\bibfnamefont {A.}~\bibnamefont {Poteryaev}}, \bibinfo {author} {\bibfnamefont {S.}~\bibnamefont {Biermann}}, \bibinfo {author} {\bibfnamefont {M.}~\bibnamefont {Posternak}}, \bibinfo {author} {\bibfnamefont {A.}~\bibnamefont {Yamasaki}},\ and\ \bibinfo {author} {\bibfnamefont {O.~K.}\ \bibnamefont {Andersen}},\ }\bibfield  {title} {\bibinfo {title} {Dynamical mean-field theory using wannier functions: A flexible route to electronic structure calculations of strongly correlated materials},\ }\href {https://doi.org/10.1103/PhysRevB.74.125120} {\bibfield  {journal} {\bibinfo  {journal} {Phys. Rev. B}\ }\textbf {\bibinfo {volume} {74}},\ \bibinfo {pages} {125120} (\bibinfo {year} {2006})}\BibitemShut {NoStop}%
\bibitem [{\citenamefont {Das}\ \emph {et~al.}(2014)\citenamefont {Das}, \citenamefont {Markiewicz},\ and\ \citenamefont {Bansil}}]{Das2014}%
  \BibitemOpen
  \bibfield  {author} {\bibinfo {author} {\bibfnamefont {T.}~\bibnamefont {Das}}, \bibinfo {author} {\bibfnamefont {R.~S.}\ \bibnamefont {Markiewicz}},\ and\ \bibinfo {author} {\bibfnamefont {A.}~\bibnamefont {Bansil}},\ }\bibfield  {title} {\bibinfo {title} {{Intermediate coupling model of the cuprates}},\ }\href {https://doi.org/10.1080/00018732.2014.940227} {\bibfield  {journal} {\bibinfo  {journal} {Adv. Phys.}\ }\textbf {\bibinfo {volume} {63}},\ \bibinfo {pages} {151} (\bibinfo {year} {2014})}\BibitemShut {NoStop}%
\bibitem [{\citenamefont {Freimuth}\ \emph {et~al.}(2023)\citenamefont {Freimuth}, \citenamefont {Blügel},\ and\ \citenamefont {Mokrousov}}]{Freimuth2023}%
  \BibitemOpen
  \bibfield  {author} {\bibinfo {author} {\bibfnamefont {F.}~\bibnamefont {Freimuth}}, \bibinfo {author} {\bibfnamefont {S.}~\bibnamefont {Blügel}},\ and\ \bibinfo {author} {\bibfnamefont {Y.}~\bibnamefont {Mokrousov}},\ }\href {https://arxiv.org/abs/2301.04734} {\bibinfo {title} {Construction of wannier functions from the spectral moments of correlated electron systems}} (\bibinfo {year} {2023}),\ \Eprint {https://arxiv.org/abs/2301.04734} {arXiv:2301.04734 [cond-mat.str-el]} \BibitemShut {NoStop}%
\bibitem [{\citenamefont {Vollhardt}(2020)}]{Vollhardt2020}%
  \BibitemOpen
  \bibfield  {author} {\bibinfo {author} {\bibfnamefont {D.}~\bibnamefont {Vollhardt}},\ }\href {https://doi.org/10.7566/JPSCP.30.011001} {\bibinfo {title} {Dynamical mean-field theory of strongly correlated electron systems}} (\bibinfo {year} {2020})\BibitemShut {NoStop}%
\bibitem [{\citenamefont {Boyack}\ \emph {et~al.}(2024)\citenamefont {Boyack}, \citenamefont {Delacr{\'{e}}taz}, \citenamefont {Dupuis},\ and\ \citenamefont {Witczak-Krempa}}]{Boyack2024}%
  \BibitemOpen
  \bibfield  {author} {\bibinfo {author} {\bibfnamefont {R.}~\bibnamefont {Boyack}}, \bibinfo {author} {\bibfnamefont {L.}~\bibnamefont {Delacr{\'{e}}taz}}, \bibinfo {author} {\bibfnamefont {{\'{E}}.}~\bibnamefont {Dupuis}},\ and\ \bibinfo {author} {\bibfnamefont {W.}~\bibnamefont {Witczak-Krempa}},\ }\bibfield  {title} {\bibinfo {title} {{Conformal field theories in a magnetic field}},\ }\href {https://doi.org/10.1103/PhysRevResearch.6.043093} {\bibfield  {journal} {\bibinfo  {journal} {Phys. Rev. Res.}\ }\textbf {\bibinfo {volume} {6}},\ \bibinfo {pages} {043093} (\bibinfo {year} {2024})}\BibitemShut {NoStop}%
\bibitem [{\citenamefont {Read}(2017)}]{Read2017}%
  \BibitemOpen
  \bibfield  {author} {\bibinfo {author} {\bibfnamefont {N.}~\bibnamefont {Read}},\ }\bibfield  {title} {\bibinfo {title} {{Compactly supported Wannier functions and algebraic K-theory}},\ }\href {https://doi.org/10.1103/PhysRevB.95.115309} {\bibfield  {journal} {\bibinfo  {journal} {Phys. Rev. B}\ }\textbf {\bibinfo {volume} {95}},\ \bibinfo {pages} {115309} (\bibinfo {year} {2017})},\ \Eprint {https://arxiv.org/abs/1608.04696} {1608.04696} \BibitemShut {NoStop}%
\bibitem [{\citenamefont {Schindler}\ \emph {et~al.}(2020)\citenamefont {Schindler}, \citenamefont {Bradlyn}, \citenamefont {Fischer},\ and\ \citenamefont {Neupert}}]{Schindler2020}%
  \BibitemOpen
  \bibfield  {author} {\bibinfo {author} {\bibfnamefont {F.}~\bibnamefont {Schindler}}, \bibinfo {author} {\bibfnamefont {B.}~\bibnamefont {Bradlyn}}, \bibinfo {author} {\bibfnamefont {M.~H.}\ \bibnamefont {Fischer}},\ and\ \bibinfo {author} {\bibfnamefont {T.}~\bibnamefont {Neupert}},\ }\bibfield  {title} {\bibinfo {title} {{Pairing Obstructions in Topological Superconductors}},\ }\href {https://doi.org/10.1103/PhysRevLett.124.247001} {\bibfield  {journal} {\bibinfo  {journal} {Phys. Rev. Lett.}\ }\textbf {\bibinfo {volume} {124}},\ \bibinfo {pages} {247001} (\bibinfo {year} {2020})},\ \Eprint {https://arxiv.org/abs/2001.02682} {2001.02682} \BibitemShut {NoStop}%
\bibitem [{\citenamefont {Wahl}\ \emph {et~al.}(2013)\citenamefont {Wahl}, \citenamefont {Tu}, \citenamefont {Schuch},\ and\ \citenamefont {Cirac}}]{Wahl2013}%
  \BibitemOpen
  \bibfield  {author} {\bibinfo {author} {\bibfnamefont {T.~B.}\ \bibnamefont {Wahl}}, \bibinfo {author} {\bibfnamefont {H.-H.}\ \bibnamefont {Tu}}, \bibinfo {author} {\bibfnamefont {N.}~\bibnamefont {Schuch}},\ and\ \bibinfo {author} {\bibfnamefont {J.~I.}\ \bibnamefont {Cirac}},\ }\bibfield  {title} {\bibinfo {title} {{Projected Entangled-Pair States Can Describe Chiral Topological States}},\ }\href {https://doi.org/10.1103/PhysRevLett.111.236805} {\bibfield  {journal} {\bibinfo  {journal} {Phys. Rev. Lett.}\ }\textbf {\bibinfo {volume} {111}},\ \bibinfo {pages} {236805} (\bibinfo {year} {2013})},\ \Eprint {https://arxiv.org/abs/1308.0316} {1308.0316} \BibitemShut {NoStop}%
\bibitem [{\citenamefont {Qi}(2011)}]{Qi2011}%
  \BibitemOpen
  \bibfield  {author} {\bibinfo {author} {\bibfnamefont {X.-L.}\ \bibnamefont {Qi}},\ }\bibfield  {title} {\bibinfo {title} {{Generic Wave-Function Description of Fractional Quantum Anomalous Hall States and Fractional Topological Insulators}},\ }\href {https://doi.org/10.1103/PhysRevLett.107.126803} {\bibfield  {journal} {\bibinfo  {journal} {Phys. Rev. Lett.}\ }\textbf {\bibinfo {volume} {107}},\ \bibinfo {pages} {126803} (\bibinfo {year} {2011})},\ \Eprint {https://arxiv.org/abs/1105.4298} {1105.4298} \BibitemShut {NoStop}%
\bibitem [{\citenamefont {Nakagawa}\ \emph {et~al.}(2020)\citenamefont {Nakagawa}, \citenamefont {Slager}, \citenamefont {Higashikawa},\ and\ \citenamefont {Oka}}]{Nakagawa2020}%
  \BibitemOpen
  \bibfield  {author} {\bibinfo {author} {\bibfnamefont {M.}~\bibnamefont {Nakagawa}}, \bibinfo {author} {\bibfnamefont {R.-J.}\ \bibnamefont {Slager}}, \bibinfo {author} {\bibfnamefont {S.}~\bibnamefont {Higashikawa}},\ and\ \bibinfo {author} {\bibfnamefont {T.}~\bibnamefont {Oka}},\ }\bibfield  {title} {\bibinfo {title} {{Wannier representation of Floquet topological states}},\ }\href {https://doi.org/10.1103/PhysRevB.101.075108} {\bibfield  {journal} {\bibinfo  {journal} {Phys. Rev. B}\ }\textbf {\bibinfo {volume} {101}},\ \bibinfo {pages} {075108} (\bibinfo {year} {2020})},\ \Eprint {https://arxiv.org/abs/1903.12197} {1903.12197} \BibitemShut {NoStop}%
\bibitem [{\citenamefont {Gupta}\ and\ \citenamefont {Bradlyn}(2022)}]{Gupta2022}%
  \BibitemOpen
  \bibfield  {author} {\bibinfo {author} {\bibfnamefont {V.}~\bibnamefont {Gupta}}\ and\ \bibinfo {author} {\bibfnamefont {B.}~\bibnamefont {Bradlyn}},\ }\bibfield  {title} {\bibinfo {title} {{Wannier-function methods for topological modes in one-dimensional photonic crystals}},\ }\href {https://doi.org/10.1103/PhysRevA.105.053521} {\bibfield  {journal} {\bibinfo  {journal} {Phys. Rev. A}\ }\textbf {\bibinfo {volume} {105}},\ \bibinfo {pages} {053521} (\bibinfo {year} {2022})}\BibitemShut {NoStop}%
\bibitem [{\citenamefont {Li}\ \emph {et~al.}(2024)\citenamefont {Li}, \citenamefont {Dong}, \citenamefont {Ledwith},\ and\ \citenamefont {Khalaf}}]{Li2024}%
  \BibitemOpen
  \bibfield  {author} {\bibinfo {author} {\bibfnamefont {Q.}~\bibnamefont {Li}}, \bibinfo {author} {\bibfnamefont {J.}~\bibnamefont {Dong}}, \bibinfo {author} {\bibfnamefont {P.~J.}\ \bibnamefont {Ledwith}},\ and\ \bibinfo {author} {\bibfnamefont {E.}~\bibnamefont {Khalaf}},\ }\href {https://arxiv.org/abs/2407.02561} {\bibinfo {title} {Constraints on real space representations of chern bands}} (\bibinfo {year} {2024}),\ \Eprint {https://arxiv.org/abs/2407.02561} {arXiv:2407.02561 [cond-mat.str-el]} \BibitemShut {NoStop}%
\bibitem [{\citenamefont {Wang}\ \emph {et~al.}(2024)\citenamefont {Wang}, \citenamefont {Shi}, \citenamefont {Liu},\ and\ \citenamefont {Wang}}]{Wang2024}%
  \BibitemOpen
  \bibfield  {author} {\bibinfo {author} {\bibfnamefont {H.}~\bibnamefont {Wang}}, \bibinfo {author} {\bibfnamefont {R.}~\bibnamefont {Shi}}, \bibinfo {author} {\bibfnamefont {Z.}~\bibnamefont {Liu}},\ and\ \bibinfo {author} {\bibfnamefont {J.}~\bibnamefont {Wang}},\ }\href {https://arxiv.org/abs/2411.13071} {\bibinfo {title} {Orbital description of landau levels}} (\bibinfo {year} {2024}),\ \Eprint {https://arxiv.org/abs/2411.13071} {arXiv:2411.13071 [cond-mat.mes-hall]} \BibitemShut {NoStop}%
\bibitem [{\citenamefont {Nakajima}\ \emph {et~al.}(2012)\citenamefont {Nakajima}, \citenamefont {Hidaka}, \citenamefont {Nakagawa}, \citenamefont {Tamegai}, \citenamefont {Nishizaki}, \citenamefont {Sasaki},\ and\ \citenamefont {Kobayashi}}]{Nakajima2012}%
  \BibitemOpen
  \bibfield  {author} {\bibinfo {author} {\bibfnamefont {Y.}~\bibnamefont {Nakajima}}, \bibinfo {author} {\bibfnamefont {H.}~\bibnamefont {Hidaka}}, \bibinfo {author} {\bibfnamefont {T.}~\bibnamefont {Nakagawa}}, \bibinfo {author} {\bibfnamefont {T.}~\bibnamefont {Tamegai}}, \bibinfo {author} {\bibfnamefont {T.}~\bibnamefont {Nishizaki}}, \bibinfo {author} {\bibfnamefont {T.}~\bibnamefont {Sasaki}},\ and\ \bibinfo {author} {\bibfnamefont {N.}~\bibnamefont {Kobayashi}},\ }\bibfield  {title} {\bibinfo {title} {{Two-band superconductivity featuring different anisotropies in the ternary iron silicide Lu2Fe3Si5}},\ }\href {https://doi.org/10.1103/PhysRevB.85.174524} {\bibfield  {journal} {\bibinfo  {journal} {Phys. Rev. B}\ }\textbf {\bibinfo {volume} {85}},\ \bibinfo {pages} {174524} (\bibinfo {year} {2012})}\BibitemShut {NoStop}%
\bibitem [{\citenamefont {Machida}\ \emph {et~al.}(2011)\citenamefont {Machida}, \citenamefont {Sakai}, \citenamefont {Izawa}, \citenamefont {Okuyama},\ and\ \citenamefont {Watanabe}}]{Machida2011}%
  \BibitemOpen
  \bibfield  {author} {\bibinfo {author} {\bibfnamefont {Y.}~\bibnamefont {Machida}}, \bibinfo {author} {\bibfnamefont {S.}~\bibnamefont {Sakai}}, \bibinfo {author} {\bibfnamefont {K.}~\bibnamefont {Izawa}}, \bibinfo {author} {\bibfnamefont {H.}~\bibnamefont {Okuyama}},\ and\ \bibinfo {author} {\bibfnamefont {T.}~\bibnamefont {Watanabe}},\ }\bibfield  {title} {\bibinfo {title} {{Enhanced Quasiparticle Heat Conduction in the Multigap Superconductor Lu2Fe3Si5}},\ }\href {https://doi.org/10.1103/PhysRevLett.106.107002} {\bibfield  {journal} {\bibinfo  {journal} {Phys. Rev. Lett.}\ }\textbf {\bibinfo {volume} {106}},\ \bibinfo {pages} {107002} (\bibinfo {year} {2011})}\BibitemShut {NoStop}%
\bibitem [{\citenamefont {Tamegai}\ \emph {et~al.}(2009)\citenamefont {Tamegai}, \citenamefont {Nakajima}, \citenamefont {Nakagawa}, \citenamefont {Li},\ and\ \citenamefont {Harima}}]{Tamegai2009}%
  \BibitemOpen
  \bibfield  {author} {\bibinfo {author} {\bibfnamefont {T.}~\bibnamefont {Tamegai}}, \bibinfo {author} {\bibfnamefont {Y.}~\bibnamefont {Nakajima}}, \bibinfo {author} {\bibfnamefont {T.}~\bibnamefont {Nakagawa}}, \bibinfo {author} {\bibfnamefont {G.~J.}\ \bibnamefont {Li}},\ and\ \bibinfo {author} {\bibfnamefont {H.}~\bibnamefont {Harima}},\ }\bibfield  {title} {\bibinfo {title} {{Two-gap superconductivity in R 2 Fe 3 Si 5 ( R = Lu and Sc)}},\ }\href {https://doi.org/10.1088/1742-6596/150/5/052264} {\bibfield  {journal} {\bibinfo  {journal} {J. Phys. Conf. Ser.}\ }\textbf {\bibinfo {volume} {150}},\ \bibinfo {pages} {052264} (\bibinfo {year} {2009})}\BibitemShut {NoStop}%
\bibitem [{\citenamefont {Nakajima}\ \emph {et~al.}(2008)\citenamefont {Nakajima}, \citenamefont {Nakagawa}, \citenamefont {Tamegai},\ and\ \citenamefont {Harima}}]{Nakajima2008}%
  \BibitemOpen
  \bibfield  {author} {\bibinfo {author} {\bibfnamefont {Y.}~\bibnamefont {Nakajima}}, \bibinfo {author} {\bibfnamefont {T.}~\bibnamefont {Nakagawa}}, \bibinfo {author} {\bibfnamefont {T.}~\bibnamefont {Tamegai}},\ and\ \bibinfo {author} {\bibfnamefont {H.}~\bibnamefont {Harima}},\ }\bibfield  {title} {\bibinfo {title} {{Specific-heat evidence for two-gap superconductivity in the ternary-iron silicide Lu2Fe3Si5}},\ }\href {https://doi.org/10.1103/PhysRevLett.100.157001} {\bibfield  {journal} {\bibinfo  {journal} {Phys. Rev. Lett.}\ }\textbf {\bibinfo {volume} {100}},\ \bibinfo {pages} {157001} (\bibinfo {year} {2008})}\BibitemShut {NoStop}%
\bibitem [{\citenamefont {Vining}\ \emph {et~al.}(1983)\citenamefont {Vining}, \citenamefont {Shelton}, \citenamefont {Braun},\ and\ \citenamefont {Pelizzone}}]{Vining1983}%
  \BibitemOpen
  \bibfield  {author} {\bibinfo {author} {\bibfnamefont {C.~B.}\ \bibnamefont {Vining}}, \bibinfo {author} {\bibfnamefont {R.~N.}\ \bibnamefont {Shelton}}, \bibinfo {author} {\bibfnamefont {H.~F.}\ \bibnamefont {Braun}},\ and\ \bibinfo {author} {\bibfnamefont {M.}~\bibnamefont {Pelizzone}},\ }\bibfield  {title} {\bibinfo {title} {{Low-temperature heat capacity of superconducting ternary iron silicides}},\ }\href {https://doi.org/10.1103/PhysRevB.27.2800} {\bibfield  {journal} {\bibinfo  {journal} {Phys. Rev. B}\ }\textbf {\bibinfo {volume} {27}},\ \bibinfo {pages} {2800} (\bibinfo {year} {1983})}\BibitemShut {NoStop}%
\bibitem [{\citenamefont {Biswas}\ \emph {et~al.}(2011)\citenamefont {Biswas}, \citenamefont {Balakrishnan}, \citenamefont {Paul}, \citenamefont {Lees},\ and\ \citenamefont {Hillier}}]{Biswas2011}%
  \BibitemOpen
  \bibfield  {author} {\bibinfo {author} {\bibfnamefont {P.~K.}\ \bibnamefont {Biswas}}, \bibinfo {author} {\bibfnamefont {G.}~\bibnamefont {Balakrishnan}}, \bibinfo {author} {\bibfnamefont {D.~M.}\ \bibnamefont {Paul}}, \bibinfo {author} {\bibfnamefont {M.~R.}\ \bibnamefont {Lees}},\ and\ \bibinfo {author} {\bibfnamefont {A.~D.}\ \bibnamefont {Hillier}},\ }\bibfield  {title} {\bibinfo {title} {{Two-gap superconductivity in Lu2Fe3Si5: A transverse-field muon spin rotation study}},\ }\href {https://doi.org/10.1103/PhysRevB.83.054517} {\bibfield  {journal} {\bibinfo  {journal} {Phys. Rev. B}\ }\textbf {\bibinfo {volume} {83}},\ \bibinfo {pages} {054517} (\bibinfo {year} {2011})}\BibitemShut {NoStop}%
\bibitem [{\citenamefont {Gordon}\ \emph {et~al.}(2008)\citenamefont {Gordon}, \citenamefont {Vannette}, \citenamefont {Martin}, \citenamefont {Nakajima}, \citenamefont {Tamegai},\ and\ \citenamefont {Prozorov}}]{Gordon2008}%
  \BibitemOpen
  \bibfield  {author} {\bibinfo {author} {\bibfnamefont {R.~T.}\ \bibnamefont {Gordon}}, \bibinfo {author} {\bibfnamefont {M.~D.}\ \bibnamefont {Vannette}}, \bibinfo {author} {\bibfnamefont {C.}~\bibnamefont {Martin}}, \bibinfo {author} {\bibfnamefont {Y.}~\bibnamefont {Nakajima}}, \bibinfo {author} {\bibfnamefont {T.}~\bibnamefont {Tamegai}},\ and\ \bibinfo {author} {\bibfnamefont {R.}~\bibnamefont {Prozorov}},\ }\bibfield  {title} {\bibinfo {title} {{Two-gap superconductivity seen in penetration-depth measurements of Lu2 Fe3 Si5 single crystals}},\ }\href {https://doi.org/10.1103/PhysRevB.78.024514} {\bibfield  {journal} {\bibinfo  {journal} {Phys. Rev. B}\ }\textbf {\bibinfo {volume} {78}},\ \bibinfo {pages} {024514} (\bibinfo {year} {2008})}\BibitemShut {NoStop}%
\bibitem [{\citenamefont {Hidaka}\ \emph {et~al.}(2010)\citenamefont {Hidaka}, \citenamefont {Nakajima},\ and\ \citenamefont {Tamegai}}]{Hidaka2010}%
  \BibitemOpen
  \bibfield  {author} {\bibinfo {author} {\bibfnamefont {H.}~\bibnamefont {Hidaka}}, \bibinfo {author} {\bibfnamefont {Y.}~\bibnamefont {Nakajima}},\ and\ \bibinfo {author} {\bibfnamefont {T.}~\bibnamefont {Tamegai}},\ }\bibfield  {title} {\bibinfo {title} {{Impurity effects in two-gap superconductor Lu2Fe3Si5}},\ }\href {https://doi.org/10.1016/j.physc.2009.10.119} {\bibfield  {journal} {\bibinfo  {journal} {Phys. C Supercond. its Appl.}\ }\textbf {\bibinfo {volume} {470}},\ \bibinfo {pages} {S619} (\bibinfo {year} {2010})}\BibitemShut {NoStop}%
\bibitem [{\citenamefont {Watanabe}\ \emph {et~al.}(2009)\citenamefont {Watanabe}, \citenamefont {Sasame}, \citenamefont {Okuyama}, \citenamefont {Takase},\ and\ \citenamefont {Takano}}]{Watanabe2009}%
  \BibitemOpen
  \bibfield  {author} {\bibinfo {author} {\bibfnamefont {T.}~\bibnamefont {Watanabe}}, \bibinfo {author} {\bibfnamefont {H.}~\bibnamefont {Sasame}}, \bibinfo {author} {\bibfnamefont {H.}~\bibnamefont {Okuyama}}, \bibinfo {author} {\bibfnamefont {K.}~\bibnamefont {Takase}},\ and\ \bibinfo {author} {\bibfnamefont {Y.}~\bibnamefont {Takano}},\ }\bibfield  {title} {\bibinfo {title} {{Disorder-sensitive superconductivity in the doped iron silicide superconductor (Lu1-x Rx) 2 Fe3 Si5 (R=Sc, Y, and Dy)}},\ }\href {https://doi.org/10.1103/PhysRevB.80.100502} {\bibfield  {journal} {\bibinfo  {journal} {Phys. Rev. B}\ }\textbf {\bibinfo {volume} {80}},\ \bibinfo {pages} {100502} (\bibinfo {year} {2009})}\BibitemShut {NoStop}%
\bibitem [{\citenamefont {Hidaka}\ \emph {et~al.}(2009)\citenamefont {Hidaka}, \citenamefont {Nakajima},\ and\ \citenamefont {Tamegai}}]{Hidaka2009}%
  \BibitemOpen
  \bibfield  {author} {\bibinfo {author} {\bibfnamefont {H.}~\bibnamefont {Hidaka}}, \bibinfo {author} {\bibfnamefont {Y.}~\bibnamefont {Nakajima}},\ and\ \bibinfo {author} {\bibfnamefont {T.}~\bibnamefont {Tamegai}},\ }\bibfield  {title} {\bibinfo {title} {{Non-magnetic impurity effect in two-gap superconductor Lu2Fe3Si5}},\ }\href {https://doi.org/10.1016/j.physc.2009.05.184} {\bibfield  {journal} {\bibinfo  {journal} {Phys. C Supercond.}\ }\textbf {\bibinfo {volume} {469}},\ \bibinfo {pages} {999} (\bibinfo {year} {2009})}\BibitemShut {NoStop}%
\bibitem [{\citenamefont {Sasame}\ \emph {et~al.}(2009)\citenamefont {Sasame}, \citenamefont {Masubuchi}, \citenamefont {Takase}, \citenamefont {Takano},\ and\ \citenamefont {Watanabe}}]{Sasame2009}%
  \BibitemOpen
  \bibfield  {author} {\bibinfo {author} {\bibfnamefont {H.}~\bibnamefont {Sasame}}, \bibinfo {author} {\bibfnamefont {T.}~\bibnamefont {Masubuchi}}, \bibinfo {author} {\bibfnamefont {K.}~\bibnamefont {Takase}}, \bibinfo {author} {\bibfnamefont {Y.}~\bibnamefont {Takano}},\ and\ \bibinfo {author} {\bibfnamefont {T.}~\bibnamefont {Watanabe}},\ }\bibfield  {title} {\bibinfo {title} {{Superconducting properties of Lu 2 Fe 3 Si 5 with non-magnetic impurities}},\ }\href {https://doi.org/10.1088/1742-6596/150/5/052226} {\bibfield  {journal} {\bibinfo  {journal} {J. Phys. Conf. Ser.}\ }\textbf {\bibinfo {volume} {150}},\ \bibinfo {pages} {052226} (\bibinfo {year} {2009})}\BibitemShut {NoStop}%
\bibitem [{\citenamefont {Karkin}\ \emph {et~al.}(2011)\citenamefont {Karkin}, \citenamefont {Yangirov}, \citenamefont {Akshentsev},\ and\ \citenamefont {Goshchitskii}}]{Karkin2011}%
  \BibitemOpen
  \bibfield  {author} {\bibinfo {author} {\bibfnamefont {A.~E.}\ \bibnamefont {Karkin}}, \bibinfo {author} {\bibfnamefont {M.~R.}\ \bibnamefont {Yangirov}}, \bibinfo {author} {\bibfnamefont {Y.~N.}\ \bibnamefont {Akshentsev}},\ and\ \bibinfo {author} {\bibfnamefont {B.~N.}\ \bibnamefont {Goshchitskii}},\ }\bibfield  {title} {\bibinfo {title} {{Superconductivity in iron silicide Lu2Fe3Si 5 probed by radiation-induced disordering}},\ }\href {https://doi.org/10.1103/PhysRevB.84.054541} {\bibfield  {journal} {\bibinfo  {journal} {Phys. Rev. B}\ }\textbf {\bibinfo {volume} {84}},\ \bibinfo {pages} {054541} (\bibinfo {year} {2011})}\BibitemShut {NoStop}%
\bibitem [{\citenamefont {Braun}(1980)}]{Braun1980}%
  \BibitemOpen
  \bibfield  {author} {\bibinfo {author} {\bibfnamefont {H.~F.}\ \bibnamefont {Braun}},\ }\bibfield  {title} {\bibinfo {title} {{Superconductivity of rare earth-iron silicides}},\ }\href {https://doi.org/10.1016/0375-9601(80)90849-X} {\bibfield  {journal} {\bibinfo  {journal} {Phys. Lett. A}\ }\textbf {\bibinfo {volume} {75}},\ \bibinfo {pages} {386} (\bibinfo {year} {1980})}\BibitemShut {NoStop}%
\bibitem [{\citenamefont {Sakuma}(2013)}]{Sakuma2013}%
  \BibitemOpen
  \bibfield  {author} {\bibinfo {author} {\bibfnamefont {R.}~\bibnamefont {Sakuma}},\ }\bibfield  {title} {\bibinfo {title} {{Symmetry-adapted Wannier functions in the maximal localization procedure}},\ }\href {https://doi.org/10.1103/PhysRevB.87.235109} {\bibfield  {journal} {\bibinfo  {journal} {Phys. Rev. B}\ }\textbf {\bibinfo {volume} {87}},\ \bibinfo {pages} {235109} (\bibinfo {year} {2013})},\ \Eprint {https://arxiv.org/abs/1306.0032} {1306.0032} \BibitemShut {NoStop}%
\bibitem [{\citenamefont {Kang}\ and\ \citenamefont {Vafek}(2018)}]{Kang2018}%
  \BibitemOpen
  \bibfield  {author} {\bibinfo {author} {\bibfnamefont {J.}~\bibnamefont {Kang}}\ and\ \bibinfo {author} {\bibfnamefont {O.}~\bibnamefont {Vafek}},\ }\bibfield  {title} {\bibinfo {title} {{Symmetry, Maximally Localized Wannier States, and a Low-Energy Model for Twisted Bilayer Graphene Narrow Bands}},\ }\href {https://doi.org/10.1103/PhysRevX.8.031088} {\bibfield  {journal} {\bibinfo  {journal} {Phys. Rev. X}\ }\textbf {\bibinfo {volume} {8}},\ \bibinfo {pages} {031088} (\bibinfo {year} {2018})},\ \Eprint {https://arxiv.org/abs/1805.04918} {1805.04918} \BibitemShut {NoStop}%
\bibitem [{\citenamefont {Ramires}(2022)}]{Ramires2022}%
  \BibitemOpen
  \bibfield  {author} {\bibinfo {author} {\bibfnamefont {A.}~\bibnamefont {Ramires}},\ }\bibfield  {title} {\bibinfo {title} {{Nonunitary superconductivity in complex quantum materials}},\ }\href {https://doi.org/10.1088/1361-648X/ac6d3a} {\bibfield  {journal} {\bibinfo  {journal} {J. Phys. Condens. Matter}\ }\textbf {\bibinfo {volume} {34}},\ \bibinfo {pages} {304001} (\bibinfo {year} {2022})},\ \Eprint {https://arxiv.org/abs/2202.12178} {2202.12178} \BibitemShut {NoStop}%
\bibitem [{\citenamefont {Yogendra}\ \emph {et~al.}(2025)\citenamefont {Yogendra}, \citenamefont {Baskaran},\ and\ \citenamefont {Das}}]{Yogendra2024}%
  \BibitemOpen
  \bibfield  {author} {\bibinfo {author} {\bibfnamefont {K.~B.}\ \bibnamefont {Yogendra}}, \bibinfo {author} {\bibfnamefont {G.}~\bibnamefont {Baskaran}},\ and\ \bibinfo {author} {\bibfnamefont {T.}~\bibnamefont {Das}},\ }\bibfield  {title} {\bibinfo {title} {Fractional wannier orbitals and tight-binding gauge fields in kitaev honeycomb superlattices with flat majorana bands},\ }\href {https://doi.org/10.1103/37kh-kc81} {\bibfield  {journal} {\bibinfo  {journal} {Phys. Rev. X}\ ,\ \bibinfo {pages} {(Accepted)}} (\bibinfo {year} {2025})}\BibitemShut {NoStop}%
\bibitem [{\citenamefont {Kanamori}(1963)}]{Kanamori1963}%
  \BibitemOpen
  \bibfield  {author} {\bibinfo {author} {\bibfnamefont {J.}~\bibnamefont {Kanamori}},\ }\bibfield  {title} {\bibinfo {title} {{Electron Correlation and Ferromagnetism of Transition Metals}},\ }\href {https://doi.org/10.1143/PTP.30.275} {\bibfield  {journal} {\bibinfo  {journal} {Prog. Theor. Phys.}\ }\textbf {\bibinfo {volume} {30}},\ \bibinfo {pages} {275} (\bibinfo {year} {1963})}\BibitemShut {NoStop}%
\bibitem [{\citenamefont {Georges}\ \emph {et~al.}(2013)\citenamefont {Georges}, \citenamefont {de' Medici},\ and\ \citenamefont {Mravlje}}]{Georges2013}%
  \BibitemOpen
  \bibfield  {author} {\bibinfo {author} {\bibfnamefont {A.}~\bibnamefont {Georges}}, \bibinfo {author} {\bibfnamefont {L.}~\bibnamefont {de' Medici}},\ and\ \bibinfo {author} {\bibfnamefont {J.}~\bibnamefont {Mravlje}},\ }\bibfield  {title} {\bibinfo {title} {{Strong Correlations from Hund's Coupling}},\ }\href {https://doi.org/10.1146/annurev-conmatphys-020911-125045} {\bibfield  {journal} {\bibinfo  {journal} {Annu. Rev. Condens. Matter Phys.}\ }\textbf {\bibinfo {volume} {4}},\ \bibinfo {pages} {137} (\bibinfo {year} {2013})}\BibitemShut {NoStop}%
\bibitem [{\citenamefont {Das}\ and\ \citenamefont {Dolui}(2015)}]{Das2015}%
  \BibitemOpen
  \bibfield  {author} {\bibinfo {author} {\bibfnamefont {T.}~\bibnamefont {Das}}\ and\ \bibinfo {author} {\bibfnamefont {K.}~\bibnamefont {Dolui}},\ }\bibfield  {title} {\bibinfo {title} {{Superconducting dome in MoS2 and TiSe2 generated by quasiparticle-phonon coupling}},\ }\href {https://doi.org/10.1103/PhysRevB.91.094510} {\bibfield  {journal} {\bibinfo  {journal} {Phys. Rev. B}\ }\textbf {\bibinfo {volume} {91}},\ \bibinfo {pages} {094510} (\bibinfo {year} {2015})},\ \Eprint {https://arxiv.org/abs/1411.3096} {1411.3096} \BibitemShut {NoStop}%
\bibitem [{\citenamefont {Kohn}\ and\ \citenamefont {Luttinger}(1965)}]{KohnLuttinger}%
  \BibitemOpen
  \bibfield  {author} {\bibinfo {author} {\bibfnamefont {W.}~\bibnamefont {Kohn}}\ and\ \bibinfo {author} {\bibfnamefont {J.~M.}\ \bibnamefont {Luttinger}},\ }\bibfield  {title} {\bibinfo {title} {New mechanism for superconductivity},\ }\href {https://doi.org/10.1103/PhysRevLett.15.524} {\bibfield  {journal} {\bibinfo  {journal} {Phys. Rev. Lett.}\ }\textbf {\bibinfo {volume} {15}},\ \bibinfo {pages} {524} (\bibinfo {year} {1965})}\BibitemShut {NoStop}%
\bibitem [{\citenamefont {Chubukov}(1993)}]{ChubukovKL}%
  \BibitemOpen
  \bibfield  {author} {\bibinfo {author} {\bibfnamefont {A.~V.}\ \bibnamefont {Chubukov}},\ }\bibfield  {title} {\bibinfo {title} {Kohn-luttinger effect and the instability of a two-dimensional repulsive fermi liquid at t=0},\ }\href {https://doi.org/10.1103/PhysRevB.48.1097} {\bibfield  {journal} {\bibinfo  {journal} {Phys. Rev. B}\ }\textbf {\bibinfo {volume} {48}},\ \bibinfo {pages} {1097} (\bibinfo {year} {1993})}\BibitemShut {NoStop}%
\bibitem [{\citenamefont {Scalapino}(2012)}]{ScalapinoRMP}%
  \BibitemOpen
  \bibfield  {author} {\bibinfo {author} {\bibfnamefont {D.~J.}\ \bibnamefont {Scalapino}},\ }\bibfield  {title} {\bibinfo {title} {A common thread: The pairing interaction for unconventional superconductors},\ }\href {https://doi.org/10.1103/RevModPhys.84.1383} {\bibfield  {journal} {\bibinfo  {journal} {Rev. Mod. Phys.}\ }\textbf {\bibinfo {volume} {84}},\ \bibinfo {pages} {1383} (\bibinfo {year} {2012})}\BibitemShut {NoStop}%
\bibitem [{\citenamefont {Pines}\ and\ \citenamefont {Bohm}(1952)}]{Pines1952}%
  \BibitemOpen
  \bibfield  {author} {\bibinfo {author} {\bibfnamefont {D.}~\bibnamefont {Pines}}\ and\ \bibinfo {author} {\bibfnamefont {D.}~\bibnamefont {Bohm}},\ }\bibfield  {title} {\bibinfo {title} {{A Collective Description of Electron Interactions: II. Collective vs Individual Particle Aspects of the Interactions}},\ }\href {https://doi.org/10.1103/PhysRev.85.338} {\bibfield  {journal} {\bibinfo  {journal} {Phys. Rev.}\ }\textbf {\bibinfo {volume} {85}},\ \bibinfo {pages} {338} (\bibinfo {year} {1952})}\BibitemShut {NoStop}%
\bibitem [{\citenamefont {Bohm}\ and\ \citenamefont {Pines}(1953)}]{Bohm1953}%
  \BibitemOpen
  \bibfield  {author} {\bibinfo {author} {\bibfnamefont {D.}~\bibnamefont {Bohm}}\ and\ \bibinfo {author} {\bibfnamefont {D.}~\bibnamefont {Pines}},\ }\bibfield  {title} {\bibinfo {title} {{A collective description of electron interactions: III. Coulomb interactions in a degenerate electron gas}},\ }\href {https://doi.org/10.1103/PhysRev.92.609} {\bibfield  {journal} {\bibinfo  {journal} {Phys. Rev.}\ }\textbf {\bibinfo {volume} {92}},\ \bibinfo {pages} {609} (\bibinfo {year} {1953})}\BibitemShut {NoStop}%
\bibitem [{\citenamefont {Pines}(2016)}]{Pines2016}%
  \BibitemOpen
  \bibfield  {author} {\bibinfo {author} {\bibfnamefont {D.}~\bibnamefont {Pines}},\ }\bibfield  {title} {\bibinfo {title} {{Emergent behavior in strongly correlated electron systems}},\ }\href {https://doi.org/10.1088/0034-4885/79/9/092501} {\bibfield  {journal} {\bibinfo  {journal} {Reports Prog. Phys.}\ }\textbf {\bibinfo {volume} {79}},\ \bibinfo {pages} {092501} (\bibinfo {year} {2016})},\ \Eprint {https://arxiv.org/abs/1601.05891} {1601.05891} \BibitemShut {NoStop}%
\bibitem [{\citenamefont {Gell-Mann}\ and\ \citenamefont {Brueckner}(1957)}]{Gell-Mann1957}%
  \BibitemOpen
  \bibfield  {author} {\bibinfo {author} {\bibfnamefont {M.}~\bibnamefont {Gell-Mann}}\ and\ \bibinfo {author} {\bibfnamefont {K.~A.}\ \bibnamefont {Brueckner}},\ }\bibfield  {title} {\bibinfo {title} {{Correlation Energy of an Electron Gas at High Density}},\ }\href {https://doi.org/10.1103/PhysRev.106.364} {\bibfield  {journal} {\bibinfo  {journal} {Phys. Rev.}\ }\textbf {\bibinfo {volume} {106}},\ \bibinfo {pages} {364} (\bibinfo {year} {1957})}\BibitemShut {NoStop}%
\bibitem [{\citenamefont {McLACHLAN}\ and\ \citenamefont {BALL}(1964)}]{McLACHLAN1964}%
  \BibitemOpen
  \bibfield  {author} {\bibinfo {author} {\bibfnamefont {A.~D.}\ \bibnamefont {McLACHLAN}}\ and\ \bibinfo {author} {\bibfnamefont {M.~A.}\ \bibnamefont {BALL}},\ }\bibfield  {title} {\bibinfo {title} {{Time-Dependent Hartree—Fock Theory for Molecules}},\ }\href {https://doi.org/10.1103/RevModPhys.36.844} {\bibfield  {journal} {\bibinfo  {journal} {Rev. Mod. Phys.}\ }\textbf {\bibinfo {volume} {36}},\ \bibinfo {pages} {844} (\bibinfo {year} {1964})}\BibitemShut {NoStop}%
\bibitem [{\citenamefont {Oddershede}(1978)}]{Oddershede1978}%
  \BibitemOpen
  \bibfield  {author} {\bibinfo {author} {\bibfnamefont {J.}~\bibnamefont {Oddershede}},\ }\bibfield  {title} {\bibinfo {title} {Polarization propagator calculations}\ }(\bibinfo  {publisher} {Academic Press},\ \bibinfo {year} {1978})\ pp.\ \bibinfo {pages} {275--352}\BibitemShut {NoStop}%
\bibitem [{\citenamefont {Szabo}\ and\ \citenamefont {Ostlund}(1977)}]{Szabo1977}%
  \BibitemOpen
  \bibfield  {author} {\bibinfo {author} {\bibfnamefont {A.}~\bibnamefont {Szabo}}\ and\ \bibinfo {author} {\bibfnamefont {N.~S.}\ \bibnamefont {Ostlund}},\ }\bibfield  {title} {\bibinfo {title} {{The correlation energy in the random phase approximation: Intermolecular forces between closed-shell systems}},\ }\href {https://doi.org/10.1063/1.434580} {\bibfield  {journal} {\bibinfo  {journal} {J. Chem. Phys.}\ }\textbf {\bibinfo {volume} {67}},\ \bibinfo {pages} {4351} (\bibinfo {year} {1977})}\BibitemShut {NoStop}%
\bibitem [{\citenamefont {Adhikary}\ and\ \citenamefont {Das}(2020)}]{Adhikary2020fwave}%
  \BibitemOpen
  \bibfield  {author} {\bibinfo {author} {\bibfnamefont {P.}~\bibnamefont {Adhikary}}\ and\ \bibinfo {author} {\bibfnamefont {T.}~\bibnamefont {Das}},\ }\bibfield  {title} {\bibinfo {title} {{Prediction of f-wave pairing symmetry in YBa2Cu3 O6+x cuprates}},\ }\href {https://doi.org/10.1103/PhysRevB.101.214517} {\bibfield  {journal} {\bibinfo  {journal} {Phys. Rev. B}\ }\textbf {\bibinfo {volume} {101}},\ \bibinfo {pages} {214517} (\bibinfo {year} {2020})},\ \Eprint {https://arxiv.org/abs/2002.07382} {2002.07382} \BibitemShut {NoStop}%
\bibitem [{\citenamefont {Das}\ \emph {et~al.}(2011)\citenamefont {Das}, \citenamefont {Zhu},\ and\ \citenamefont {Graf}}]{Das2011}%
  \BibitemOpen
  \bibfield  {author} {\bibinfo {author} {\bibfnamefont {T.}~\bibnamefont {Das}}, \bibinfo {author} {\bibfnamefont {J.-X.}\ \bibnamefont {Zhu}},\ and\ \bibinfo {author} {\bibfnamefont {M.~J.}\ \bibnamefont {Graf}},\ }\bibfield  {title} {\bibinfo {title} {{Local suppression of the superfluid density of PuCoGa5 by strong onsite disorder}},\ }\href {https://doi.org/10.1103/PhysRevB.84.134510} {\bibfield  {journal} {\bibinfo  {journal} {Phys. Rev. B}\ }\textbf {\bibinfo {volume} {84}},\ \bibinfo {pages} {134510} (\bibinfo {year} {2011})}\BibitemShut {NoStop}%
\bibitem [{\citenamefont {Sticlet}\ \emph {et~al.}(2012)\citenamefont {Sticlet}, \citenamefont {Bena},\ and\ \citenamefont {Simon}}]{Sticlet2012}%
  \BibitemOpen
  \bibfield  {author} {\bibinfo {author} {\bibfnamefont {D.}~\bibnamefont {Sticlet}}, \bibinfo {author} {\bibfnamefont {C.}~\bibnamefont {Bena}},\ and\ \bibinfo {author} {\bibfnamefont {P.}~\bibnamefont {Simon}},\ }\bibfield  {title} {\bibinfo {title} {{Spin and Majorana Polarization in Topological Superconducting Wires}},\ }\href {https://doi.org/10.1103/PhysRevLett.108.096802} {\bibfield  {journal} {\bibinfo  {journal} {Phys. Rev. Lett.}\ }\textbf {\bibinfo {volume} {108}},\ \bibinfo {pages} {096802} (\bibinfo {year} {2012})},\ \Eprint {https://arxiv.org/abs/1109.5697} {1109.5697} \BibitemShut {NoStop}%
\bibitem [{\citenamefont {Pal}(2018)}]{Pal2018}%
  \BibitemOpen
  \bibfield  {author} {\bibinfo {author} {\bibfnamefont {B.}~\bibnamefont {Pal}},\ }\bibfield  {title} {\bibinfo {title} {{Nontrivial topological flat bands in a diamond-octagon lattice geometry}},\ }\href {https://doi.org/10.1103/PhysRevB.98.245116} {\bibfield  {journal} {\bibinfo  {journal} {Phys. Rev. B}\ }\textbf {\bibinfo {volume} {98}},\ \bibinfo {pages} {245116} (\bibinfo {year} {2018})}\BibitemShut {NoStop}%
\bibitem [{\citenamefont {Nie}\ \emph {et~al.}(2015)\citenamefont {Nie}, \citenamefont {Song}, \citenamefont {Weng},\ and\ \citenamefont {Fang}}]{Nie2015}%
  \BibitemOpen
  \bibfield  {author} {\bibinfo {author} {\bibfnamefont {S.~M.}\ \bibnamefont {Nie}}, \bibinfo {author} {\bibfnamefont {Z.}~\bibnamefont {Song}}, \bibinfo {author} {\bibfnamefont {H.}~\bibnamefont {Weng}},\ and\ \bibinfo {author} {\bibfnamefont {Z.}~\bibnamefont {Fang}},\ }\bibfield  {title} {\bibinfo {title} {{Quantum spin Hall effect in two-dimensional transition-metal dichalcogenide haeckelites}},\ }\href {https://doi.org/10.1103/PhysRevB.91.235434} {\bibfield  {journal} {\bibinfo  {journal} {Phys. Rev. B}\ }\textbf {\bibinfo {volume} {91}},\ \bibinfo {pages} {235434} (\bibinfo {year} {2015})}\BibitemShut {NoStop}%
\bibitem [{\citenamefont {Bao}\ \emph {et~al.}(2015)\citenamefont {Bao}, \citenamefont {Tao}, \citenamefont {Liu}, \citenamefont {Zhang},\ and\ \citenamefont {Liu}}]{Bao2014}%
  \BibitemOpen
  \bibfield  {author} {\bibinfo {author} {\bibfnamefont {A.}~\bibnamefont {Bao}}, \bibinfo {author} {\bibfnamefont {H.-S.}\ \bibnamefont {Tao}}, \bibinfo {author} {\bibfnamefont {H.-D.}\ \bibnamefont {Liu}}, \bibinfo {author} {\bibfnamefont {X.}~\bibnamefont {Zhang}},\ and\ \bibinfo {author} {\bibfnamefont {W.-M.}\ \bibnamefont {Liu}},\ }\bibfield  {title} {\bibinfo {title} {{Quantum magnetic phase transition in square-octagon lattice}},\ }\href {https://doi.org/10.1038/srep06918} {\bibfield  {journal} {\bibinfo  {journal} {Sci. Rep.}\ }\textbf {\bibinfo {volume} {4}},\ \bibinfo {pages} {6918} (\bibinfo {year} {2015})}\BibitemShut {NoStop}%
\bibitem [{\citenamefont {Thouless}\ \emph {et~al.}(1982)\citenamefont {Thouless}, \citenamefont {Kohmoto}, \citenamefont {Nightingale},\ and\ \citenamefont {den Nijs}}]{Thouless1982}%
  \BibitemOpen
  \bibfield  {author} {\bibinfo {author} {\bibfnamefont {D.~J.}\ \bibnamefont {Thouless}}, \bibinfo {author} {\bibfnamefont {M.}~\bibnamefont {Kohmoto}}, \bibinfo {author} {\bibfnamefont {M.~P.}\ \bibnamefont {Nightingale}},\ and\ \bibinfo {author} {\bibfnamefont {M.}~\bibnamefont {den Nijs}},\ }\bibfield  {title} {\bibinfo {title} {{Quantized Hall Conductance in a Two-Dimensional Periodic Potential}},\ }\href {https://doi.org/10.1103/PhysRevLett.49.405} {\bibfield  {journal} {\bibinfo  {journal} {Phys. Rev. Lett.}\ }\textbf {\bibinfo {volume} {49}},\ \bibinfo {pages} {405} (\bibinfo {year} {1982})}\BibitemShut {NoStop}%
\bibitem [{\citenamefont {Zhu}\ and\ \citenamefont {Ting}(2001)}]{supercellZhu}%
  \BibitemOpen
  \bibfield  {author} {\bibinfo {author} {\bibfnamefont {J.-X.}\ \bibnamefont {Zhu}}\ and\ \bibinfo {author} {\bibfnamefont {C.~S.}\ \bibnamefont {Ting}},\ }\bibfield  {title} {\bibinfo {title} {Quasiparticle states at a $\mathit{d}$-wave vortex core in high- ${T}_{c}$ superconductors: Induction of local spin density wave order},\ }\href {https://doi.org/10.1103/PhysRevLett.87.147002} {\bibfield  {journal} {\bibinfo  {journal} {Phys. Rev. Lett.}\ }\textbf {\bibinfo {volume} {87}},\ \bibinfo {pages} {147002} (\bibinfo {year} {2001})}\BibitemShut {NoStop}%
\bibitem [{\citenamefont {Han}\ \emph {et~al.}(2002)\citenamefont {Han}, \citenamefont {Wang}, \citenamefont {Zhang},\ and\ \citenamefont {Li}}]{supercellHan}%
  \BibitemOpen
  \bibfield  {author} {\bibinfo {author} {\bibfnamefont {Q.}~\bibnamefont {Han}}, \bibinfo {author} {\bibfnamefont {Z.~D.}\ \bibnamefont {Wang}}, \bibinfo {author} {\bibfnamefont {L.-y.}\ \bibnamefont {Zhang}},\ and\ \bibinfo {author} {\bibfnamefont {X.-G.}\ \bibnamefont {Li}},\ }\bibfield  {title} {\bibinfo {title} {Electronic structure of the vortex lattice of $d\ensuremath{-},$ $d+is\ensuremath{-},$ and ${d}_{{x}^{2}\ensuremath{-}{y}^{2}}{+id}_{\mathrm{xy}}$-wave superconductors},\ }\href {https://doi.org/10.1103/PhysRevB.65.064527} {\bibfield  {journal} {\bibinfo  {journal} {Phys. Rev. B}\ }\textbf {\bibinfo {volume} {65}},\ \bibinfo {pages} {064527} (\bibinfo {year} {2002})}\BibitemShut {NoStop}%
\bibitem [{\citenamefont {Bergman}\ \emph {et~al.}(2008)\citenamefont {Bergman}, \citenamefont {Wu},\ and\ \citenamefont {Balents}}]{Bergman2008}%
  \BibitemOpen
  \bibfield  {author} {\bibinfo {author} {\bibfnamefont {D.~L.}\ \bibnamefont {Bergman}}, \bibinfo {author} {\bibfnamefont {C.}~\bibnamefont {Wu}},\ and\ \bibinfo {author} {\bibfnamefont {L.}~\bibnamefont {Balents}},\ }\bibfield  {title} {\bibinfo {title} {{Band touching from real-space topology in frustrated hopping models}},\ }\href {https://doi.org/10.1103/PhysRevB.78.125104} {\bibfield  {journal} {\bibinfo  {journal} {Phys. Rev. B}\ }\textbf {\bibinfo {volume} {78}},\ \bibinfo {pages} {125104} (\bibinfo {year} {2008})},\ \Eprint {https://arxiv.org/abs/0803.3628} {0803.3628} \BibitemShut {NoStop}%
\bibitem [{\citenamefont {Chubukov}(2012)}]{Chubukov2012}%
  \BibitemOpen
  \bibfield  {author} {\bibinfo {author} {\bibfnamefont {A.}~\bibnamefont {Chubukov}},\ }\bibfield  {title} {\bibinfo {title} {Pairing mechanism in fe-based superconductors},\ }\href {https://doi.org/10.1146/annurev-conmatphys-020911-125055} {\bibfield  {journal} {\bibinfo  {journal} {Annu. Rev. Condens. Matter Phys.}\ }\textbf {\bibinfo {volume} {3}},\ \bibinfo {pages} {57} (\bibinfo {year} {2012})}\BibitemShut {NoStop}%
\bibitem [{\citenamefont {Mazin}\ \emph {et~al.}(2008)\citenamefont {Mazin}, \citenamefont {Singh}, \citenamefont {Johannes},\ and\ \citenamefont {Du}}]{Mazin2008}%
  \BibitemOpen
  \bibfield  {author} {\bibinfo {author} {\bibfnamefont {I.~I.}\ \bibnamefont {Mazin}}, \bibinfo {author} {\bibfnamefont {D.~J.}\ \bibnamefont {Singh}}, \bibinfo {author} {\bibfnamefont {M.~D.}\ \bibnamefont {Johannes}},\ and\ \bibinfo {author} {\bibfnamefont {M.~H.}\ \bibnamefont {Du}},\ }\bibfield  {title} {\bibinfo {title} {{Unconventional Superconductivity with a Sign Reversal in the Order Parameter of LaFeAsO1-xFx}},\ }\href {https://doi.org/10.1103/PhysRevLett.101.057003} {\bibfield  {journal} {\bibinfo  {journal} {Phys. Rev. Lett.}\ }\textbf {\bibinfo {volume} {101}},\ \bibinfo {pages} {057003} (\bibinfo {year} {2008})}\BibitemShut {NoStop}%
\bibitem [{\citenamefont {Das}\ and\ \citenamefont {Balatsky}(2012)}]{Das2012JoP}%
  \BibitemOpen
  \bibfield  {author} {\bibinfo {author} {\bibfnamefont {T.}~\bibnamefont {Das}}\ and\ \bibinfo {author} {\bibfnamefont {A.~V.}\ \bibnamefont {Balatsky}},\ }\bibfield  {title} {\bibinfo {title} {Testing the sign-changing superconducting gap in iron-based superconductors with quasiparticle interference and neutron scattering},\ }\href {https://doi.org/10.1088/0953-8984/24/18/182201} {\bibfield  {journal} {\bibinfo  {journal} {J. Condens. Matter Phys.}\ }\textbf {\bibinfo {volume} {24}},\ \bibinfo {pages} {182201} (\bibinfo {year} {2012})}\BibitemShut {NoStop}%
\bibitem [{\citenamefont {Das}\ \emph {et~al.}(2015)\citenamefont {Das}, \citenamefont {Zhu},\ and\ \citenamefont {Graf}}]{Das2015SR}%
  \BibitemOpen
  \bibfield  {author} {\bibinfo {author} {\bibfnamefont {T.}~\bibnamefont {Das}}, \bibinfo {author} {\bibfnamefont {J.-X.}\ \bibnamefont {Zhu}},\ and\ \bibinfo {author} {\bibfnamefont {M.~J.}\ \bibnamefont {Graf}},\ }\bibfield  {title} {\bibinfo {title} {Theory of nodal s±-wave pairing symmetry in the pu-based 115 superconductor family},\ }\href {https://doi.org/10.1038/srep08632} {\bibfield  {journal} {\bibinfo  {journal} {Sci. Rep.}\ }\textbf {\bibinfo {volume} {5}},\ \bibinfo {pages} {8632} (\bibinfo {year} {2015})}\BibitemShut {NoStop}%
\bibitem [{\citenamefont {Black-Schaffer}\ and\ \citenamefont {Honerkamp}(2014)}]{Black-Schaffer_2014}%
  \BibitemOpen
  \bibfield  {author} {\bibinfo {author} {\bibfnamefont {A.~M.}\ \bibnamefont {Black-Schaffer}}\ and\ \bibinfo {author} {\bibfnamefont {C.}~\bibnamefont {Honerkamp}},\ }\bibfield  {title} {\bibinfo {title} {Chiral d-wave superconductivity in doped graphene},\ }\href {https://doi.org/10.1088/0953-8984/26/42/423201} {\bibfield  {journal} {\bibinfo  {journal} {Journal of Physics: Condensed Matter}\ }\textbf {\bibinfo {volume} {26}},\ \bibinfo {pages} {423201} (\bibinfo {year} {2014})}\BibitemShut {NoStop}%
\bibitem [{\citenamefont {Ray}\ \emph {et~al.}(2019)\citenamefont {Ray}, \citenamefont {Jung},\ and\ \citenamefont {Das}}]{Ray2019}%
  \BibitemOpen
  \bibfield  {author} {\bibinfo {author} {\bibfnamefont {S.}~\bibnamefont {Ray}}, \bibinfo {author} {\bibfnamefont {J.}~\bibnamefont {Jung}},\ and\ \bibinfo {author} {\bibfnamefont {T.}~\bibnamefont {Das}},\ }\bibfield  {title} {\bibinfo {title} {{Wannier pairs in superconducting twisted bilayer graphene and related systems}},\ }\href {https://doi.org/10.1103/PhysRevB.99.134515} {\bibfield  {journal} {\bibinfo  {journal} {Phys. Rev. B}\ }\textbf {\bibinfo {volume} {99}},\ \bibinfo {pages} {134515} (\bibinfo {year} {2019})},\ \Eprint {https://arxiv.org/abs/1804.09674} {1804.09674} \BibitemShut {NoStop}%
\bibitem [{\citenamefont {Tinkham}(2004)}]{tinkham2004}%
  \BibitemOpen
  \bibfield  {author} {\bibinfo {author} {\bibfnamefont {M.}~\bibnamefont {Tinkham}},\ }\href {https://books.google.co.in/books?id=VpUk3NfwDIkC} {\emph {\bibinfo {title} {Introduction to Superconductivity}}},\ Dover Books on Physics Series\ (\bibinfo  {publisher} {Dover Publications},\ \bibinfo {year} {2004})\BibitemShut {NoStop}%
\bibitem [{\citenamefont {Sigrist}\ and\ \citenamefont {Ueda}(1991)}]{Sigrist1991}%
  \BibitemOpen
  \bibfield  {author} {\bibinfo {author} {\bibfnamefont {M.}~\bibnamefont {Sigrist}}\ and\ \bibinfo {author} {\bibfnamefont {K.}~\bibnamefont {Ueda}},\ }\bibfield  {title} {\bibinfo {title} {{Phenomenological theory of unconventional superconductivity}},\ }\href {https://doi.org/10.1103/RevModPhys.63.239} {\bibfield  {journal} {\bibinfo  {journal} {Rev. Mod. Phys.}\ }\textbf {\bibinfo {volume} {63}},\ \bibinfo {pages} {239} (\bibinfo {year} {1991})}\BibitemShut {NoStop}%
\bibitem [{\citenamefont {Mandal}\ \emph {et~al.}(2022)\citenamefont {Mandal}, \citenamefont {Kataria}, \citenamefont {Patra}, \citenamefont {Singh}, \citenamefont {Biswas}, \citenamefont {Hillier}, \citenamefont {Das},\ and\ \citenamefont {Singh}}]{Mandal2022}%
  \BibitemOpen
  \bibfield  {author} {\bibinfo {author} {\bibfnamefont {M.}~\bibnamefont {Mandal}}, \bibinfo {author} {\bibfnamefont {A.}~\bibnamefont {Kataria}}, \bibinfo {author} {\bibfnamefont {C.}~\bibnamefont {Patra}}, \bibinfo {author} {\bibfnamefont {D.}~\bibnamefont {Singh}}, \bibinfo {author} {\bibfnamefont {P.~K.}\ \bibnamefont {Biswas}}, \bibinfo {author} {\bibfnamefont {A.~D.}\ \bibnamefont {Hillier}}, \bibinfo {author} {\bibfnamefont {T.}~\bibnamefont {Das}},\ and\ \bibinfo {author} {\bibfnamefont {R.~P.}\ \bibnamefont {Singh}},\ }\bibfield  {title} {\bibinfo {title} {{Time-reversal symmetry breaking in frustrated superconductor Re2Hf}},\ }\href {https://doi.org/10.1103/PhysRevB.105.094513} {\bibfield  {journal} {\bibinfo  {journal} {Phys. Rev. B}\ }\textbf {\bibinfo {volume} {105}},\ \bibinfo {pages} {094513} (\bibinfo {year} {2022})}\BibitemShut {NoStop}%
\bibitem [{\citenamefont {Neha}\ \emph {et~al.}(2019)\citenamefont {Neha}, \citenamefont {Biswas}, \citenamefont {Das},\ and\ \citenamefont {Patnaik}}]{Neha2019}%
  \BibitemOpen
  \bibfield  {author} {\bibinfo {author} {\bibfnamefont {P.}~\bibnamefont {Neha}}, \bibinfo {author} {\bibfnamefont {P.~K.}\ \bibnamefont {Biswas}}, \bibinfo {author} {\bibfnamefont {T.}~\bibnamefont {Das}},\ and\ \bibinfo {author} {\bibfnamefont {S.}~\bibnamefont {Patnaik}},\ }\bibfield  {title} {\bibinfo {title} {Time-reversal symmetry breaking in topological superconductor sr0.1bi2se3},\ }\href {https://doi.org/10.1103/PhysRevMaterials.3.074201} {\bibfield  {journal} {\bibinfo  {journal} {Phys. Rev. Mater.}\ }\textbf {\bibinfo {volume} {3}},\ \bibinfo {pages} {074201} (\bibinfo {year} {2019})}\BibitemShut {NoStop}%
\bibitem [{\citenamefont {Pizzi}\ \emph {et~al.}(2020)\citenamefont {Pizzi}, \citenamefont {Vitale}, \citenamefont {Arita}, \citenamefont {Blügel}, \citenamefont {Freimuth}, \citenamefont {Géranton}, \citenamefont {Gibertini}, \citenamefont {Gresch}, \citenamefont {Johnson}, \citenamefont {Koretsune}, \citenamefont {Ibañez-Azpiroz}, \citenamefont {Lee}, \citenamefont {Lihm}, \citenamefont {Marchand}, \citenamefont {Marrazzo}, \citenamefont {Mokrousov}, \citenamefont {Mustafa}, \citenamefont {Nohara}, \citenamefont {Nomura}, \citenamefont {Paulatto}, \citenamefont {Poncé}, \citenamefont {Ponweiser}, \citenamefont {Qiao}, \citenamefont {Thöle}, \citenamefont {Tsirkin}, \citenamefont {Wierzbowska}, \citenamefont {Marzari}, \citenamefont {Vanderbilt}, \citenamefont {Souza}, \citenamefont {Mostofi},\ and\ \citenamefont {Yates}}]{Wannier90}%
  \BibitemOpen
  \bibfield  {author} {\bibinfo {author} {\bibfnamefont {G.}~\bibnamefont {Pizzi}}, \bibinfo {author} {\bibfnamefont {V.}~\bibnamefont {Vitale}}, \bibinfo {author} {\bibfnamefont {R.}~\bibnamefont {Arita}}, \bibinfo {author} {\bibfnamefont {S.}~\bibnamefont {Blügel}}, \bibinfo {author} {\bibfnamefont {F.}~\bibnamefont {Freimuth}}, \bibinfo {author} {\bibfnamefont {G.}~\bibnamefont {Géranton}}, \bibinfo {author} {\bibfnamefont {M.}~\bibnamefont {Gibertini}}, \bibinfo {author} {\bibfnamefont {D.}~\bibnamefont {Gresch}}, \bibinfo {author} {\bibfnamefont {C.}~\bibnamefont {Johnson}}, \bibinfo {author} {\bibfnamefont {T.}~\bibnamefont {Koretsune}}, \bibinfo {author} {\bibfnamefont {J.}~\bibnamefont {Ibañez-Azpiroz}}, \bibinfo {author} {\bibfnamefont {H.}~\bibnamefont {Lee}}, \bibinfo {author} {\bibfnamefont {J.-M.}\ \bibnamefont {Lihm}}, \bibinfo {author} {\bibfnamefont {D.}~\bibnamefont {Marchand}}, \bibinfo {author} {\bibfnamefont {A.}~\bibnamefont {Marrazzo}}, \bibinfo {author} {\bibfnamefont
  {Y.}~\bibnamefont {Mokrousov}}, \bibinfo {author} {\bibfnamefont {J.~I.}\ \bibnamefont {Mustafa}}, \bibinfo {author} {\bibfnamefont {Y.}~\bibnamefont {Nohara}}, \bibinfo {author} {\bibfnamefont {Y.}~\bibnamefont {Nomura}}, \bibinfo {author} {\bibfnamefont {L.}~\bibnamefont {Paulatto}}, \bibinfo {author} {\bibfnamefont {S.}~\bibnamefont {Poncé}}, \bibinfo {author} {\bibfnamefont {T.}~\bibnamefont {Ponweiser}}, \bibinfo {author} {\bibfnamefont {J.}~\bibnamefont {Qiao}}, \bibinfo {author} {\bibfnamefont {F.}~\bibnamefont {Thöle}}, \bibinfo {author} {\bibfnamefont {S.~S.}\ \bibnamefont {Tsirkin}}, \bibinfo {author} {\bibfnamefont {M.}~\bibnamefont {Wierzbowska}}, \bibinfo {author} {\bibfnamefont {N.}~\bibnamefont {Marzari}}, \bibinfo {author} {\bibfnamefont {D.}~\bibnamefont {Vanderbilt}}, \bibinfo {author} {\bibfnamefont {I.}~\bibnamefont {Souza}}, \bibinfo {author} {\bibfnamefont {A.~A.}\ \bibnamefont {Mostofi}},\ and\ \bibinfo {author} {\bibfnamefont {J.~R.}\ \bibnamefont {Yates}},\ }\bibfield  {title}
  {\bibinfo {title} {Wannier90 as a community code: new features and applications},\ }\href {https://doi.org/10.1088/1361-648X/ab51ff} {\bibfield  {journal} {\bibinfo  {journal} {J. Condens. Matter Phys.}\ }\textbf {\bibinfo {volume} {32}},\ \bibinfo {pages} {165902} (\bibinfo {year} {2020})}\BibitemShut {NoStop}%
\bibitem [{\citenamefont {Adhikary}\ \emph {et~al.}(2020)\citenamefont {Adhikary}, \citenamefont {Bandyopadhyay}, \citenamefont {Das}, \citenamefont {Dasgupta},\ and\ \citenamefont {Saha-Dasgupta}}]{Adhikary2020}%
  \BibitemOpen
  \bibfield  {author} {\bibinfo {author} {\bibfnamefont {P.}~\bibnamefont {Adhikary}}, \bibinfo {author} {\bibfnamefont {S.}~\bibnamefont {Bandyopadhyay}}, \bibinfo {author} {\bibfnamefont {T.}~\bibnamefont {Das}}, \bibinfo {author} {\bibfnamefont {I.}~\bibnamefont {Dasgupta}},\ and\ \bibinfo {author} {\bibfnamefont {T.}~\bibnamefont {Saha-Dasgupta}},\ }\bibfield  {title} {\bibinfo {title} {Orbital-selective superconductivity in a two-band model of infinite-layer nickelates},\ }\href {https://doi.org/10.1103/PhysRevB.102.100501} {\bibfield  {journal} {\bibinfo  {journal} {Phys. Rev. B}\ }\textbf {\bibinfo {volume} {102}},\ \bibinfo {pages} {1} (\bibinfo {year} {2020})}\BibitemShut {NoStop}%
\bibitem [{\citenamefont {Das}\ and\ \citenamefont {Balatsky}(2011)}]{Das2011PRL}%
  \BibitemOpen
  \bibfield  {author} {\bibinfo {author} {\bibfnamefont {T.}~\bibnamefont {Das}}\ and\ \bibinfo {author} {\bibfnamefont {A.~V.}\ \bibnamefont {Balatsky}},\ }\bibfield  {title} {\bibinfo {title} {Two energy scales in the magnetic resonance spectrum of electron and hole doped pnictide superconductors},\ }\bibfield  {journal} {\bibinfo  {journal} {Phys. Rev. Lett.}\ }\textbf {\bibinfo {volume} {106}},\ \href {https://doi.org/10.1103/PhysRevLett.106.157004} {10.1103/PhysRevLett.106.157004} (\bibinfo {year} {2011})\BibitemShut {NoStop}%
\bibitem [{\citenamefont {Zhou}\ \emph {et~al.}(2008)\citenamefont {Zhou}, \citenamefont {Chen},\ and\ \citenamefont {Zhang}}]{Zhou2008}%
  \BibitemOpen
  \bibfield  {author} {\bibinfo {author} {\bibfnamefont {Y.}~\bibnamefont {Zhou}}, \bibinfo {author} {\bibfnamefont {W.-Q.}\ \bibnamefont {Chen}},\ and\ \bibinfo {author} {\bibfnamefont {F.-C.}\ \bibnamefont {Zhang}},\ }\bibfield  {title} {\bibinfo {title} {{Symmetry of superconducting states with two orbitals on a tetragonal lattice: Application to LaFeAsO1-xFx}},\ }\href {https://doi.org/10.1103/PhysRevB.78.064514} {\bibfield  {journal} {\bibinfo  {journal} {Phys. Rev. B}\ }\textbf {\bibinfo {volume} {78}},\ \bibinfo {pages} {064514} (\bibinfo {year} {2008})},\ \Eprint {https://arxiv.org/abs/0806.0712} {0806.0712} \BibitemShut {NoStop}%
\bibitem [{\citenamefont {Venderbos}\ \emph {et~al.}(2018)\citenamefont {Venderbos}, \citenamefont {Savary}, \citenamefont {Ruhman}, \citenamefont {Lee},\ and\ \citenamefont {Fu}}]{Venderbos2018}%
  \BibitemOpen
  \bibfield  {author} {\bibinfo {author} {\bibfnamefont {J.~W.}\ \bibnamefont {Venderbos}}, \bibinfo {author} {\bibfnamefont {L.}~\bibnamefont {Savary}}, \bibinfo {author} {\bibfnamefont {J.}~\bibnamefont {Ruhman}}, \bibinfo {author} {\bibfnamefont {P.~A.}\ \bibnamefont {Lee}},\ and\ \bibinfo {author} {\bibfnamefont {L.}~\bibnamefont {Fu}},\ }\bibfield  {title} {\bibinfo {title} {{Pairing States of Spin- 3/2 Fermions: Symmetry-Enforced Topological Gap Functions}},\ }\href {https://doi.org/10.1103/PhysRevX.8.011029} {\bibfield  {journal} {\bibinfo  {journal} {Phys. Rev. X}\ }\textbf {\bibinfo {volume} {8}},\ \bibinfo {pages} {011029} (\bibinfo {year} {2018})},\ \Eprint {https://arxiv.org/abs/1709.04487} {1709.04487} \BibitemShut {NoStop}%
\bibitem [{\citenamefont {Brydon}\ \emph {et~al.}(2016)\citenamefont {Brydon}, \citenamefont {Wang}, \citenamefont {Weinert},\ and\ \citenamefont {Agterberg}}]{Brydon2016}%
  \BibitemOpen
  \bibfield  {author} {\bibinfo {author} {\bibfnamefont {P.~M.~R.}\ \bibnamefont {Brydon}}, \bibinfo {author} {\bibfnamefont {L.}~\bibnamefont {Wang}}, \bibinfo {author} {\bibfnamefont {M.}~\bibnamefont {Weinert}},\ and\ \bibinfo {author} {\bibfnamefont {D.~F.}\ \bibnamefont {Agterberg}},\ }\bibfield  {title} {\bibinfo {title} {{Pairing of j=3 /2 Fermions in Half-Heusler Superconductors}},\ }\href {https://doi.org/10.1103/PhysRevLett.116.177001} {\bibfield  {journal} {\bibinfo  {journal} {Phys. Rev. Lett.}\ }\textbf {\bibinfo {volume} {116}},\ \bibinfo {pages} {177001} (\bibinfo {year} {2016})},\ \Eprint {https://arxiv.org/abs/1603.03376} {1603.03376} \BibitemShut {NoStop}%
\bibitem [{\citenamefont {Das}(2012)}]{Das2012}%
  \BibitemOpen
  \bibfield  {author} {\bibinfo {author} {\bibfnamefont {T.}~\bibnamefont {Das}},\ }\bibfield  {title} {\bibinfo {title} {{Pairing symmetries of several iron-based superconductor families and some similarities with cuprates and heavy-fermions}},\ }\href {https://doi.org/10.1051/epjconf/20122300014} {\bibfield  {journal} {\bibinfo  {journal} {EPJ Web Conf.}\ }\textbf {\bibinfo {volume} {23}},\ \bibinfo {pages} {00014} (\bibinfo {year} {2012})}\BibitemShut {NoStop}%
\bibitem [{\citenamefont {Linder}\ and\ \citenamefont {Balatsky}(2019)}]{Linder2019}%
  \BibitemOpen
  \bibfield  {author} {\bibinfo {author} {\bibfnamefont {J.}~\bibnamefont {Linder}}\ and\ \bibinfo {author} {\bibfnamefont {A.~V.}\ \bibnamefont {Balatsky}},\ }\bibfield  {title} {\bibinfo {title} {{Odd-frequency superconductivity}},\ }\href {https://doi.org/10.1103/RevModPhys.91.045005} {\bibfield  {journal} {\bibinfo  {journal} {Rev. Mod. Phys.}\ }\textbf {\bibinfo {volume} {91}},\ \bibinfo {pages} {045005} (\bibinfo {year} {2019})},\ \Eprint {https://arxiv.org/abs/1709.03986} {1709.03986} \BibitemShut {NoStop}%
\bibitem [{\citenamefont {Suh}\ \emph {et~al.}(2023)\citenamefont {Suh}, \citenamefont {Yu}, \citenamefont {Shishidou}, \citenamefont {Weinert}, \citenamefont {Brydon},\ and\ \citenamefont {Agterberg}}]{Suh2023}%
  \BibitemOpen
  \bibfield  {author} {\bibinfo {author} {\bibfnamefont {H.~G.}\ \bibnamefont {Suh}}, \bibinfo {author} {\bibfnamefont {Y.}~\bibnamefont {Yu}}, \bibinfo {author} {\bibfnamefont {T.}~\bibnamefont {Shishidou}}, \bibinfo {author} {\bibfnamefont {M.}~\bibnamefont {Weinert}}, \bibinfo {author} {\bibfnamefont {P.~M.~R.}\ \bibnamefont {Brydon}},\ and\ \bibinfo {author} {\bibfnamefont {D.~F.}\ \bibnamefont {Agterberg}},\ }\bibfield  {title} {\bibinfo {title} {{Superconductivity of anomalous pseudospin in nonsymmorphic materials}},\ }\href {https://doi.org/10.1103/PhysRevResearch.5.033204} {\bibfield  {journal} {\bibinfo  {journal} {Phys. Rev. Res.}\ }\textbf {\bibinfo {volume} {5}},\ \bibinfo {pages} {033204} (\bibinfo {year} {2023})}\BibitemShut {NoStop}%
\bibitem [{\citenamefont {Basak}\ \emph {et~al.}(2012)\citenamefont {Basak}, \citenamefont {Das}, \citenamefont {Lin}, \citenamefont {Hasan}, \citenamefont {Markiewicz},\ and\ \citenamefont {Bansil}}]{cuprate3band}%
  \BibitemOpen
  \bibfield  {author} {\bibinfo {author} {\bibfnamefont {S.}~\bibnamefont {Basak}}, \bibinfo {author} {\bibfnamefont {T.}~\bibnamefont {Das}}, \bibinfo {author} {\bibfnamefont {H.}~\bibnamefont {Lin}}, \bibinfo {author} {\bibfnamefont {M.~Z.}\ \bibnamefont {Hasan}}, \bibinfo {author} {\bibfnamefont {R.~S.}\ \bibnamefont {Markiewicz}},\ and\ \bibinfo {author} {\bibfnamefont {A.}~\bibnamefont {Bansil}},\ }\bibfield  {title} {\bibinfo {title} {Coexisting pseudogap, charge-transfer-gap, and mott-gap energy scales in the resonant inelastic x-ray scattering spectra of electron-doped cuprate superconductors},\ }\href {https://doi.org/10.1103/PhysRevB.85.075104} {\bibfield  {journal} {\bibinfo  {journal} {Phys. Rev. B}\ }\textbf {\bibinfo {volume} {85}},\ \bibinfo {pages} {075104} (\bibinfo {year} {2012})}\BibitemShut {NoStop}%
\bibitem [{\citenamefont {Kresse}\ and\ \citenamefont {Furthm\"uller}(1996)}]{vasp1}%
  \BibitemOpen
  \bibfield  {author} {\bibinfo {author} {\bibfnamefont {G.}~\bibnamefont {Kresse}}\ and\ \bibinfo {author} {\bibfnamefont {J.}~\bibnamefont {Furthm\"uller}},\ }\bibfield  {title} {\bibinfo {title} {Efficient iterative schemes for ab initio total-energy calculations using a plane-wave basis set},\ }\href {https://doi.org/10.1103/PhysRevB.54.11169} {\bibfield  {journal} {\bibinfo  {journal} {Phys. Rev. B}\ }\textbf {\bibinfo {volume} {54}},\ \bibinfo {pages} {11169} (\bibinfo {year} {1996})}\BibitemShut {NoStop}%
\bibitem [{\citenamefont {Kresse}\ and\ \citenamefont {Furthmüller}(1996)}]{vasp2}%
  \BibitemOpen
  \bibfield  {author} {\bibinfo {author} {\bibfnamefont {G.}~\bibnamefont {Kresse}}\ and\ \bibinfo {author} {\bibfnamefont {J.}~\bibnamefont {Furthmüller}},\ }\bibfield  {title} {\bibinfo {title} {Efficiency of ab-initio total energy calculations for metals and semiconductors using a plane-wave basis set},\ }\href {https://doi.org/https://doi.org/10.1016/0927-0256(96)00008-0} {\bibfield  {journal} {\bibinfo  {journal} {Comput. Mater. Sci.}\ }\textbf {\bibinfo {volume} {6}},\ \bibinfo {pages} {15} (\bibinfo {year} {1996})}\BibitemShut {NoStop}%
\bibitem [{\citenamefont {Perdew}\ \emph {et~al.}(1996)\citenamefont {Perdew}, \citenamefont {Burke},\ and\ \citenamefont {Ernzerhof}}]{pbe}%
  \BibitemOpen
  \bibfield  {author} {\bibinfo {author} {\bibfnamefont {J.~P.}\ \bibnamefont {Perdew}}, \bibinfo {author} {\bibfnamefont {K.}~\bibnamefont {Burke}},\ and\ \bibinfo {author} {\bibfnamefont {M.}~\bibnamefont {Ernzerhof}},\ }\bibfield  {title} {\bibinfo {title} {Generalized gradient approximation made simple},\ }\href {https://doi.org/10.1103/PhysRevLett.77.3865} {\bibfield  {journal} {\bibinfo  {journal} {Phys. Rev. Lett.}\ }\textbf {\bibinfo {volume} {77}},\ \bibinfo {pages} {3865} (\bibinfo {year} {1996})}\BibitemShut {NoStop}%
\bibitem [{\citenamefont {Bl\"ochl}(1994)}]{paw1}%
  \BibitemOpen
  \bibfield  {author} {\bibinfo {author} {\bibfnamefont {P.~E.}\ \bibnamefont {Bl\"ochl}},\ }\bibfield  {title} {\bibinfo {title} {Projector augmented-wave method},\ }\href {https://doi.org/10.1103/PhysRevB.50.17953} {\bibfield  {journal} {\bibinfo  {journal} {Phys. Rev. B}\ }\textbf {\bibinfo {volume} {50}},\ \bibinfo {pages} {17953} (\bibinfo {year} {1994})}\BibitemShut {NoStop}%
\bibitem [{\citenamefont {Kresse}\ and\ \citenamefont {Joubert}(1999)}]{paw2}%
  \BibitemOpen
  \bibfield  {author} {\bibinfo {author} {\bibfnamefont {G.}~\bibnamefont {Kresse}}\ and\ \bibinfo {author} {\bibfnamefont {D.}~\bibnamefont {Joubert}},\ }\bibfield  {title} {\bibinfo {title} {From ultrasoft pseudopotentials to the projector augmented-wave method},\ }\href {https://doi.org/10.1103/PhysRevB.59.1758} {\bibfield  {journal} {\bibinfo  {journal} {Phys. Rev. B}\ }\textbf {\bibinfo {volume} {59}},\ \bibinfo {pages} {1758} (\bibinfo {year} {1999})}\BibitemShut {NoStop}%
\bibitem [{\citenamefont {Monkhorst}\ and\ \citenamefont {Pack}(1976)}]{Monkhorst1976}%
  \BibitemOpen
  \bibfield  {author} {\bibinfo {author} {\bibfnamefont {H.~J.}\ \bibnamefont {Monkhorst}}\ and\ \bibinfo {author} {\bibfnamefont {J.~D.}\ \bibnamefont {Pack}},\ }\bibfield  {title} {\bibinfo {title} {Special points for brillouin-zone integrations},\ }\href {https://doi.org/10.1103/PhysRevB.13.5188} {\bibfield  {journal} {\bibinfo  {journal} {Phys. Rev. B}\ }\textbf {\bibinfo {volume} {13}},\ \bibinfo {pages} {5188} (\bibinfo {year} {1976})}\BibitemShut {NoStop}%
\bibitem [{\citenamefont {Dudarev}\ \emph {et~al.}(1998)\citenamefont {Dudarev}, \citenamefont {Botton}, \citenamefont {Savrasov}, \citenamefont {Humphreys},\ and\ \citenamefont {Sutton}}]{Dudarev1998}%
  \BibitemOpen
  \bibfield  {author} {\bibinfo {author} {\bibfnamefont {S.~L.}\ \bibnamefont {Dudarev}}, \bibinfo {author} {\bibfnamefont {G.~A.}\ \bibnamefont {Botton}}, \bibinfo {author} {\bibfnamefont {S.~Y.}\ \bibnamefont {Savrasov}}, \bibinfo {author} {\bibfnamefont {C.~J.}\ \bibnamefont {Humphreys}},\ and\ \bibinfo {author} {\bibfnamefont {A.~P.}\ \bibnamefont {Sutton}},\ }\bibfield  {title} {\bibinfo {title} {Electron-energy-loss spectra and the structural stability of nickel oxide: An lsda+u study},\ }\href {https://doi.org/10.1103/PhysRevB.57.1505} {\bibfield  {journal} {\bibinfo  {journal} {Phys. Rev. B}\ }\textbf {\bibinfo {volume} {57}},\ \bibinfo {pages} {1505} (\bibinfo {year} {1998})}\BibitemShut {NoStop}%
\bibitem [{\citenamefont {Guo}\ \emph {et~al.}(2022)\citenamefont {Guo}, \citenamefont {Zhao}, \citenamefont {Zhou},\ and\ \citenamefont {Zhao}}]{Guo2022}%
  \BibitemOpen
  \bibfield  {author} {\bibinfo {author} {\bibfnamefont {Y.}~\bibnamefont {Guo}}, \bibinfo {author} {\bibfnamefont {Y.}~\bibnamefont {Zhao}}, \bibinfo {author} {\bibfnamefont {S.}~\bibnamefont {Zhou}},\ and\ \bibinfo {author} {\bibfnamefont {J.}~\bibnamefont {Zhao}},\ }\bibfield  {title} {\bibinfo {title} {{Oxidation behavior of layered Fe n GeTe 2 ( n = 3, 4, 5) and Cr 2 Ge 2 Te 6 governed by interlayer coupling}},\ }\href {https://doi.org/10.1039/D2NR02375J} {\bibfield  {journal} {\bibinfo  {journal} {Nanoscale}\ }\textbf {\bibinfo {volume} {14}},\ \bibinfo {pages} {11452} (\bibinfo {year} {2022})}\BibitemShut {NoStop}%
\bibitem [{\citenamefont {Momma}\ and\ \citenamefont {Izumi}(2011)}]{Momma2011}%
  \BibitemOpen
  \bibfield  {author} {\bibinfo {author} {\bibfnamefont {K.}~\bibnamefont {Momma}}\ and\ \bibinfo {author} {\bibfnamefont {F.}~\bibnamefont {Izumi}},\ }\bibfield  {title} {\bibinfo {title} {{VESTA 3 for three-dimensional visualization of crystal, volumetric and morphology data}},\ }\href {https://doi.org/10.1107/S0021889811038970} {\bibfield  {journal} {\bibinfo  {journal} {J. Appl. Crystallogr.}\ }\textbf {\bibinfo {volume} {44}},\ \bibinfo {pages} {1272} (\bibinfo {year} {2011})}\BibitemShut {NoStop}%
\bibitem [{\citenamefont {Winiarski}\ and\ \citenamefont {Samsel-Czeka{\l}a}(2013)}]{Winiarski2013}%
  \BibitemOpen
  \bibfield  {author} {\bibinfo {author} {\bibfnamefont {M.}~\bibnamefont {Winiarski}}\ and\ \bibinfo {author} {\bibfnamefont {M.}~\bibnamefont {Samsel-Czeka{\l}a}},\ }\bibfield  {title} {\bibinfo {title} {{The electronic structure of rare-earth iron silicide R2Fe3Si5 superconductors}},\ }\href {https://doi.org/10.1016/j.solidstatesciences.2013.10.006} {\bibfield  {journal} {\bibinfo  {journal} {Solid State Sci.}\ }\textbf {\bibinfo {volume} {26}},\ \bibinfo {pages} {134} (\bibinfo {year} {2013})}\BibitemShut {NoStop}%
\bibitem [{\citenamefont {Samsel-Czeka{\l}a}\ and\ \citenamefont {Winiarski}(2012)}]{Samsel-Czekaa2012}%
  \BibitemOpen
  \bibfield  {author} {\bibinfo {author} {\bibfnamefont {M.}~\bibnamefont {Samsel-Czeka{\l}a}}\ and\ \bibinfo {author} {\bibfnamefont {M.}~\bibnamefont {Winiarski}},\ }\bibfield  {title} {\bibinfo {title} {{Electronic structure and Fermi surface of iron-based superconductors R2Fe3Si5 (R = Lu;Y;Sc) from first principles}},\ }\href {https://doi.org/10.1016/j.intermet.2012.07.003} {\bibfield  {journal} {\bibinfo  {journal} {Intermetallics}\ }\textbf {\bibinfo {volume} {31}},\ \bibinfo {pages} {186} (\bibinfo {year} {2012})}\BibitemShut {NoStop}%
\bibitem [{\citenamefont {Damle}\ \emph {et~al.}(2015)\citenamefont {Damle}, \citenamefont {Lin},\ and\ \citenamefont {Ying}}]{Damle2015}%
  \BibitemOpen
  \bibfield  {author} {\bibinfo {author} {\bibfnamefont {A.}~\bibnamefont {Damle}}, \bibinfo {author} {\bibfnamefont {L.}~\bibnamefont {Lin}},\ and\ \bibinfo {author} {\bibfnamefont {L.}~\bibnamefont {Ying}},\ }\bibfield  {title} {\bibinfo {title} {Compressed representation of kohn–sham orbitals via selected columns of the density matrix},\ }\href {https://doi.org/10.1021/ct500985f} {\bibfield  {journal} {\bibinfo  {journal} {J. Chem. Theory Comput.}\ }\textbf {\bibinfo {volume} {11}},\ \bibinfo {pages} {1463} (\bibinfo {year} {2015})}\BibitemShut {NoStop}%
\bibitem [{\citenamefont {Damle}\ and\ \citenamefont {Lin}(2018)}]{Damle2018}%
  \BibitemOpen
  \bibfield  {author} {\bibinfo {author} {\bibfnamefont {A.}~\bibnamefont {Damle}}\ and\ \bibinfo {author} {\bibfnamefont {L.}~\bibnamefont {Lin}},\ }\bibfield  {title} {\bibinfo {title} {Disentanglement via entanglement: A unified method for wannier localization},\ }\href {https://doi.org/10.1137/17M1129696} {\bibfield  {journal} {\bibinfo  {journal} {Multiscale Modeling and Simulation}\ }\textbf {\bibinfo {volume} {16}},\ \bibinfo {pages} {1392} (\bibinfo {year} {2018})}\BibitemShut {NoStop}%
\end{thebibliography}%

\end{document}